\documentclass[aps,prr,twocolumn,superscriptaddress]{revtex4-2}

\usepackage{dcolumn}
\usepackage{bm}
\usepackage[T1]{fontenc}
\usepackage{graphicx}
\usepackage{xcolor}
\usepackage{tabularx}
\usepackage{multirow}
\usepackage{amsmath}
\usepackage{colortbl}
\usepackage{hyperref}
\hypersetup{colorlinks=true, linkcolor=blue, citecolor=red, urlcolor=magenta, pdftitle={Structural chirality in the extended/periodic inorganic solids: Review and perspectives}, pdfauthor={E. Bousquet}}

\usepackage{natbib}
\definecolor{gray}{rgb}{0.8,0.8,0.8}

\usepackage[normalem]{ulem}

\begin{document}

\title{Structural chirality and related properties in the periodic inorganic solids: Review and perspectives  }

\author{Eric Bousquet}
\affiliation{Physique Th\'eorique des Mat\'eriaux, QMAT, Universit\'e de Li\`ege, B-4000 Sart-Tilman, Belgium}

\author{Mauro Fava}
\affiliation{Physique Th\'eorique des Mat\'eriaux, QMAT, Universit\'e de Li\`ege, B-4000 Sart-Tilman, Belgium}

\author{Zachary Romestan}
\affiliation{Department of Physics and Astronomy, West Virginia University, Morgantown, WV 26505-6315, USA}

\author{Fernando G\'omez-Ortiz}
\affiliation{Physique Th\'eorique des Mat\'eriaux, QMAT, Universit\'e de Li\`ege, B-4000 Sart-Tilman, Belgium}

\author{Emma E. McCabe}
\affiliation{Department of Physics, Durham University, South Road, Durham, DH1 3LE, U. K.}

\author{Aldo H. Romero}
\affiliation{Department of Physics and Astronomy, West Virginia University, Morgantown, WV 26505-6315, USA}

\begin{abstract}
Chirality refers to the asymmetry of objects that cannot be superimposed on their mirror image.
It is a concept that exists in various scientific fields and has profound consequences.
Although these are perhaps most widely recognized within biology, chemistry, and pharmacology, recent advances in chiral phonons, topological systems, crystal enantiomorphic materials, and magneto-chiral materials have brought this topic to the forefront of condensed matter physics research. 
Our review discusses the symmetry requirements and the features associated with structural chirality in inorganic materials. 
This allows us to explore the nature of phase transitions in these systems, the coupling between order parameters, and their impact on the material's physical properties.
We highlight essential contributions to the field, particularly recent progress in the study of chiral phonons, altermagnetism, magnetochirality between others. 
Despite the rarity of naturally occurring inorganic chiral crystals, this review also highlights a significant knowledge gap, presenting challenges and opportunities for structural chirality mostly at the fundamental level, e.g., chiral displacive phse transitions and ferrochirality, possibilities of tuning and switching structural chirality by external means (electric, magnetic, or strain fields), whether chirality could be an independent order parameter, and whether structural chirality could be quantified, etc. 
Beyond simply summarising this field of research, this review aims to inspire further research in materials science by addressing future challenges, encouraging the exploration of chirality beyond traditional boundaries, and seeking the development of innovative materials with superior or new properties. 
\end{abstract}

\maketitle

\tableofcontents


\

\textit{“C’est la dissymm\'etrie qui cr\'ee le ph\'enom\`ene} (dissymmetry creates the phenomenon)”, Pierre Curie~\cite{Curie1894}

\ 

Our left and right hands are prototypical examples of chiral objects, as they are non-superimposable mirror images of each other. 
It is from this observation that the name “handedness” was historically given to define this mirror image property and the term chirality was coined for the first time by Lord Kelvin in 1894 \cite{cintas2007tracing,dezarnauddandine2007,Thompson1894} as it is constructed from the Greek word \textit{keir}=hand. 
Chirality emerges from a structure's specific symmetry characteristics, or the lack of them, which reflects the impossibility of an object coincident with its mirror reflection. 

\begin{figure}
    \centering
    \includegraphics[width=8.5cm,keepaspectratio=true]{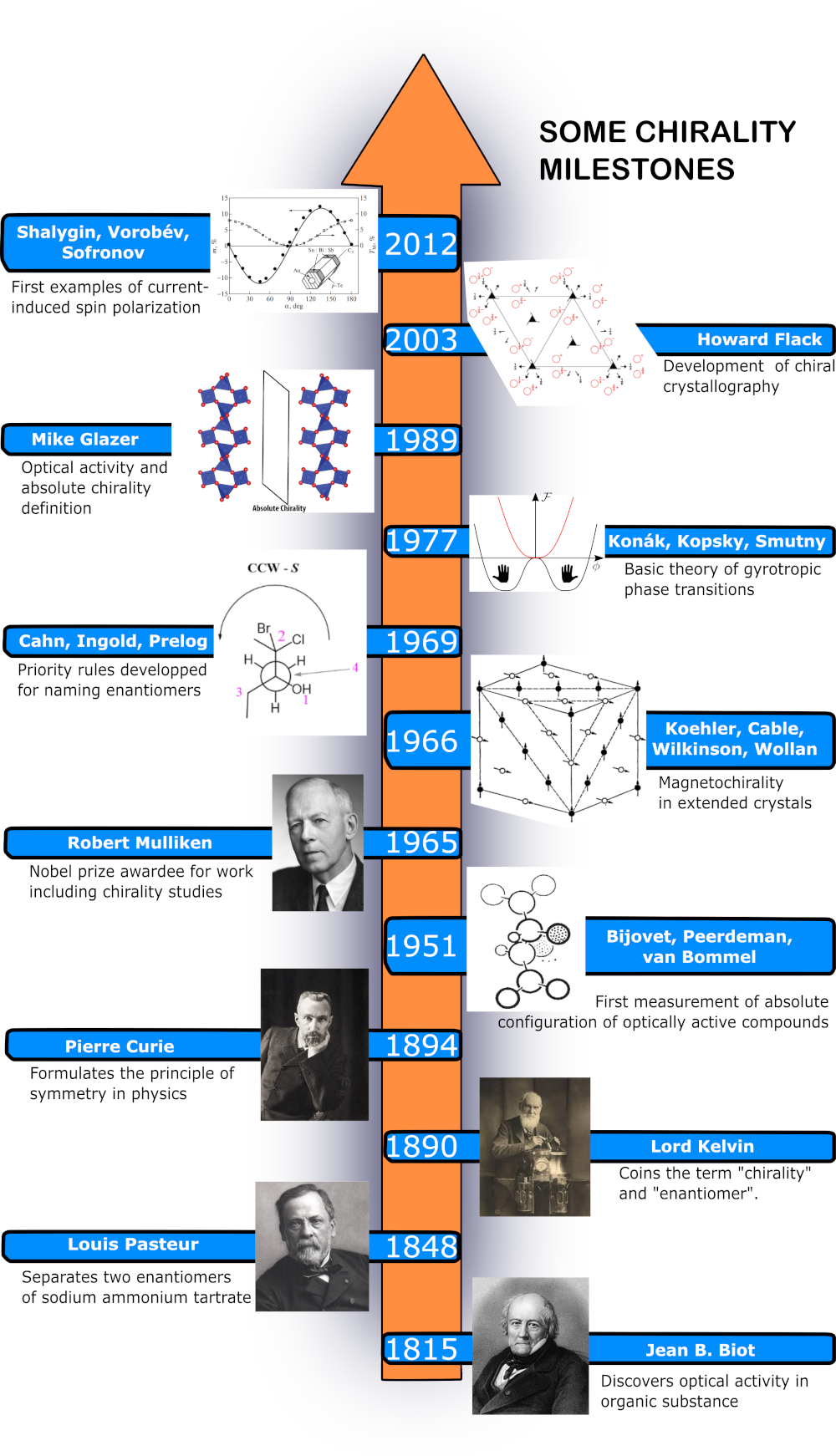}
    \caption{\label{figTL}{Timeline sketch highlighting important contributions to chirality and structural chirality.}}
\end{figure}

\begin{figure*}[hbt!]
    \centering
    \includegraphics[trim={0. 0cm 0 0cm},clip,width=1.0\textwidth]{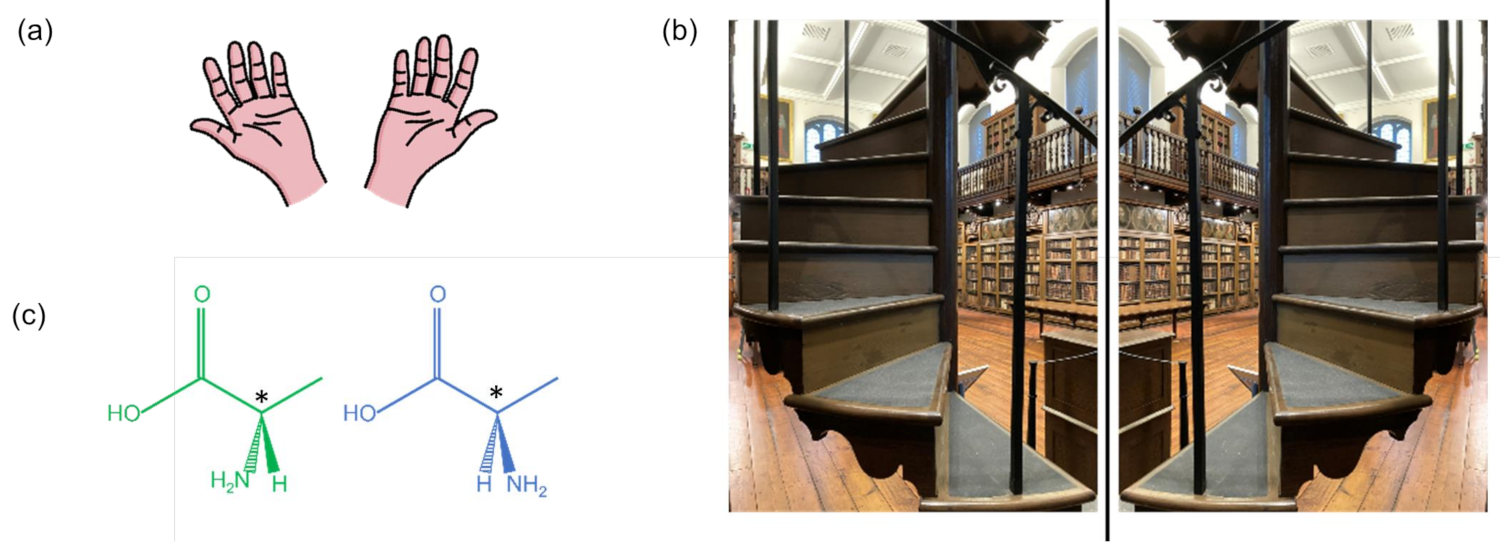}
    \caption{\label{fig1}{Illustration of the chiral nature of (a) left and right hands, (b) a spiral (helical) staircase (left) and its mirror image (right) (Cosin’s Library, Durham University); helices are hallmarks of structural chirality (Section \ref{optical_activity}), and (c) L- (green) and D- (blue) enantiomorphs of alanine (an essential amino acid) with chiral carbon noted *.}}
\end{figure*}

The history of chirality is marked by significant milestones, from the early observations of optical activity to its profound implications in various scientific disciplines, including chemistry, biology, and medicine~\cite{yamamoto2012comprehensive, mauskopf2006history}, some selected events are shown in Figure~\ref{figTL}. 
Jean-Baptiste Biot first observed the rotation of plane-polarized light by chiral substances in 1812~\cite{biot1812} and in 1848, Louis Pasteur suggested a molecular origin for this phenomenon~\cite{pasteur1848relations, cintas2007tracing,dezarnauddandine2007}. 
After this observation on crystal chirality, Pierre Curie highlighted the fundamental link between physical properties and the crystal structure and symmetry of materials~\cite{Curie1894}. 
“Dissymmetry” refers to the lack of improper rotation symmetry elements, which is the formal requirement for chirality; this was perhaps expressed more tangibly by Lord Kelvin in terms of the non-superimposable nature of mirror images of chiral objects~\cite{Thompson1894} (such as left and right hands, specific molecules and crystals, see Figure~\ref{fig1}).
Hence, Pasteur’s crucial breakthrough became the foundation of molecular and crystal chirality and led to many discoveries in material science, physics, chemistry, and biology. 
The term "chirality" itself was coined by Lord Kelvin in 1894~\cite{kelvin1894molecular} while different enantiomers or diastereomers of a compound were formerly called optical isomers due to their different optical properties. 
Initially, chirality was considered restricted to organic chemistry, but further advancements in the field overthrew this misconception. 
Chirality was again discussed mainly around 1891 when Emil Fischer published his famous paper concerning the configuration of glucose and its isomers, an article dealing with the stereochemistry of the monomeric building block glucose, a real milestone~\cite{fischer1909konfiguration}. 
This led to a recognition of the importance of molecular chirality in biological systems and the concept of
homochirality, a term introduced by Kelvin in 1904~\cite{kelvin1904baltimore}. Homochirality describes a process or system's preference for one enantiomer over the other, and it has significant implications for the origin of life and the emergence of biological homochirality~\cite{weller2024}.
Indeed, the concept of chirality is so crucial and fundamental in natural sciences that it appears at the root of many diverse fields like cosmology, particle physics, astrophysics, chemistry reactions, liquid crystals, optics, medicine or life itself; chirality is a trans-disciplinary property which manifests at any length scale in nature~\cite{morrow2017}.

Despite its broad interest and extensive study in biology, chemistry, and pharmaceutical medicines, chirality is relatively unexplored in solid-state extended crystals. 
Some seminal works exploring the origins of optical activity explained chirality in terms of helices in the crystal structures of extended solids~\cite{Glazer1986}. 
This insight gave a much deeper understanding of the optical activity and absolute configuration of chiral materials (e.g., $\alpha$-TeO$_2$~\cite{thomas1988crystal},
KLiSO$_4$~\cite{stadnicka1993} or 
$\alpha$-LiIO$_3$~\cite{stadnicka1985}) but also illustrated the challenges in determining the absolute configuration of a material and therefore understanding its consequences for physical properties.
A recent revival of chirality studies has taken place in high-impact topics like topological insulators~\cite{schroter2019,chang2018topological, sanchez2019topological}, spin-Hall effect~\cite{zhou2023}, spin/orbit chirality~\cite{mazzola2024signatures}, Skyrmions~\cite{muhlbauer2009, zhang2023chiral} or valleytronics~\cite{schaibley2016}, i.e. mostly linked to the electrons and their spins (e.g. fermions). 
This clearly shows that the shape of the electronic structure of chiral materials can be indirectly linked to chiral structural distortions~\cite{sanchez2019topological}, i.e., the amplitude of the structural distortion from achiral to chiral phases can dictate the electronic band dispersion and density of states. 
Still, these latter are not dependent on the handedness and, hence, are not chiral~\cite{fecher2022}.
On the other hand, in topological materials, the Berry curvature of the electronic structure is a hint of the topology and it is linked to handedness, though it cannot be a measure of the chirality itself~\cite{fecher2022}.
Another aspect of this revival is the very recent observations of chiral phonons (bosons) in 2D materials \cite{chen2019b,zhu2018, zhang2015}, in kagome lattices \cite{chen2019} or in the potential for orbital magnetic moments to be carried by chiral phonons~\cite{juraschek2019}. 
Studies of bulk chiral multiferroics \cite{johnson2012, johnson2013} or superlattices \cite{shafer2018emergent} have been carried out, but their findings have yet to be applied widely concerning chirality. 

Chirality is critical to developing and understanding the properties of these new and technologically important functional materials.
These recent developments have motivated this review in which we aim to discuss the state of the art of crystal chirality at the structural level and the consequences for physical properties, highlighting the related shortcomings in our understandings/widespread applications and posing open questions of high interest for materials science and technological applications.
This paper is focused on bulk inorganic periodic chiral crystals, e.g., we will not address chirality in molecular, hybrid~\cite{Lou2020,Jana2020}, or biological systems or at surfaces~\cite{jenkins2018} 
or in nano-hetero structures~\cite{junquera2023, shao2023, mccarter2022, prosandeev2013,yananose2021}, despite the importance of chirality in these fields.
Here, we focus on chiral bulk periodic inorganic crystals and their related response properties. 

We will start this review with a first section looking at how chirality is defined in different fields of research (Section~\ref{Sec:chirality-definitions}).
In the second section (Section~\ref{Sec:symmetry}), we give an overview of the general symmetry requirements in periodic solids as viewed through group theory and crystallography, with the distinction between crystals described by a chiral and enantiomorphic space group, and chiral crystals that are not described by one of these groups (Sohncke groups).
Because chirality is intimately linked to optical activity, we dedicate the following section to the field of optical activity (Section~\ref{optical_activity}).
After these general overviews, we will give a few examples of structural chirality and associated structural phase transitions (Section~\ref{Sec:Examples_Struct_chiral}) to illustrate the key concepts described in these sections. We also review the chiral crystals reported in existing inorganic crystal structure databases, demonstrating the relative scarcity of these inorganic chiral crystals and possible explanations for this.
This leads us on to explore the different responses of chiral crystals under applied magnetic fields (magneto chirality, chiral magnetostriction, chiral magneto gyration, etc), applied electric fields (electroporation, gyrotropy), and strain in Section~\ref{Sec:tuning-chirality-external-means}.
In each case, we will describe several crystal examples to illustrate the different responses that can be observed.
Structural chirality can naturally be extended to phonons that carry chirality, for which we dedicate an entire section (Section~\ref{sec:chiral-phonons}) with all the associated and unique properties that emerge from them. 
This allows us to consider displacive (and order-disorder) chiral phase transitions and, hence, how a chiral order parameter could be defined together with possibilities to control and flip chirality (ferrochirality) in Section~\ref{order-parameter}.
Afterward (Section~\ref{Sec:Quantifying chirality}), this notion of a chiral order parameter and displacive chiral phase transitions will bring us to discuss the open question regarding the quantification and measure of chirality in a periodic solid.
Lastly, before concluding remarks, we give some original and unexplored perspectives and properties of chiral crystals that would be promising and appealing to explore further in Section~\ref{Sec:conclusion}.

We note that a review on chirality in the solid state was recently published by Fecher, K\"ubler and Felser~\cite{fecher2022}. 
Its focus on chiral vs achiral space groups gives an excellent overview of the chiral structures of some elements and binary systems (as well as some more complex materials) and some consequences for electronic structure. 
Our review builds on this and goes beyond it, discussing structural phase transitions, the possibility of tuning chirality by applied fields, chiral phonons, quantification of chirality, and the consequences for topological and electronic structures. 
As such, these reviews are somewhat complementary, and their timing reflects the current interest in chiral extended solids.

\section{Definitions of chirality}
\label{Sec:chirality-definitions}
Given the wide-ranging significance of chirality across numerous branches of physics, multiple definitions have been established, each tailored to the particular field of study. 
While historically intertwined, it is now clear that chirality and handedness convey separate concepts. As outlined in the introduction, chirality refers to the lack of mirror symmetry in an object. 
Handedness, however, deals with categorizing these non-superimposable chiral entities into right or left orientations. 
To respect this distinction, mainly when referring to the traditional use of ``handedness'' raised by Lord Kelvin it is advisable to use quotation marks, as we will do in this review.

In this section, we shall discuss the different interpretations of chirality used in physics aiming to clarify and differentiate them, thus outlining the specific focus and scope of this review.

For instance, chirality is often encountered in magnetism and magnetic materials~\cite{togawa2012chiral, volkov2018mesoscale}. 
A paradigmatic example is the monoaxial helimagnetic order (along the $c$ axis) which associated Dzyaloshinskii-Moriya energy gain reads 
$\mathbf{D}\cdot(\mathbf{S}\times\mathbf{\nabla}_j\mathbf{S})$\cite{Kishine2022}. Since $\mathbf{D}$ and $\mathbf{S}\times\mathbf{\nabla}_j\mathbf{S}$ are (time-even) axial and polar vectors respectively, their product constitutes a chiral quantity.
Similarly, a vector chirality $\kappa$ can be defined as the sum of the cross products of neighbouring spins,
\begin{equation}
    \kappa_{ij}=S_{i}\times S_{j}
\end{equation}
The resulting vector is helpful to quantify the chirality of a group or network of spins e.g., in a ferrochiral system (with an overall chiral vector) compared with an antiferrochiral system (with alternating chiral vectors, which overall gives a “staggered” vector chirality (as observed for RbFe(MoO$_{4}$)$_{2}$ with staggered triangular chirality)~\cite{hearmon2012electric}. However, this vector chirality may not be a good indication of proper chirality ~\cite{cheong2022, simonet2012magnetic, TANYGIN2011, Barron1986} (and instead reflects ``false chirality'' in the notation of Barron; it is even concerning time-reversal symmetry.)~\cite{Barron1986}. Nevertheless, since it reflects the spin dynamics, it can give insight into the spin current and other transport properties (see below)~\cite{simonet2012magnetic, Ishizuka2018, Ishizuka2020}.
On the other hand, scalar spin chirality $\chi_{ijk}$ is defined as follows: 
\begin{equation}
    \chi_{ijk}=S_{i}\cdot (S_{j} \times S_{k})
\end{equation}
and can be thought of in terms of the solid angle between three neighboring spins. 
It is non-zero for non-co-planar spin arrangements and changes signs under time reversal but is invariant under spin rotation~\cite{Kawamura2011} (important for coupling to polarization in multiferroics)~\cite{chan2022neutron}.
This language is useful for describing, for example, the spin arrangement with uniform scalar chirality of the spin ice arrangement of Pr$^{3+}$ moments Pr$_{2}$Ir$_{2}$O$_{7}$ and its anomalous Hall effect~\cite{Udagawa2013}.

In other disciplines like particle physics, chirality assumes a more mathematical characterization. 
There, particle chirality is discerned through its transformation within either a right- or left-handed representation of the Minkowski spacetime isometries~\cite{Schopper-20}. 
Furthermore, a clear differentiation between helicity, defined as the projection of a particle's spin in the direction of its motion, and chirality is delineated. 
For massless particles, both definitions coincide, and handedness and chirality are thus equivalent. 
Recently, some authors have leveraged this precise definition of helicity, employing Pauli matrices to mathematically describe chirality in odd-frequency Cooper pairs~\cite{Tamura-19}, as well as to quantify chirality polarizations in topological insulators and superconductors~\cite{daido2019}. 
Furthermore, in Ref.~\cite{Vavilin-22} this definition of the helicity operator has been extended to the space of solutions of Fourier-transformed Maxwell's equations expressed as $\frac{1}{\textbf{k}}\nabla\times$. 
Subsequently, they achieved a general distinction between enantiomers through an investigation of light-matter interaction.

Lastly, in topological materials, chiral crystals are identified as belonging to one of the 65 Sohncke space groups, which exclusively contain orientation-preserving operations, allowing them to be classified by a specific handedness. In these space groups, structural chirality imbues the materials with various intriguing properties, including natural optical activity, negative refraction in metamaterials, non-reciprocal effects like magneto chiral birefringence of light or electronic magneto chiral anisotropy, and the formation of chiral magnetic textures such as helices and skyrmions. As we discuss in this review, the idea of handedness should not be identified as absolute chirality and the existence of enantiomorphic pairs. For example, Schroter \textit{et. al.},~\cite{schroter2019} the presence of Weyl fermions in recently discovered Weyl semimetals are discussed, where the crystal structure and the electronic wavefunctions exhibit chirality at point-like two-band crossings of the quasiparticle dispersion, and chirality is used in the sense of handedness.

On the other hand, in crystallography and condensed matter physics, which will be the focus of this review, a more geometric approach akin to that employed in chemistry has been adopted. 
Under this framework, a crystal is deemed chiral if it cannot be perfectly overlaid onto its mirror image. This requires the object to be ``dissymmetric'', i.e., lacking proper rotation symmetry elements, as discussed in Section ~\ref{Sec:symmetry}. Additionally, a distinction between handedness, rotational direction, and chirality has been elucidated, as extensively discussed in Section~\ref{Sec:Quantifying chirality}.
Subsequently in Section~\ref{Sec:symmetry}, we will explore the prerequisites of symmetry for a crystal to exhibit chirality and detail experimental methodologies for their characterization in the ensuing sections. In Section~\ref{order-parameter} we shall also discuss different sources of chirality as according to Hlinka, chirality may also arise as the product between a vector $\mathbf{V}$ and an axial $\mathbf{A}$ representation. Consequently, it is reasonable to consider such a product as an interesting quantity to track chirality.

\section{Crystallography and symmetry requirements for chirality in crystals}
\label{Sec:symmetry}

Given the importance of chirality in extended solids for functionality, relatively little attention has been paid to this “structural chirality” especially compared with molecular systems with discrete point symmetry. 
The formal definition of chirality in terms of dissymmetry given by Curie~\cite{Curie1894} is worth expanding upon: whether considering point group or space group symmetry, this requirement means the object must possess no improper rotation symmetry elements. 
Mirror planes and centers of inversion symmetry can be described by improper rotations (e.g. considering the operations of crystallographic space groups, the inversion operation could be described by a $\bar{1}$ rotation-inversion operation; the mirror plane by a $\bar{2}$ roto-inversion operation),~\cite{BurnsGlazer1990} leading to the (often used but incomplete) statement that a requirement for chirality is the absence of mirror planes and inversion symmetry. 
A consequence of this lack of improper rotation symmetry elements is that when two chiral objects (referred to as enantiomorphs in extended systems) are considered, their mirror images are non-superimposable, hence the “handedness” referred to by Lord Kelvin.~\cite{Thompson1894}

Crystallographic symmetry operations can be categorized into those that preserve handedness (i.e., map a right-handed helix onto a right-handed helix) referred to as “operations of the first kind” and those that reverse handedness (i.e., map a right-handed helix onto a left-handed helix) – “operations of the second kind”.~\cite{nespolo2018crystallographic} 
Of the 230 crystallographic space groups, 165 contain improper symmetry operations (operations of the second kind) and describe achiral crystals. 
An achiral crystal described by one of these 165 space groups could be grown from a racemic mixture (i.e., a mixture containing both enantiomer forms), and the improper rotation operations of the space group would relate the two enantiomeric molecules in the crystal structure, giving a structure with no overall chirality.~\cite{Nespolo2021}

The Sohncke groups include 11 pairs of enantiomorphic space groups, i.e., 22 chiral space groups (Table \ref{spacegroups}). These chiral space groups are characterized by each containing a screw axis of the form $n_{p}$ or $n_{n-p}$ (but not both!) where $p\ne \frac{n}{2}$.~\cite{Nespolo2021} 

\begin{table}[h!]
\begin{center}
\begin{tabular}{ |p{2.5cm}|p{3cm}|p{2.6cm} | }
\hline
\multicolumn{3}{|l|}{\textbf{Classes compatible with optical activity}} \\
\hline 
\multicolumn{3}{|l|}{$\overline{4}$} \\
\hline
\multicolumn{3}{|l|}{$m$} \\
\hline
\multicolumn{3}{|l|}{$mm2$} \\
\hline
\multicolumn{3}{|l|}{$\overline{4}2m$} \\
\hline 
\textbf{ Classes compatible} & \textbf{Non-enantiomorphic} &  \textbf{Enantiomorphic (chiral)} \\
\textbf{with chirality} & \textbf{space groups} & \textbf{space groups} \\
\hline
\centering 1 & \centering \cellcolor{gray}$P1$ 
      &   \\
\hline
\centering 2 & \centering\cellcolor{gray}$P2$, $P2_{1}$, $C2$ 
      &  \\
\hline
\centering 3 & \centering\cellcolor{gray}$P3$, $R3$ 
     & \cellcolor{gray}$P3_{1}$, $P3_{2}$ \\
\hline
\centering 4 &\cellcolor{gray}$P4$, $P4_{2}$, $I4$, $I4_{1}$
   & \cellcolor{gray} $P4_{1}$, $P4_{3}$  \\
\hline
\centering 6 & \centering \cellcolor{gray}$P6$, $P6_{3}$        
   & \cellcolor{gray}$P6_{1}$, $P6_{5}$,  \\
  & \cellcolor{gray}     
       & \cellcolor{gray}$P6_{2}$, $P6_{4}$ \\
\hline
\centering 422  & \centering\cellcolor{gray}$P422$, $P42_{1}2$,  $P4_{2}22$, 
        & \cellcolor{gray}$P4_{1}22$, $P4_{3}22$ \\
    & \cellcolor{gray}$P4_22_12$,  $I422$, $I4_122$ 
        & \cellcolor{gray}$P4_12_12$, $P4_32_12$ \\
\hline
\centering 222 & \centering\cellcolor{gray}$P222$, $P222_{1}$,   
                & \cellcolor{gray}   \\
    & \centering\cellcolor{gray}$P2_{1}2_{1}2$, $P2_{1}2_{1}2_{1}$, 
                 & \cellcolor{gray}\\
    &  \centering\cellcolor{gray} $C222$, $C222_{1}$,  
                 & \cellcolor{gray}\\
    &  \centering\cellcolor{gray}$F222$, $I222$,  
                 & \cellcolor{gray}\\
    &   \centering\cellcolor{gray}$I2_{1}2_{1}2_{1}$  
                 &  \cellcolor{gray}\\
\hline
\centering 622 & \centering\cellcolor{gray}$P622$, $P6_{3}22$ 
         & \cellcolor{gray}$P6_{1}22$, $P6_{5}22$,  \\
    &  \cellcolor{gray} 
         &  \cellcolor{gray}$P6_{2}22$, $P6_{4}22$ \\ 
\hline
\centering 32 & \centering\cellcolor{gray}$P312$, $P321$, $R32$  
         &  \cellcolor{gray}$P3_112$, $P3_212$\\
   &   \cellcolor{gray}                    
         & \cellcolor{gray}$P3_121$, $P3_221$\\
\hline
\centering \it{T} & \centering\cellcolor{gray} $P23$, $F23$, $I23$, 
          &  \cellcolor{gray}  \\
       &  \centering\cellcolor{gray} $P2_{1}3$, $I2_{1}3$ 
           & \cellcolor{gray} \\ 
\hline
\centering \it{O} & \centering\cellcolor{gray} $P432$, $F432$, $I432$,  
        & \cellcolor{gray} $P4_{1}32$, $P4_{3}32$ \\
       & \centering\cellcolor{gray} $P4_{2}32$, $F4_{1}32$, 
         & \cellcolor{gray} \\
       & \centering\cellcolor{gray} $I4_{1}32$ 
        & \cellcolor{gray} \\
\hline
\end{tabular}
\caption{Acentric crystal classes compatible with optical activity; the 65 Sohncke space groups are highlighted in grey; the 22 enantiomorphic (chiral) space groups are listed in the right-hand column.}\label{spacegroups}
\end{center}
\end{table}

For example, of the cubic class $O$, space groups $P4_1 32$ and $P4_3 32$ are an enantiomorphic pair containing $4_1$ and $4_3$ screw axes, respectively. Space group $P4_2 32$ does not have an enantiomorphic pair: it contains a $4_2$ screw axis and $4_{(4-2)}=4_2$, and so opposite enantiomorphic structures would be described by the same space group. Combining the $4_1$ screw axis with centering (e.g., $F, I$) gives the $4_3$ axis, so these two screw axes of opposite handedness coexist, incompatible with chirality. Hence, space groups $I4_1 32$ and $F4_1 32$ are not enantiomorphic.~\cite{Nespolo2021} 

These screw axes of opposite handedness, giving helices in the crystal structure, are a hallmark of the structures described by these 22 chiral structures – so much so that developing spirals at the nanoscale in inorganic materials is a recent development towards preparing functional chiral materials.~\cite{Lu2023}

A necessary consequence for functional materials using chirality is that the crystal packing in structures described by one of the 22 enantiomorphic space groups is chiral, meaning that either chiral or achiral objects (molecules or structural building units) can be packed in a chiral fashion (i.e., about one of the chiral helices) to give a chiral crystal.~\cite{nespolo2018crystallographic, Nespolo2021} 
The term “structural chirality” describes this chirality due to the crystal packing, distinguishing it from any chirality inherent in a discrete object (e.g., a molecular crystal made of chiral molecular building blocks).~\cite{Glazer1989} 
A pair of enantiomorphic space groups will describe the two enantiomorphs of opposite chirality for the same chemical composition and stoichiometry.

The remaining 43 Sohncke space groups that are not enantiomorphic will preserve the chirality of a chiral object (e.g., pack a chiral object such as a molecular enantiomorph) without symmetry operations generating the opposite enantiomorph, giving a chiral crystal. However, the packing in these 43 Sohnke space groups will not give chirality for achiral objects. 
Switching the sense of chirality of a crystal described by one of these 43 non-enantiomorphic Sohncke space groups does not change the space group symmetry – two equivalent crystals of opposite chirality are described by the same space group (see, for example, Pb$_{5}$Ge$_{3}$O$_{11}$ described in Section~\ref{Sec:Examples_Struct_chiral}).

\section{Chirality and the origins of optical activity}
\label{optical_activity}
Chirality and optical activities are intimately linked.
Indeed, the history of chirality dates back to the early 19th century, when the first observation of optical activity, the ability of a chiral substance to rotate the plane of polarization of light, was made by the French physicist Jean-Baptiste Biot in 1812~\cite{mason1986origin}. 
He discovered that certain organic substances, such as tartaric acid and its salts, could rotate the plane of polarized light~\cite{biot1812}. 
A similar observation was reported the year before by Arago~\cite{arago1811},
who noticed that polarized sunlight gave color changes on an image when an analyzer crystal was rotated.  
However, the significance of this discovery was not recognized at the time because of the lack of a microscopic mechanism for the phenomenon. 
Indeed, at that time, the atomic theory of matter was still under development, and the concept of molecular structure was not yet fully elaborated. 
The significance of optical activity went beyond a mere curiosity in the interaction of light with matter, pointing towards a fundamental asymmetry in nature at the molecular level. 
Pasteur's work, some decades later, demonstrated the connection between the rotation of polarized light and the molecular structure of the substances causing this rotation~\cite{mason1986origin}. 
Pasteur's separation of the two forms of tartaric acid crystals and his subsequent experiments showed that the ability to rotate light was related to the shape of the molecules and their three-dimensional arrangement.  
He manually separated the two types of crystals with a microscope and found that solutions of each of the two types of crystal rotated polarized light in opposite directions.
This experiment proved to be the first time someone had separated two enantiomers or mirror-image forms of a molecule.

\begin{figure}
    \centering
    \includegraphics[width=8.5cm,keepaspectratio=true]{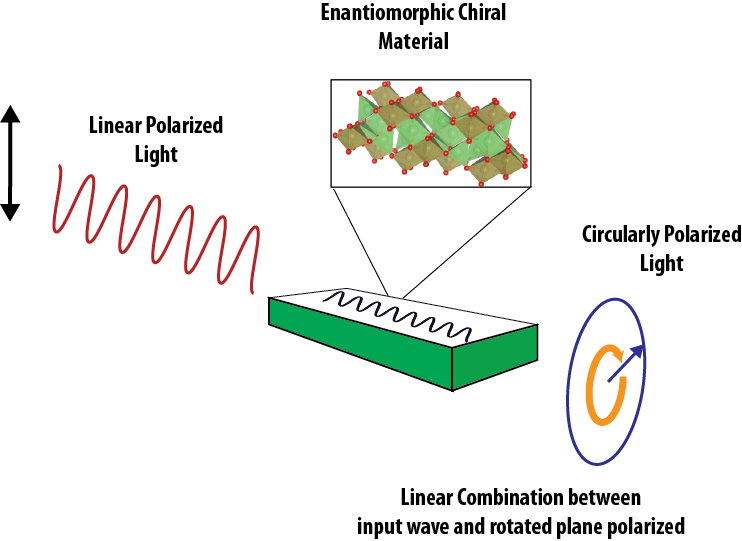}
    \caption{\label{figOC}{Illustration of Optical Activity through Light Rotation in Chiral Crystals. }}
\end{figure}
Pasteur's work laid the foundation for understanding that the molecules in living organisms are chiral and that this chirality is critical to how they function.
For example, in biochemistry, the chirality of amino acids is crucial because only one enantiomer of amino acids is used to make proteins in living organisms~\cite{maddox1998remains,Soai2008}.
Thus, while Biot's discovery was significant, its full importance was not appreciated until the underlying principles of molecular chirality were understood.

The theoretical explanation for optical activity and chirality was developed further in the late 19th and early 20th centuries, which we depict in Figure~\ref{figOC}. 

Natural optical activity (NOA) arises from the spatial dispersion of the dielectric response or optical conductivity for conductors. This results from finite light speed, allowing currents generated at nonzero wavevectors to affect the local response dependent on frequency. By expanding the non-local dielectric response in powers of $q$ to first order, the natural optical activity can be seen to arise as a third-rank tensor as
\begin{equation}
    \epsilon_{ij}(\omega,\mathbf{q}) = \epsilon_{ij}(\omega,\mathbf{q}=0) + i\lambda_{ijk}(\omega)q_{k} + ...
\end{equation}
Thus the NOA can be obtained from
\begin{equation}
    \lambda_{ijk}(\omega) =-i \left .\frac{\partial \epsilon(\omega,\mathbf{q})}{\partial \mathbf{q}}\right\vert_{\mathbf{q}=0}.
\end{equation}

Sometimes the NOA is written as a rank two tensor called the gyration tensor as

\begin{equation}
    g_{ij}(\omega) = \frac{1}{2}\varepsilon_{kli}\lambda_{klj}(\omega).
\end{equation}

Considering the non-negligible contribution to the dielectric response, the crucial role in stereochemistry, and the potential for applications, several theoretical efforts have been made to provide accurate predictive tools to compute the natural optical activity. 
Early theoretical efforts to understand the phenomenon based on semiclassical theories of electromagnetism~\cite{Condon1937}. 
First principles methods have been developed from these theories using sum over states formulas~\cite{Pederson1995}, linear response~\cite{Varsano2009}, and real-time propagation~\cite{Yabana1999}. 
While these approaches are successful in molecules, the origin dependence of the theory is unsuitable for crystalline materials. 
Band theory formalism in the long-wavelength limit and neglecting local field effects~\cite{Zhong1992,Zhong1993,Malashevich2010,Pozo2023} combined with a first principles sum over states approach~\cite{Wang2023OpticalActivity} extended the theory of optical activity to crystals. While this approach is successful for some materials, it significantly underestimates the optical activity when local field effects cannot be ignored~\cite{Jonnson1996}. A framework based in density functional perturbation theory has been proposed recently~\cite{zabalo2023} to address these issues.

In 1874, Jacobus Henricus van 't Hoff and Joseph Le Bel independently proposed that the carbon atom could form four bonds directed toward the corners of a tetrahedron with the potential to bond to four different species). 
This pseudo-tetrahedral arrangement could create two different spatial arrangements (isomers) of atoms, which explains the existence of enantiomers~\cite{ramberg2017chemical}.
The recognition that many biological molecules are chiral led to the development of stereochemistry, which studies the spatial arrangement of atoms in molecules and their impact on their behavior and reactions.
In pharmacology, the chirality of drugs is paramount because different enantiomers of a drug can have very different biological activities (one may be therapeutic, while the other could be inactive or even harmful, as was tragically demonstrated in the case of thalidomide in the 1950s and 1960s)~\cite{ceramella2022}. 
In the case of periodic systems, the history of optical activity is strongly tied to the development of crystallography, and an understanding of symmetry through the application of group theory (see Section ~\ref{Sec:symmetry}). Experimental crystallographic methods allowed scientists to determine the arrangement of atoms within a crystal and understand how this arrangement affects the crystal's properties, including optical activity.
A good review of the approaches to describe optical activity can be found in Ref.~\cite{Glazer1986}.

Following the Neumann~\cite{authier2013international} symmetry principle, optical activity can only be exhibited by non-centrosymmetric structures. 
This is because an inversion center in a crystal structure would negate the conditions necessary to manifest optical activity. 
Therefore, optical activity can demonstrate the lack of inversion symmetry in a crystal structure (and these measurements can be complemented by anomalous X-ray scattering to determine the absolute configuration of the structure~\cite{caticha1981anomalous}).
Of the 21 acentric crystal classes, 15 are compatible with optical activity, but four of these classes ($\bar{4}$, $m$, $mm2$, $\bar{4}2m$) contain symmetry operations that are not compatible with chirality. 
Of the 15 classes consistent with optical activity, 11 are compatible with chirality and give the 65 Sohncke space groups, which contain only operations of the first kind (Table 1).~\cite{Halasyamani1998, Newnham2005, IUCr-online}, and so the optical activity we associate with chirality is not confined to only symmetries compatible with enantiomorphism,~\cite{claborn2008, Halasyamani1998}.
Crystals possessing D$_{2d}$ ($\overline{4}2m$) symmetry, or any of its non-enantiomorphic subgroups such as S$_4$ 
($\overline{4}$), C$_{2v}$ ($2mm$), or C$_s$ ($m$), can rotate the plane of polarized light when it is incident in a general direction~\cite{claborn2008}.

Optically active materials interact with light and have distinct refractive indices for left-handed circularly polarized light (LCP) and right-handed circularly polarized light (RCP).
Generally, the refractive index of a material can be expressed as 
$n = n_0 \pm \gamma_1 + i(n_2 \pm \gamma_2)$, where 
$\gamma_1$ and $\gamma_2$ depend on the product of the light's frequency and the gyrotropic constant of the material. 
Consequently, each circular polarization state, LCP, and RCP is characterized by different refractive indices, denoted as 
$n^L$ and $n^R$, respectively.
From these refractive indices, one can define the optical rotation angle 
$\phi = (2 \pi/\lambda) (n_0^L - n_0^R)d$, where $\lambda$
represents the wavelength of the incident light, and $d$
is the optical path length through the material. 
The terms $n_0^L$ and $n_0^R$ refer to the real parts of the refractive indices for LCP and RCP light, respectively. 
Similarly, the ellipticity $\Psi$, 
which quantifies the extent to which the polarization of light becomes elliptical after passing through an optically active material, can be defined as 
$\Psi = (\pi/\lambda)(n_2^L-n_2^R)$, with $n_2^L$ and $n_2^R$
being the imaginary parts of the refractive indices corresponding to LCP and RCP light absorption.
In most cases, the ellipticity is small enough for the rotation
to be measured as if the light were still plane-polarized, and what is measured experimentally is the optical rotatory dispersion (ORD). 
The effects of the imaginary parts of the index of refraction are usually determined
by measuring the absorption of LCP and RCP
radiation separately and calculating the ratio as
\[
\frac{I_R-I_L}{I_R+I_L},
\]
where $I_R$ and $I_L$ are the intensities of right and left handed polarized light, respectively.
As a function of frequency, this quantity is known as circular dichroism (CD). 
ORD measures the change in a substance's optical rotation as a function of the wavelength of light. At the same time, CD is a spectroscopic technique that measures the difference in the absorption of LCP versus RCP as it passes through a chiral substance.

Therefore, optical activity can be used to distinguish between enantiomorphs of chiral materials as opposite enantiomorphs will give opposite rotations of the plane of polarization, as in SiO$_2$ (Quartz), HgS (Cinnabar), NaClO$_3$ (which is a chiral crystal), AlPO$_4$ (Berlinite)~\cite{Glazer1986}, Te~\cite{Nomura1960}, Bi and many different chiral magnetic crystals such as MnSi, FeGe, Cu$_2$OSeO$_3$~\cite{Versteeg2016}, 
Fe$_{(1-x)}$Co$_x$Si, CrNb$_3$S$_6$, $\beta$-Mn, ~\cite{Glazer1986} etc. 
Chiral crystals are being actively researched for their nonlinear optical properties, which are of significant interest in the field of nonlinear optics, in  particular, due to the recent observation of large chiral-specific second harmonic generation (SHG) response~\cite{byers1994second, haupert2009chirality}.
One of the most innovative applications of optical activity lies in chiral catalysis~\cite{baiker1998chiral, song2002supported,  mallat2007asymmetric}.
Here, heterogeneous catalysts exploit the chiral surfaces of crystal materials to drive enantioselective reactions, which are directly influenced by the material's optical activity. 
This underscores the importance of optical activity in catalysis and provides further motivation for synthesizing enantiomerically pure compounds, a critical aspect of drug development and various chemical manufacturing processes.

In industrial contexts, the measurement of optical activity is employed for quality control, particularly in the manufacturing of optical devices \cite{nafie2011vibrational, zheng2020photoprogrammable, bahar2008detection}. 
Furthermore, manipulating light's polarization state in optoelectronic devices leverages the unique properties of chiral crystals, enhancing the performance of optical sensors in communication devices and other applications that empower a two-fold improvement of the rate of information transport through polarization multiplexed fiber optic waveguides~\cite{Rochat2004}, quantum communication~\cite{Gisin2007,Liao2018,Humphreys2018}, optical integrated circuits~\cite{Quan2019,Fang2019}, on chip photodetectors~\cite{Yang2013,Li2015Circularly,Shulz2019} and quantum computing~\cite{Service2001,Lodahl2017}. Furthermore, these magneto-optical (MO) effects and the use of CPL form the basis for technological breakthroughs, such as all-optical switching in spintronic devices~\cite{Wang2019SpinOptoelectronic,Abendroth2019,Long2018Spin}, CPL LEDs~\cite{Han2018,Wu2019,Zinna2017,zinna2020,Yang2013Induction}, photoresponsive displays~\cite{Bisoyi2014,Heffern2014}, and bioresponsive imaging~\cite{kissick2011second,Lee2013}. Other devices also use magneto-optical effects like Faraday rotation and Kerr rotation. 
While these rotate the plane of polarization, these effects are unrelated to the material structure or symmetries (see Section ~\ref{magnetism} for magnetic-symmetry dependent effects).

\subsection{Chiral axis and absolute chirality}
\label{sec:absolute_chir}

The optical activity of a material (its ability to rotate the plane of polarised light) is defined by its magnitude and its sign i.e., the direction of rotation: viewed along the beam direction towards the light source, if the sense of rotation of the polarisation is clockwise, the material is described as left-handed (levorotatory), and right-handed (dextrorotatory) if the sense of rotation is anticlockwise, as discussed in Section ~\ref{optical_activity}.
The sense of rotation should be related to the structural helix of the most polarisable atoms in the crystal structure.~\cite{Glazer1989, Glazer1986} 
However, it can be challenging to determine the true chirality (i.e. the "absolute configuration" of molecular systems, or "absolute structure" or "absolute chirality" for crystalline materials~\cite{IUCr-online, Glazer1989}) of a crystal structure~\cite{Flack2008}, leading to some ambiguity in reported structures and confusion in the link between the chiral crystal structure and its measured physical properties (including optical activity, and polarization where relevant).~\cite{Glazer1989, Glazer1986} 
This has undoubtedly contributed to the difficulty of quantifying chirality (see Section \ref{Sec:Quantifying chirality}). 
Glazer and Stadnicka reported that chiral axes for relevant crystal classes were analogous to polar directions in polar structures~\cite{Glazer1989}.
This concept is perhaps most relevant when exploring electrogyration, the change in the optical activity of a ferroelectric crystal by the application of an electric field (see Section \ref{sec:Gyroelectricity})~\cite{Konak1978}.
This practical meaning of absolute chirality and the resulting link between structure (and its handedness) and the resulting physical properties should not be confused with the mathematical concepts of ``absolute'' and ``relative'' chirality derived from graph theory by Le Guennec~\cite{leguennec1998,leguennec2000a,leguennec2000b}. 
See also Ref.~\cite{simonet2012magnetic} for a detailed discussion about the experimental determination of the absolute chirality via neutron scattering probes.

\section{Examples of structural chirality and associated phase transitions}
\label{Sec:Examples_Struct_chiral}
In this section, we give specific examples of crystals exhibiting non-reconstructive phase transitions involving structural chirality, discussing their  optical activity where relevant.
We discuss MgTi$_2$O$_4$, K$_3$NiO$_2$,  CsCuCl$_3$, Na$_2$Ca$_2$Si$_3$O$_9$ and Ag$_4$P$_2$O$_7$ with low symmetry (chiral) structures described by enantiomorphic space groups, and Pb$_5$Ge$_3$O$_{11}$ and Ba(TiO)Cu$_4$(PO$_4$)$_4$ with low symmetry structure described by a non-enantiomorphic Sohncke space group.

\subsection{\texorpdfstring{MgTi\textsubscript{2}O\textsubscript{4}}{MgTi2O4}}
\label{MgTiO4_text}
\begin{figure}
\centering
\includegraphics[width=8.5cm,keepaspectratio=true]{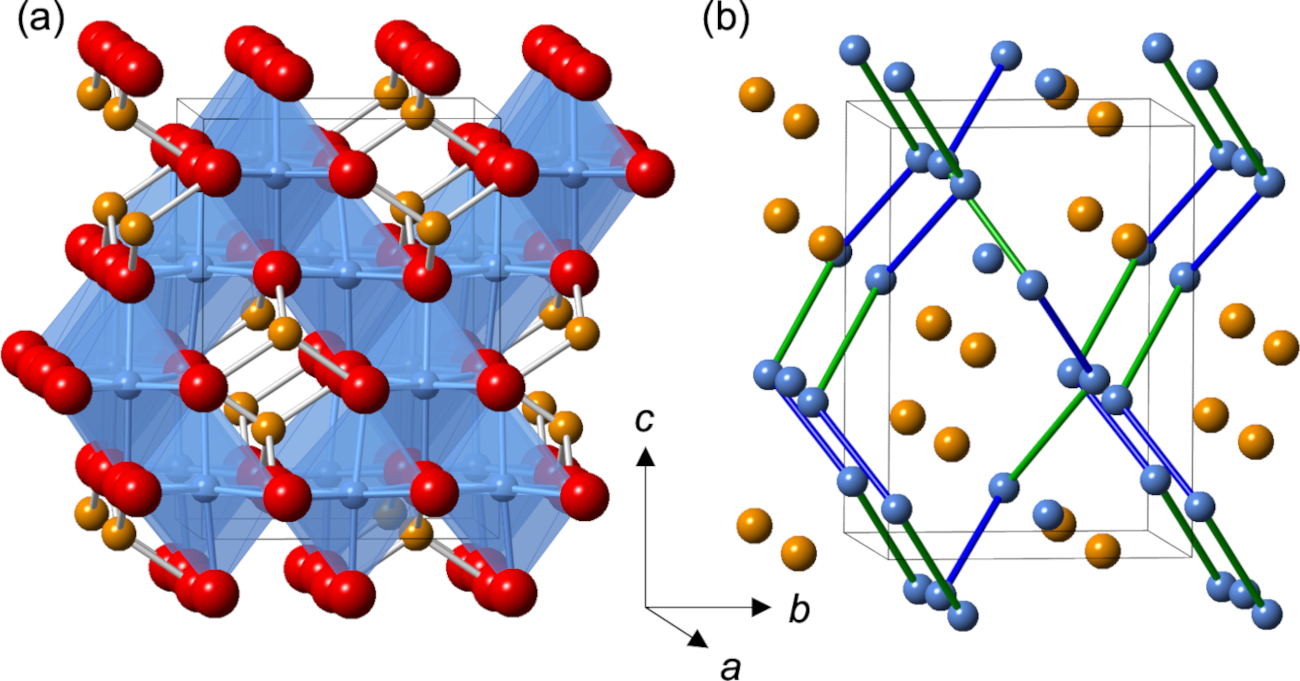}
 \caption{Schematic showing the crystal structure of the chiral $P4_{1}2_{1}2$ phase of spinel MgTi$_{2}$O$_{4}$ showing (a) TiO$_{6}$ octahedra (blue) and MgO$_{4}$ units and (b) highlighting shortest (dimerized) and longest Ti – Ti bonds in blue and green (with oxide ions omitted for clarity); Mg, Ti and O sites are shown in orange, blue and red, respectively.  Adapted from ~\cite{Schmidt2004}.}
 \label{fig:MgTi2O4}
\end{figure}

Above 260 K, MgTi$_{2}$O$_{4}$ adopts a normal spinel structure of cubic symmetry ($Fd\bar{3}m$). On cooling, it undergoes a transition from a metallic state to a spin-singlet insulating state, driven by the dimerization of pairs of $3d^{1}$ Ti$^{3+}$ ions. 
This metal-insulator transition is accompanied by a cubic to tetragonal phase transition~\cite{Isobe2002}.
Variable temperature synchrotron and neutron diffraction experiments indicate that the symmetry is lowered to $P4_{1}2_{1}2$ (or its enantiomorphic pair $P4_{3}2_{1}2$), with a change in the size of the unit cell to accommodate ordering of Ti–Ti dimers with shorter bonds. 
This symmetry lowering gives helical chains of long and short Ti – Ti bonds (Figure ~\ref{fig:MgTi2O4}). 
Variable temperature neutron powder diffraction data indicates that this is an abrupt transition, suggesting that it is of a first-order kind~\cite{Schmidt2004}.

The structural phase transition can be described by the higher-dimensional $X_{4}$ irrep and as both enantiomorphs must be degenerate, the transition from the achiral phase to either enantiomorph must be described by the same irrep. 
The transitions to the two enantiomorphs differ by the order parameter direction: $(0,a; 0,0; 0,0)$ gives the $P4_{1}2_{1}2$ enantiomorph, whilst $(a,0; 0,0; 0,0)$ gives the $P4_{3}2_{1}2$ enantiomorph~\cite{Campbell2006, stokes2007isotropy}.
In summary for MgTi$_{2}$O$_{4}$:
\begin{itemize}
\item{The phase transition from achiral structure to a structure described by an enantiomorphic space group is first order and is accompanied by a change in unit cell volume;}
\item{The phase transition is described by an irrep ($X_4$) that is at least two-dimensional;}~\cite{Ivanov2011}
\item{The transition to the two enantiomorphs (described by a pair of enantiomorphic space groups) are described by the same irrep but with different order-parameter directions.}
\end{itemize}

\subsection{\texorpdfstring{K\textsubscript{3}NiO\textsubscript{2}}{K3NiO2}}
\label{KNO_section}

\begin{figure}
\centering
\includegraphics[width=8.5cm,keepaspectratio=true]{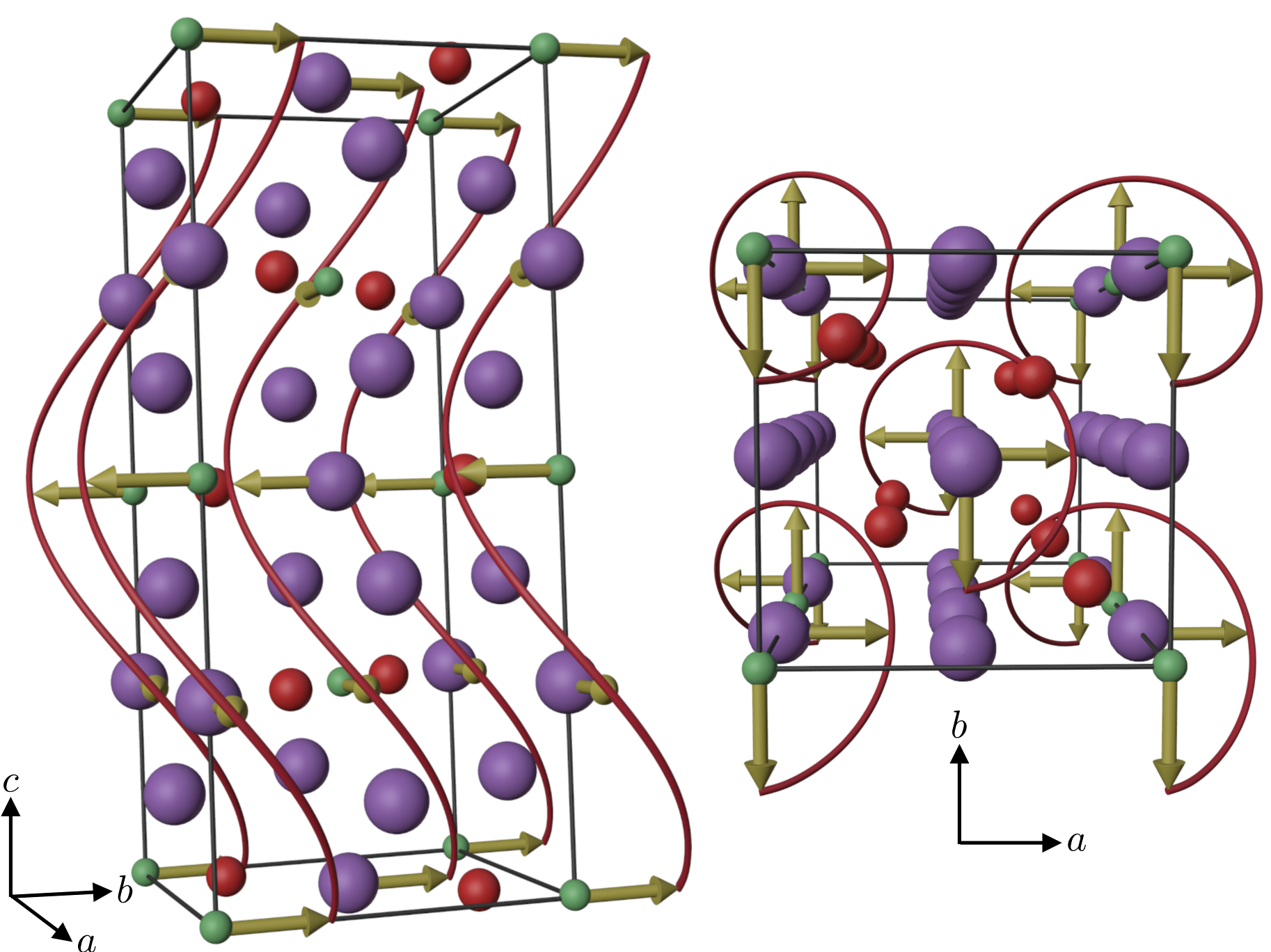}
 \caption{
 Schematic view of the symmetry allowed atom eigendirection of the (a,a) chiral domain coming from the $Z_3$ irrep of the $P4_{2}/mnm$ space group of $A_3$$B$O$_2$ crystal family. 
 The purple, green and red balls corresponds to $A$, $B$ and O atoms, the displacements are represented by gold arrows and the spirals are highlighted by the red lines connecting each displacement arrows.
 Only the atom displacements creating the spiral are shown and all the arrows of allowed displacements are normalized to the same length for clarity. In the case of K$_3$NiO$_2$, the Ni does not contribute much to the distortion, i.e. it is reduced to mostly K motion. From \cite{favaKNO}.
 }
 \label{fig:K3NiO2}
\end{figure}

On cooling below 410 K, K$_{3}$NiO$_2$ undergoes a first-order phase transition from a centric structure of $P4_{2}/mnm$ symmetry to a non-centrosymmetric and chiral phase of $P4_{1}2_{1}2$ (or $P4_{3}2_{1}2$) symmetry. 
As for MgTi$_{2}$O$_{4}$ described above, the structures of the two enantiomorphs are described by an enantiomorphic pair of chiral space groups~\cite{Duris2012}.
For these two structures to be degenerate, the distortion from the $P4_{2}/mnm$ phase to each enantiomorph must be described by the same two-dimensional $Z_3$ irrep: the phase transition involves slight rotations of the NiO$_2^{3-}$ units about the [110]/[1$\bar{1}$0] directions and small displacements of potassium ions which breaks the linearity of the O-Ni-O-K-O-Ni-O chains~\cite{Duris2012} (see Figure ~\ref{fig:K3NiO2}).
The two enantiomorphs differ in the direction of this 2D irrep: the $(a,a)$ and $(\bar{a},\bar{a})$ order parameter directions give the $P4_{1}2_{1}2$ enantiomorph whilst $(a,\bar{a})$ and $(\bar{a},a)$ give the $P4_{3}2_{1}2$ enantiomorph~\cite{favaKNO}. 
For these enantiomorphic structures, whilst $(a,a)$ and $(a,\bar{a})$ domains are related by a mirror plane, they are non-superimposable (they contain opposite structural helices) and so are described by different space groups. 
Recently, Fava et al.~\cite{favaKNO} have shown from first-principles calculations that the transition from the $P4_{2}/mnm$ high symmetry phase to the low chiral phases $P4_{1}2_{1}2$ and $P4_{3}2_{1}2$ can be explained from the condensation of a soft zone boundary mode of the $Z_3$ symmetry label.
In summary, for K$_{3}$NiO$_{2}$:
\begin{itemize}
\item{The phase transition from achiral structure, to a structure described by enantiomorphic space group is first order and is accompanied by a change in unit cell volume;}
\item{The phase transition is described by an two-dimensional irrep ($Z_3$), with most of the displacement due to potassium and oxygen motion.}
\item{The transition to the two enantiomorphs (described by a pair of enantiomorphic space groups) are described by the same irrep but with different order-parameter directions.}
\end{itemize}

\subsection{\texorpdfstring{CsCuCl\textsubscript{3}}{CsCuCl3}}
\label{CsCuCl3}

\begin{figure}
\centering
\includegraphics[width=8.5cm,keepaspectratio=true]{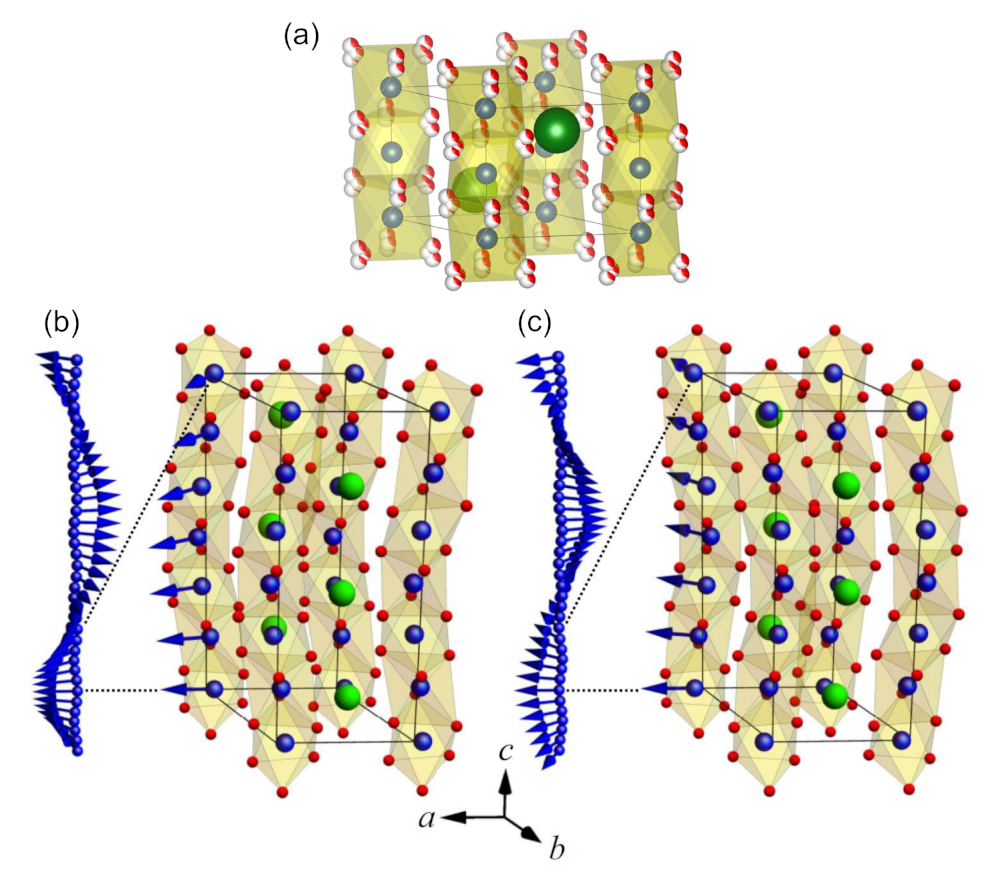}
 \caption{Crystal and magnetic structures of CsCuCl$_{3}$ showing (a) high-temperature achiral phase of $P6_{3}/mmc$ symmetry and below, low-temperature chiral phases of (b) $P6_{1}22$ symmetry and (c) $P6_{5}22$ symmetry, showing Cs, Cu and Cl sites in green, blue and red, with CuCl$_{6}$ octahedra in yellow, and Cu$^{2+}$ moments in blue. The right-handed and left-handed helimagnetic structures for the right-handed ($P6_{1}22$) and left-handed ($P6_{5}22$) enantiomorphs are shown by the chains of blue arrows in panels (b) and (c). The lower panel is reproduced from reference ~\cite{kousaka2017CsCuCl3}.}
 \label{fig:CsCuCl3}
\end{figure}

At room temperature CsCuCl$_{3}$ adopts a chiral crystal structure described by enantiomorphic $P6_{1}22$ or $P6_{5}22$ symmetries~\cite{Schlueter1966}.
Optical measurements reveal the domain structure of samples, with domains of the two opposite enantiomorphs rotating the plane of polarised light in opposite directions. 
On warming through 423 K, CsCuCl$_3$ undergoes a first-order phase transition to an achiral phase of $P6_{3}/mmc$ symmetry where the unit cell $c$ axis is reduced by a third compared with the chiral crystal structure (see Figure ~\ref{fig:CsCuCl3}), and in which the optical rotatory power is lost ~\cite{Hirotsu1975} (a “gyrotropic” phase transition). 
The $P6_{3}/mmc$ $\rightarrow$ $P6_{1}22 / P6_{5}22$ phase transition is due to a cooperative Jahn-Teller distortion of the CuCl$_{6}$ octahedra which share faces along $c$, leading to helical displacements of ions ~\cite{hirotsu1977} to give either right-handed helices ($P6_{1}22$ symmetry) or left-handed helices ($P6_{5}22$ symmetry).
The structural phase transition involves displacements of Cl$^{-}$ ions described by the $\Delta_6$ irrep with $(a, 0, 0, 0)$ order parameter direction giving the $P6_{5}22$ enantiomorph, whilst the $(0, a, 0, 0)$ order parameter direction gives the $P6_{1}22$ enantiomorph~\cite{hirotsu1977,Campbell2006, stokes2007isotropy}.

The Cu$^{2+}$ $S=\frac{1}{2}$ moments order antiferromagnetically below $T_{\rm{N}}=10.5$ K with a triangular helical spin structure: moments lie in the $ab$ plane but the direction within the plane modulates along $c$ ~\cite{Adachi1980} (with a period of $\sim 210$ {\AA}), Figure ~\ref{fig:CsCuCl3} (b) and (c).
This helical structure results from intrachain ferromagnetic interactions between Cu$^{2+}$ nearest-neighbours and antisymmetric Dzyaloshinskii–Moriya (DM) interactions, ~\cite{Adachi1980} allowed for this non-centrosymmetric structure. ~\cite{Yosida1996, Moskvin2019} 
The crystal chirality determines the DM axial vector (the sense of the Cu$^{2+}$ helix along $c$). So the magnetic chirality depends directly on the crystal chirality:~\cite{Adachi1980} the chiral vector (along $c$) has opposite directions for the two enantiomorphs~\cite{Syromyatnikov2005}.

\begin{itemize}
\item{The gyrotropic phase transition from the achiral $P6_{3}/mmc$ phase to the enantiomorphic phases ($P6_{1}22 / P6_{5}22$ symmetries) is first order and is accompanied by a change in unit cell volume;}
\item{The phase transition is described by an irrep ($\Delta_6$) that is at least two-dimensional;}
\item{The chirality of the magnetic structure is determined by the crystal chirality.} 
\end{itemize}

\subsection{\texorpdfstring{Na\textsubscript{2}Ca\textsubscript{2}Si\textsubscript{3}O\textsubscript{9} and Ag\textsubscript{4}P\textsubscript{2}O\textsubscript{7}}{Na2Ca2Si3O9 and Ag4P2O7}}
The combeite Na$_2$Ca$_2$Si$_3$O$_9$ crystal has been also reported to have a phase transition around 758 K from an achiral high symmetry $R\bar{3}m$ space group to a low symmetry phase of either $P3_121$ or $P3_221$ enantiomorphic space groups~\cite{maki1968,fischer1987, ohsato1990, kahlenberg2023}.
The crystal structure of combeite can be seen to be mostly structured around rings of SiO$_4$ tetrahedra in its high symmetry phase, which are distorted into ellipses in the low symmetry phase.
However, few studies have been carried out on the combeite crystal regarding its chiral phase transition.

The phosphate crystal Ag$_4$P$_2$O$_7$ undergoes a similar structural chiral phase transition from the trigonal $R\bar{3}c$ space group to the enantiomorphic $P3_121$/$P3_221$ or $P3_112$/$P3_212$ space groups at 623 K~\cite{yamada1983}.
Although this crystal has not been further characterized concerning its phase transition or its chiral properties, it is of interest as one of very few materials (including Na$_2$Ca$_2$Si$_3$O$_9$) to undergo a ferrochiral transition to an enantiomorphic structure.

\subsection{\texorpdfstring{Pb\textsubscript{5}Ge\textsubscript{3}O\textsubscript{11}}{Pb5Ge3O11}}
\label{PGO}
\begin{figure}[hbt!]
    \centering
     \includegraphics[width=8.5cm,keepaspectratio=true]{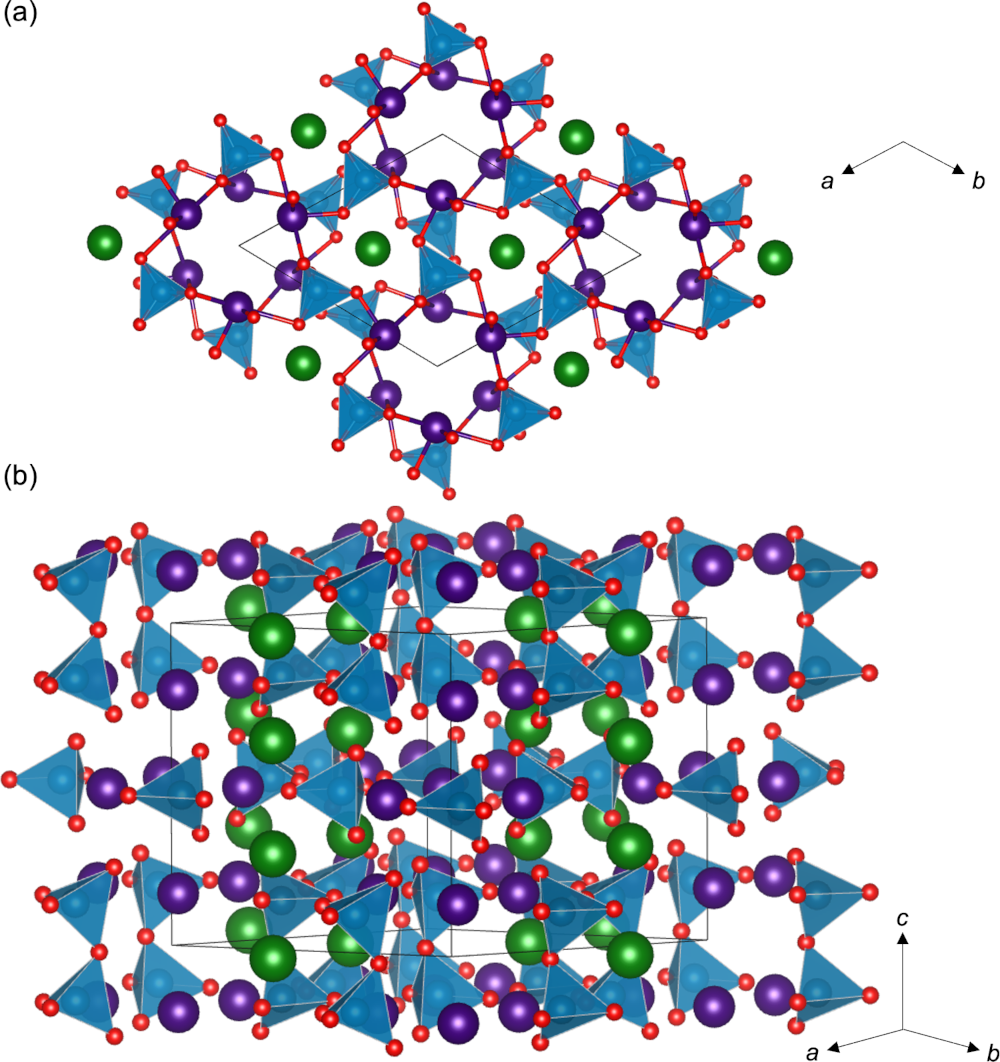}
    \caption{\label{pgo}{Illustration of the the high-temperature $P\bar{6}$ symmetry crystal structures of Pb$_{5}$Ge$_{3}$O$_{11}$ showing projection down hexagonal channels (a) and perpendicular to the channels (b) illustrating the connectivity of GeO$_{4}$ tetrahedra. O$^2-$ sites are shown in red, with Pb$^2+$ ions on the sites arranged around the hexagonal channels in purple and lower symmetry Pb$^2+$ sites shown in green, with GeO$_{4}$ tetrahedra in blue.}}
\end{figure}

The high-temperature ($T>450$ K) crystal structure of Pb$_{5}$Ge$_{3}$O$_{11}$ is described by $P\bar{6}$ symmetry, and although it is non-centrosymmetric (it is piezoelectric and second-harmonic generation active), it is paraelectric. 
Its structure is related to the structures of apatite and nasonite, composed of GeO$_{4}$ tetrahedra and Ge$_{2}$O$_{7}$ dimers, with Pb$^{2+}$ ions in interstitial sites arranged around a $6_{3}$ screw axis (see Figure~\ref{pgo} a)~\cite{Newnham1973}.
At 450 K, Pb$_{5}$Ge$_{3}$O$_{11}$ undergoes a second-order (displacive) phase transition to a polar structure of $P3$ symmetry (Figure ~\ref{pgo} b). 
This structural phase transition involves rotations of GeO$_{4}$ units and displacements of Pb$^{2+}$ ions (including along the polar $c$ axis)~\cite{Newnham1973}, described by the 1D irreducible representation (irrep)~\cite{Campbell2006, stokes2007isotropy} $\Gamma_{2}$.

In addition to being polar, the low-temperature $P3$ phase of Pb$_{5}$Ge$_{3}$O$_{11}$ is chiral and optically active: the GeO$_{4}$ rotations are thought to contribute to the optical activity, whilst the polarisation is predominantly due to the Pb$^{2+}$ and Ge$_{2}$O$_{7}$ displacements~\cite{Iwata1977}.
The $P3$ space group is one of the 43 achiral Sohncke groups, and so both enantiomorphs are described by the same symmetry.
Interestingly, as the sign of the polarisation is changed by an applied electric field, similarly, the sense of chirality and the sign of optical rotation are also reversed by an applied electric field (phenomenon called gyroelectricity, see Section~\ref{sec:Gyroelectricity})~\cite{Iwasaki2003a, Iwasaki2003b}.
Recently, it has been shown from density functional theory calculations and continuous symmetry measures~\cite{Zabrodsky1995} that the soft polar mode condensing into the $P\bar{6}$ structure which is responsible for the ferroelectric phase transition is also chiral. This explains why the optical activity and the predicted spin-momentum locking~\cite{PhysRevB.108.L201112} 
follow the change of polarization under an applied electric field as the polar distortions are also chiral in Pb$_5$Ge$_3$O$_{11}$~\cite{fava2023ferroelectricity}.
This study has also shown that when quantifying the chirality of each phonon mode of Pb$_5$Ge$_3$O$_{11}$, the mode chirality increases with the mode frequency.

In summary for Pb$_{5}$Ge$_{3}$O$_{11}$:
\begin{itemize}
\item{The phase transition from achiral structure to a structure described by a Sohncke group is second order and there's little change in unit cell volume associated with this phase transition;}
\item{The phase transition is described by a singly-degenerate polar irrep ($\Gamma_2$);}
\item{The same space group symmetry describes both enantiomorphs (Sohncke group);}
\item{The polarisation and optical rotation are coupled, such that the ferroelectric polarization hysteresis loop is also observed for the optical activity, i.e., the gyroelectric effect;}
\end{itemize}

\subsection{\texorpdfstring{Ba(TiO)Cu\textsubscript{4}(PO\textsubscript{4})\textsubscript{4}}{Ba(TiO)Cu4(PO4)4}}
\begin{figure}
\centering
\includegraphics[width=8.5cm,keepaspectratio=true]{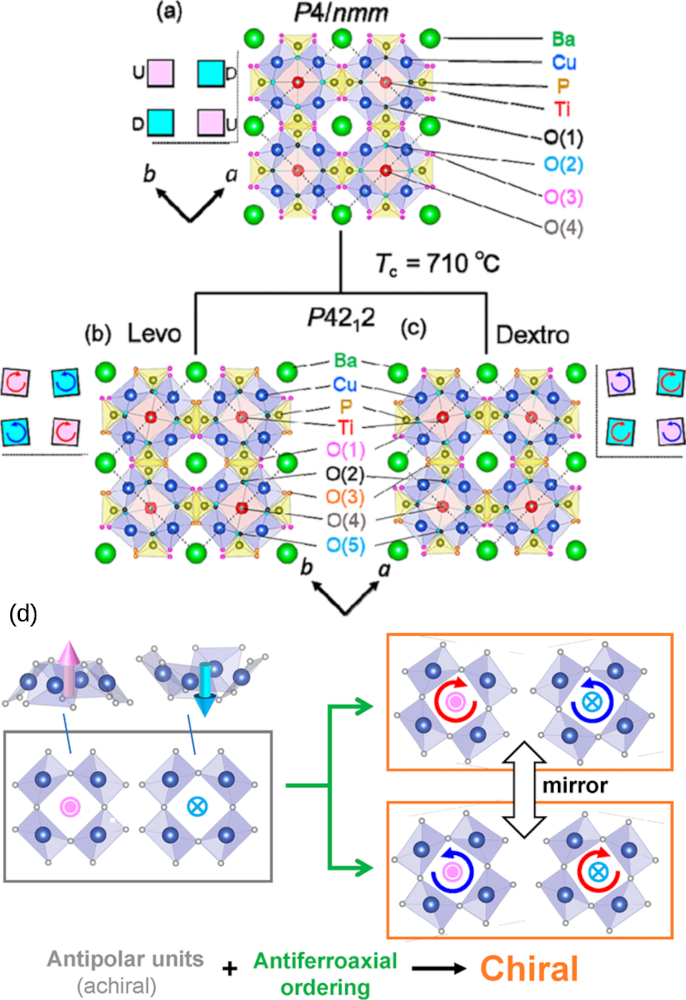}
 \caption{Illustration of the crystal structure of Ba(TiO)Cu$_4$(PO$_4$)$_4$: (a) high symmetry phase above the critical temperature (around 983 K) viewed along the c axis. 
 (b) Crystal structures of levo- and dextro- (c) domains below the critical temperature as  viewed along the c axis. 
 The squares in the insets of parts a-c schematically represent the axial rotations of Cu$_4$O$_{12}$ units of up (magenta) and down (cyan). 
 (d) shows how the combination of the antipolar distortions (left side with pink and light blue arrows) and the antiferroaxial distortions (right side with red and dark blue arrows) drives chirality in Ba(TiO)Cu$_4$(PO$_4$)$_4$. Adapted from Ref.~\cite{hayashida2021b}.}
 \label{fig:BaTiOCu4PO4_4}
\end{figure}

Ba(TiO)Cu$_4$(PO$_4$)$_4$ crystallizes in the  high symmetry tetragonal space group $P4/nmm$ and below 983 K it undergoes a structural phase transition to the low symmetry tetragonal space group $P42_12$. 
This phase transition is chiral where the low symmetry phase does not belong to one of the enantiomorphic space groups (the two chiral phases belong to the same space group).
The crystal structure of Ba(TiO)Cu$_4$(PO$_4$)$_4$ (see Figure~\ref{fig:BaTiOCu4PO4_4}) is made of a Cu$_4$O$_{12}$ building block composed of four corner shared square-planar CuO$_4$ units that are connected by PO$_4$ tetrahedra. 
In contrast to Pb$_{5}$Ge$_{3}$O$_{11}$, Ba(TiO)Cu$_4$(PO$_4$)$_4$ does not involve ferroelectric polarization nor any other ferroic order such that the phase transition of Ba(TiO)Cu$_4$(PO$_4$)$_4$ has been coined a pure ferrochiral phase transition~\cite{hayashida2021b}.
This phase transition is of the second order kind and involves a combination of an antiferroaxial order together with an antipolar order, the combination of which 
induces chirality at the zone center as no cell doubling is at play.
Other isostructural A(BO)Cu$_4$(PO$_4$)$_4$ (A=Sr, Pb and B=Ti, V) crystals were reported to crystallize in the same low symmetry $P42_12$ space group but no high symmetry phase were reported at high temperatures such that they might not exhibit the ferrochiral phase   transition~\cite{misawa2021,kato2019,Kimura2016}.
An experimental study looking at the difference between the case of a non-accessible achiral phase of Sr(TiO)Cu$_4$(PO$_4$)$_4$ and the case of Ba(TiO)Cu$_4$(PO$_4$)$_4$ where it is observed
when increasing the temperature shows that the domain formations are not the same, i.e., the Sr case shows predominant monodomain states while the Ba case shows predominant multidomain states~\cite{Kimura2016}.

In summary for Ba(TiO)Cu$_4$(PO$_4$)$_4$:
\begin{itemize}
\item{The phase transition from achiral structure to a structure described by a Sohncke group is second order;}
\item{The same space group symmetry describes both enantiomorphs (Sohncke group);}
\item{The phase transition is described by coupled antiferroaxial and antipolar distortions that induce chirality;}
\item{It is the unique zone center ferrochiral structural phase transition that has been reported so far.}
\end{itemize}

\subsection{\texorpdfstring{CsSnBr\textsubscript{3}}{CsSnBr3}}
Whilst preparing this review article, a very elegant study of the polar and chiral phase of CsSnBr$_3$ was published~\cite{fabini2024}.
At intermediate temperatures (85 K <T<269 K) it adopts the tilted perovskite structure ($a^-b^+a^-$ tilts of SnBr$_6$ octahedra) of $Pnma$ symmetry.
At low temperatures, CsSnBr$_3$ adopts a chiral and polar structure of $P2_1$ symmetry (one of the non-enantiomorphic Sohncke groups), but it reaches this low temperature phase via two continuous (second order) phase transitions.
From the $Pnma$ phase, there’s a second order phase transition at 85 K to a ferroaxial phase of $P2_1/m$ symmetry, with similar tilts to the $Pnma$ phase but with additional antipolar displacements of Sn$^{2+}$ cations~\cite{fabini2024} (due to the 5s$^2$ pairs of electrons)~\cite{walsh2011}.
On cooling a little further to 77 K, the $P2_1/m$ phase undergoes a second order phase transition to the chiral and polar $P2_1$ phase, which now allows polar displacements of the Sn$^{2+}$ cations. 
As observed for Pb$_{5}$Ge$_{3}$O$_{11}$ and Ba(TiO)Cu$_4$(PO$_4$)$_4$, the achiral – chiral (non-enantiomorphic) phase transition is second order and driven by a zone-centre polar distortion mode ($\Gamma_1^-$).
In summary:
\begin{itemize}
    \item The achiral to chiral phase transition is second order and involves little change in unit cell volume;
    \item The same space group symmetry (a non-enantiomorphic Sohncke group) describes both enantiomorphs;
    \item The transition is described by the polar zone-centre irrep $\Gamma_1^-$.
\end{itemize}

\subsection{Crystallographic databases and data mining for chiral crystals}
\label{databases}
After having presented a few specific examples of chiral crystals, we would like to discuss the following observation: of the 140,000 structures recorded in the Inorganic Crystal Structure Database (ICSD), about $20\%$ (i.e., around 8,000 structures) are described by one of the 65 Sohncke space groups, and of these, only around 900 are described by one of the 22 chiral and enantiomorphic space groups.
Hence, we can question whether the small number of crystals reported for the 11 pairs of enantiomorphic space groups are rare for a physical or chemical reason or have simply been overlooked. 

The CSD is one of the pioneering digital databases, encompassing a comprehensive collection of organic and metal-organic crystal structures~\cite {groom2014cambridge,groom2016cambridge}. 
In the 1990s, the Inorganic Crystal Structure Database (ICSD) was established, focusing on experimentally determined inorganic and mineral structures with around 200k compositions today~\cite{hellenbrandt2004inorganic, allmann2007introduction}. 
These crystal structure databases (alongside others including Materials Project~\cite{jain2013commentary}, Crystallography Open Database (COD)~\cite{gravzulis2012crystallography}, AFLOW database~\cite{calderon2015aflow}, NOMAD repository~\cite{draxl2019nomad})
allow the materials community to search for, analyze, and use this information to design new materials with desired properties. 
There has been a notable upsurge in efforts to use machine learning and data mining methodologies to examine and unveil novel crystal materials by harnessing existing crystal databases. This concept entails training a model using established structures to predict
their chemical stability. With a suitable dataset, machine learning can anticipate the formation energies of hypothetical structures, presenting a powerful instrument for materials exploration and design. 
This methodology strives to predict properties or generate new crystal structures utilizing generative models, as exemplified in recent research endeavors.~\cite{merchant2023scaling,szymanski2023autonomous}
The methods used to populate the computational databases are high-throughput techniques that screen many systems fairly rapidly~\cite{pizzi2016aiida, calderon2015aflow, ong2013python}. 
This is performed using Density Functional Theory (DFT) calculations to simulate crystal structures' electronic structure and stability. 
High-throughput DFT calculations can screen large numbers of crystal structures for desirable properties.
This means that thousands of materials are computed and then analyzed to identify those with desirable properties such as stability or specific optical, electrical, or magnetic properties. 
This allows for a much more efficient and rapid identification of promising materials. 
Researchers can use these methods to screen large libraries of compounds for chiral properties or to systematically explore the structural and compositional variations of a particular class of chiral materials.

\begin{figure*}[hbt!]
    \centering
    \includegraphics[trim={0. 0cm 0 0cm},clip,width=0.45\textwidth]{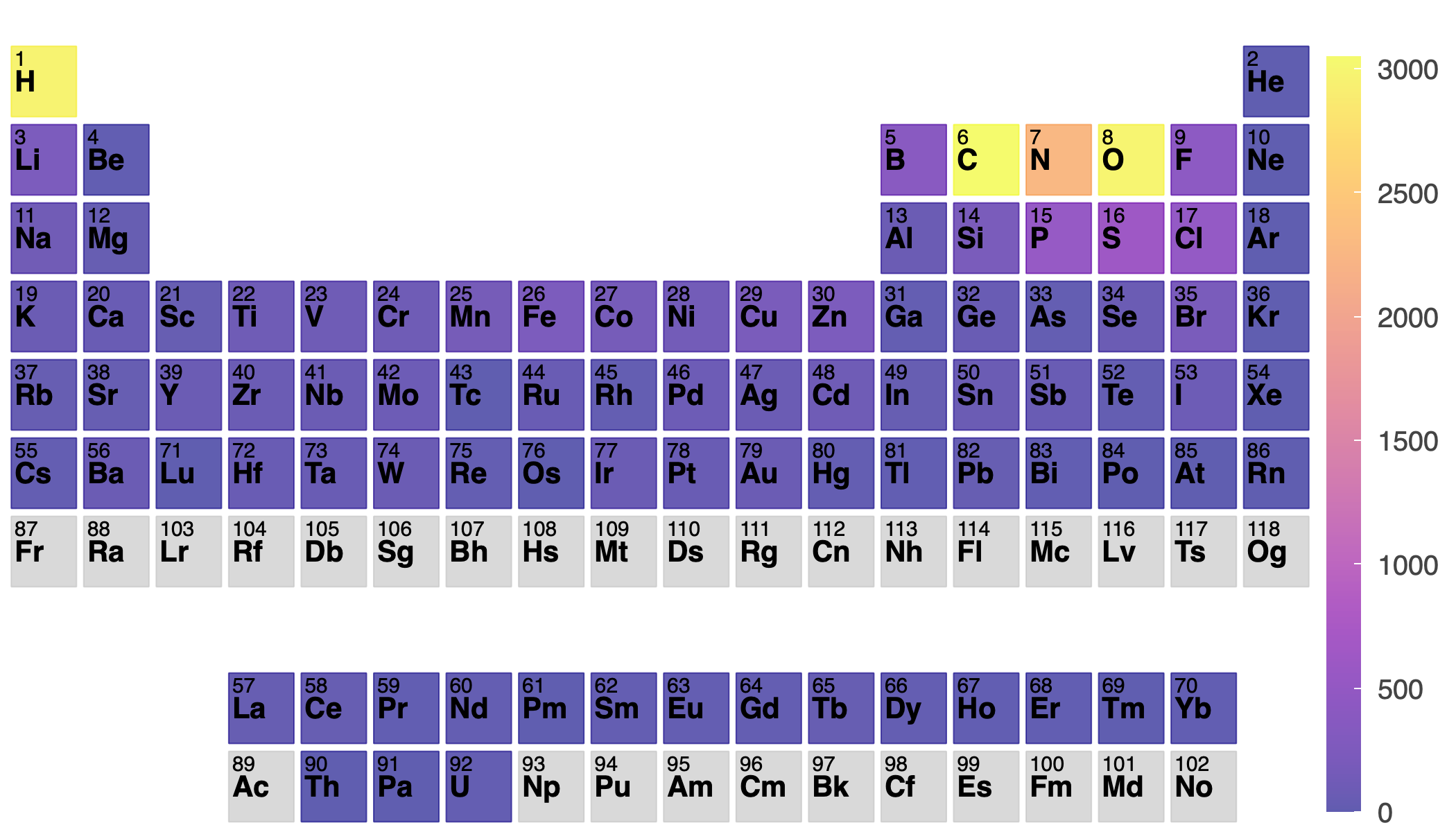}
    \includegraphics[trim={0. 0cm 0 0cm},clip,width=0.45\textwidth]
    {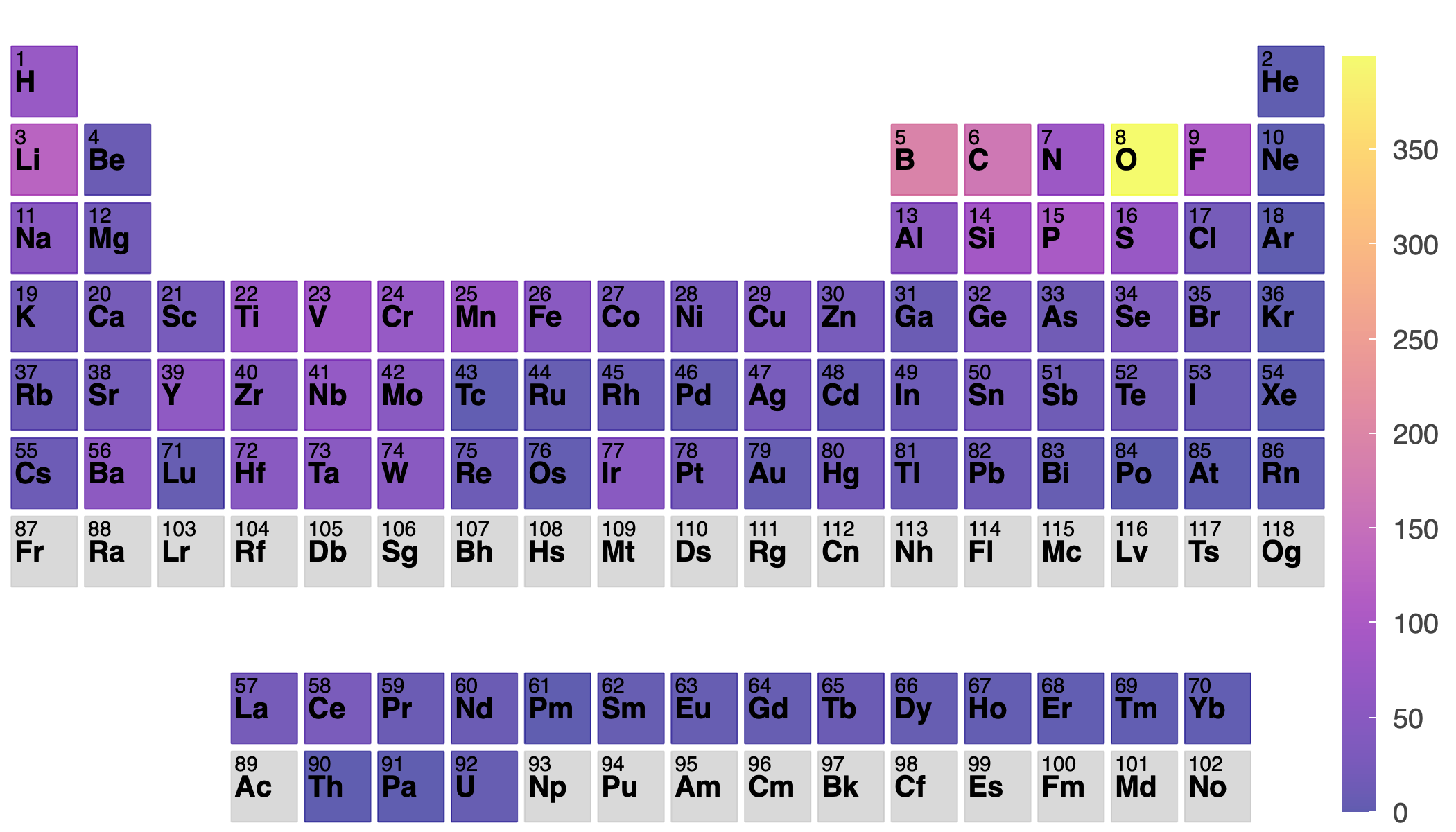}
    \caption{\label{fig:PeriodicTable}{Color trend over the different periodic table elements as a function of the frequency of chiral crystals' appearance in the COD database. The top figure is 22 chiral materials overall, and the bottom figure only includes those with less than 50 atoms in the unit cell.}}
\end{figure*}

Here, we analyze the available chiral crystal structures from the following three prominent databases: Materials Project (MP)~\cite{jain2013commentary}, Crystallographic Open Database (COD)~\cite{gravzulis2009crystallography}, and AFLOW~\cite{calderon2015aflow}. The foundational data for these databases originates from the experimental Inorganic Crystal Structure Database (ICSD)~\cite{belsky2002new, hellenbrandt2004inorganic}. These three databases have since been substantially expanded through the application of high-throughput methods.
Our study encompassed data reported up until May 2023. 
Table~\ref{Table1} summarizes our screening and shows that fewer structures are reported for chiral space groups than for achiral space groups and even for other Sohncke crystals.
COD contains structures with larger unit cells compared to the other databases. When constrained to 50 atoms per unit cell, the number of chiral structures in COD becomes similar to the other databases. To compile a comprehensive dataset, we merged all three databases into a single master database, eliminating repeated structures and excluding those in COD with incorrect Crystallographic Information File (CIF) format.  Following this process, we obtained 4,285 structures that belong to one of the 22 enantiomorphic space groups, of which only 1,034 have unit cells with less than 50 atoms. For a visual representation, Figure~\ref{fig:PeriodicTable} illustrates the distribution of different chemical species present in the reported structures.  In unrestricted unit cells, carbon emerges as the most frequently occurring element, with 3,050 structures, followed by hydrogen and oxygen, 
with 2,965 and 2,990 structures, respectively. 
This is related to the abundance of organic crystals with large unit cells in the consolidated database. 
However, in unit cells with fewer than 50 atoms, the composition shifts, with oxygen constituting 38.5\% of the reported structures in the joint database, followed by boron at 18.6\% and carbon at 16.0\%. 
Notably, 70\% of the structures with less than 50 atoms contain transition metals. 
An analysis similar to that carried out by Halasyamani and Poeppelmeier ~\cite {Halasyamani1998}, to seek to identify structural features that might be incorporated into design strategies for chiral materials, would be of interest. The relatively small number of chiral structures reported to date makes this challenging. However, proposed chiral structures from high-throughput and machine-learning computational work might soon make this more feasible.

We observed an unequal distribution of structures between enantiomorphic pairs, implying that
either achieving synthesis control over specific enantiomorphic space groups has not yet been fully realized, or that experimentally there may be challenges with determining the absolute configuration of a system (see Section ~\ref{sec:absolute_chir}).
This discrepancy highlights the complexity and challenges in controlling chiral structure formation and determination, prompting further research and innovation. 
Using a high-throughput method to search for chiral materials would be an excellent strategy for rapidly identifying new chiral materials with specific functionalities.
These high-throughput methods allow researchers to quickly test large numbers of potential chiral materials in parallel rather than individually. Such methods have the potential to uncover new chiral functional materials, adding to our understanding and meeting the demands of emerging technologies.~\cite{yang2021chiral}

\begin{table}[!h]
\begin{center}
\begin{tabular}{ |c|c|c|c|c|c|c| } 
 \hline
  Space Group & MP & AFLOW & COD & Consolidated & Octa.& Tetra.\\ 
\hline \hline
76  & 55	 &     79	  &   317 &  421 & 85 &  218 \\
78  & 22	 &      2	  &   295 &  309 & 67 & 171\\
91  & 37	 &     23	  &    31 &   78 & 49 & 46\\
92  & 152	 &     64	  &   544 &   604 & 159 & 356\\
95  & 63	 &      5	  &    42 &   98 & 74 & 72\\
96  & 72	 &     22	  &   466 &   503 & 139 & 303\\
144  & 55	 &     17	  &   236 &   262 & 49 & 146\\
145  & 63	 &      7	  &   234 &  284 & 69 & 191\\
151  & 15	 &     10	  &    12 &  25 & 13 & 13\\
152  & 144	 &    141	  &   284 &  410 & 129 & 217\\
153  & 8	 &      2	  &     7 &  14  & 3 & 4\\
154  & 43	 &     18	  &   225 &  241 & 78& 134\\
169  & 24	 &     10	  &   199 &   219 & 43 & 131\\
170  & 10	 &      2	  &   212 &   219 & 37& 143\\
171  & 3	 &      0	  &    13 &   16 & 2 & 7\\
172  & 2	 &      0	  &    11 &   13 & 3 & 9\\
178  & 22	 &      8	  &   70  &   87 & 32 & 46\\
179  & 2	 &      2	  &   65  &   65 & 18 & 39\\
180  & 50	 &    171	  &   59  &   217 & 10 & 67\\
181  & 39	 &     18	  &    29 &   54 & 5 & 39\\
212  & 2	 &     35	  &    27 &   54 & 15 & 30\\
213  & 67	 &     37	  &    33 &   86 & 40 & 28\\
 \hline
\end{tabular}
\caption{Number of structures existing in Materials Project (MP), AFLOW, and Crystallographic Open Database (COD). The consolidated column is a constructed consolidated database without repeated structures. The last two columns refer to the number of structures with octahedra or tetrahedra. 
}
\label{Table1}
\end{center}
\end{table}

\section{Chirality and its relationship to other properties} \label{Sec:tuning-chirality-external-means}

In this section, we will discuss how chirality can be tuned externally, i.e., by applied electric and magnetic fields or with strain.
While chirality is not the property directly investigated, most of the responses that we discuss below are related to tuning optical activity as an indirect probe of change in chirality in the materials. 
A dedicated discussion about flipping chirality will be done further in Section~\ref{order-parameter}.
We will also discuss the cases of topological materials and altermagnets as their properties can be related to chirality.

\subsection{Magnetism}
\label{magnetism}

Harnessing the influence of an applied magnetic field on properties associated with chirality (including optical responses) holds promise for a range of functional materials applications, including electric power systems, magnetic field sensors, all-dielectric measurement devices, and broadband frequency responses~\cite{lenz2006magnetic,li2017optical}. This has motivated research into materials capable of demonstrating magnetic-optical interactions, including magneto-chiral materials that have appropriate symmetry (either as a result of their crystal or magnetic structures) for optical activity~\cite{nye1985physical}.
 
\subsubsection{Magneto-optical phenomena}
Several magneto-optical phenomena do not rely on specific symmetries of the light-transmitting media (including the Faraday effect and magnetic circular dichroism), but instead rely on an applied magnetic field (or a magnetisation of the medium)~\cite{Newnham2005}. These phenomena contrast with natural optical activity (see Section ~\ref{optical_activity}), which depends on the symmetry of the spatial dispersion of the electric dipoles in the medium (alongside a generally weaker effect due to magnetic dipoles~\cite{caldwell1966, Pershan1967}), giving rise to differences in propagation of oppositely polarised light. The Faraday rotation $\Psi$ is given by
\begin{equation}
    \Psi = V_{ij}tN_{i}H_{j}
\end{equation}
where $t$ is the path length, $H_j$ are the components of the magnetic field, $N_i$ the directions of the wave normal, and $V_{ij}$ is the Faraday tensor (a polar second-rank tensor).  
In the inverse Faraday effect, a solid can be magnetized when exposed to intense circularly-polarised light~\cite{Newnham2005}. 
Magnetic circular dichroism is related to the Faraday effect but relies on the different absorption coefficients for oppositely circularly polarised light in an applied magnetic field. This results in a phase difference between left- and right waves propagating through a medium~\cite{Newnham2005}.

As noted above, the magnetic optical activity observed in both the Faraday effect and magnetic circular dichroism result from the applied magnetic field breaking time-reversal symmetry and not from the intrinsic symmetry of the light-transmitting medium~\cite{rikken1997observation}.
However, a magnetic phenomenon analogous to natural optical activity exists, referred to as “magnetochiral dichroism” (MChD), which relies on the symmetry of magnetic dipoles in the light-transmitting medium~\cite{rikken1997observation, Wagniere1982, Wagniere1984}. 
Barron and Vrbancich demonstrated that chiral systems should show different optical properties depending on whether the light propagates through the chiral medium either parallel or antiparallel to the magnetic field direction, with opposite effects for the two enantiomers~\cite{Barron1984}.
Atzori {\it{et al.}} give excellent overviews of the experimental work in this field and introduce the implications of this effect for sound propagation, photochemistry, and electrochemistry (and even possible explanations for the origins of homochiral life on earth!)~\cite{atzori2020, atzori2021}.
The MChD effect depends on the relative orientations of the wave vector and magnetic field and the different absorptions of light by opposite enantiomorphs arising from magneto-chiral anisotropy (MChA)~\cite{rikken1997observation, atzori2020}.
There’s a growing body of work exploring the MChD effect in molecular systems and complexes, which typically takes advantage of molecular chirality to give a magnetized chiral system~\cite{atzori2019, Pointillart2024, Ishii2020}.
However, in this review focusing on structural chirality, it’s relevant to consider possible magnetic and chiral extended solids, and the following sections explore “magnetic chirality,” i.e., chirality in spin-ordered states or textures~\cite{cheong2022}.

\subsubsection{Magnetic chirality} 
\label{subsec:magnetic-chirality}
In addition to the structural chirality described in Section~\ref{Sec:symmetry}, which arises from the crystal structure's symmetry, the symmetry of the magnetic order can give rise to additional physical properties~\cite{Perez-Mato2015, Perez-Mato2016}. 
In terms of chirality, several possibilities can be considered:
\begin{itemize}
\item Magnetic spins on a chiral crystal structure order with the magnetic chirality fixed by the chirality of the crystal structure (e.g. langasite Ba$_{3}$NbFe$_{3}$Si$_{2}$O$_{14}$);
\item Magnetic spins on a chiral crystal structure order in a non-centrosymmetric and chiral fashion (MnSi);
\item Magnetic spins on an achiral crystal structure order in a chiral fashion (e.g. Mn$_{3}$Sn).
\end{itemize}

\begin{figure}
\centering
\includegraphics[width=8.2cm,keepaspectratio=true]{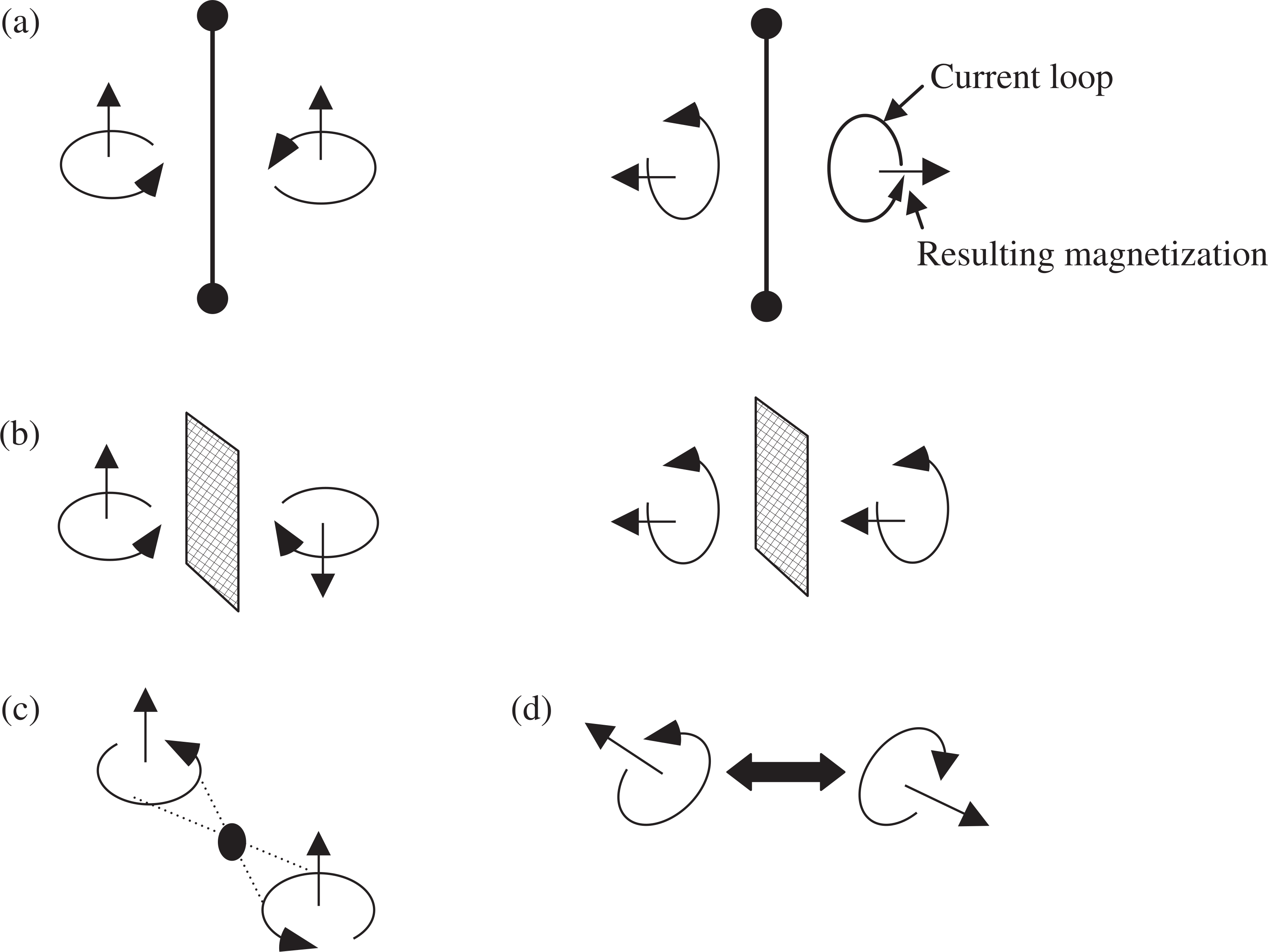}
 \caption{ Visualising magnetic moments as axial vectors resulting from a current loop. Panels show the effects of (a) a two-fold rotation, (b) reflection, (c) spatial inversion and (d) time reversal. Reproduced from~\cite{Newnham2005}. }
 \label{fig:current_loop}
\end{figure}

Before discussing these phenomena in more detail, it is helpful to consider some aspects of magnetic symmetry that are distinct from purely structural crystallography, as well as their consequences. 
Although magnetic moments are often represented by arrows (similar to electric dipoles), it’s important to note that they are described by axial vectors (rather than the polar vectors used to describe electric dipoles). 
The image of the magnetic moment arising from a current loop (Figure ~\ref{fig:current_loop}) helps to visualize the effects of various symmetry operations acting on an axial vector (and the resulting magnetic moment), which is often quite different from the effects on polar vectors.
Magnetic symmetry groups (to describe ordered magnetic structures) combine the geometric symmetry elements discussed in Section~\ref{Sec:symmetry} in addition to the non-spatial symmetry operation of “time reversal”: reversing time would reverse the current flow in a loop and, therefore reverse the direction of the magnetic moment (time reversal switches the sign of an axial vector)~\cite{Newnham2005, zheludev1986}.
However, magnetic chirality does not change under time reversal~\cite{cheong2022}.

This understanding allows us to consider two types of chirality that emerge from non-collinear magnetic structures often associated with magnetic chirality: vector chirality and scalar chirality, as discussed in Section ~\ref{Sec:chirality-definitions}.

Collinear magnetic structures are not chiral, and so in the search for magnetic chirality, non-collinear magnetic structures are needed,~\cite{inoue2021chiral} and the microscopic mechanisms that stabilize these more complex magnetic arrangements should be considered. 
Non-collinear magnetic structures arise from combinations of symmetric exchange with, for example, geometric frustration of symmetric (Heisenberg) exchange interactions~\cite{Kaplan2007} (e.g. langasite Ba$_{3}$NbFe$_{3}$Si$_{2}$O$_{14}$,~\cite{Qureshi2020, marty2008single, Stock2011} magnetocrystalline anisotropy, and Dzyaloshinskii–Moriya interaction (DMI, an antisymmetric exchange via a spin-orbit interaction, which may be non-zero if there’s no inversion symmetry between the pair of magnetic sites)~\cite{Yosida1996, Moskvin2019} (e.g. CsCuCl$_{3}$)~\cite{Adachi1980}. Several broad classes of non-collinear magnetic structures are frequently reported and its useful to consider them concerning chirality:
\begin{itemize}
\item Cycloidal spin structures (e.g. Ca$_{3}$Ru$_{2}$O$_{7}$~\cite{Faure2023}): although polar they are achiral.
\item Helical spin structures (e.g., MnSi, Ba$_{3}$NbFe$_{3}$Si$_{2}$O$_{14}$): chiral (does not break time-reversal symmetry and so an example of true chirality)~\cite{cheong2022}.
\item Canted and toroidal spin structures (e.g. BaCoSiO$_{4}$): chiral (does not break time-reversal symmetry and so an example of true chirality)~\cite{cheong2022}.
\end{itemize}

\subsubsection{Electronic magneto chiral anisotropy (eMChA)}

A fundamental condition for observing eMChA is the absence of inversion symmetry and presence of spin orbit coupling, rendering the material non-centrosymmetric. While various transport parameters may undergo modifications in such materials, a particularly noteworthy consequence of this anisotropy is its effect on electrical resistance. This influence on resistance arises from the interaction between the orientation of magnetization and the direction of current flow within the material. Consequently, alterations in the relative alignment of these two vectors can influence the material's electrical characteristics, underscoring the intricate interplay between its structural asymmetry and electrical behavior. In enantiomorphic crystals, this difference can be written as:
\begin{equation}
R^{L/R}(\mathbf{B},\mathbf{I})=R_0 \left( 1 + \mu^2 R^2 + \gamma ^{L/R} \mathbf{B} \cdot \mathbf{I} \right),
\end{equation}
with $\gamma^R = - \gamma^L$ referring to the conductor's right- and left-handed enantiomer, $\mu$ is the electron mobility, and $\mathbf{I}$ is the electrical current. This effect leads to the so-called electronic magneto chiral anisotropy, most commonly detected by the associated second-harmonic voltage generation under low-frequency a.c. as reported in chiral metals~\cite{pop2014electrical}. Though several applications can be overseen on materials with this effect (such as in rectifiers), the problem is that usually, $\gamma$ is very small. Therefore, there is a need to search for new materials with larger electronic magnetochiral anisotropy. A potential path is to use Weyl materials, where chirality can induce a charge imbalance between Weyl fermions of different chirality as reported in TaAs~\cite{morimoto2016chiral}. A different direction is concerning identifying chiral magnetic materials with larger anisotropy, as in the case of CuB$_2$O$_4$~\cite{saito2008magnetic} or LiCoPO$_4$~\cite{van2007observation}.

Exploring some well-characterised magnetochiral materials may best illustrate these aspects of magnetic chirality and the origins of these phenomena.

\subsubsection{Langasite \texorpdfstring{Ba\textsubscript{3}NbFe\textsubscript{3}Si\textsubscript{2}O\textsubscript{14}}{Langasite Ba3NbFe3Si2O14}}
Ba$_{3}$NbFe$_{3}$Si$_{2}$O$_{14}$ crystallises with a chiral structure of $P321$ symmetry adopted by the mineral langasite La$_{3}$Ga$_{5}$SiO$_{14}$. 
This space group is consistent with optical activity and piezoelectricity. 
It is one of the non-enantiomorphic Sohnke groups, meaning that crystals of opposite handedness are described by the same space group symmetry. 
Samples of opposite handedness (of their crystal structures) have been prepared, and their absolute chirality determined by anomalous X-ray scattering experiments~\cite{Qureshi2020, marty2008single}. 
The structure consists of layers of equilateral triangles of FeO$_{4}$ tetrahedra with antiferromagnetic superexchange interactions between Fe$^{3+}$ cations via oxide ions within the triangles (and super-superexchange interactions between triangles). Ba$_{3}$NbFe$_{3}$Si$_{2}$O$_{14}$ undergoes a (second order) magnetic ordering transition below $T_{\rm{N}} = 27$ K to an (approximately) co-planar antiferromagnetic structure. 
However, there is suggested to be a small out-of-plane ferromagnetic moment~\cite{marty2008single}.

\begin{figure}
\centering
\includegraphics[width=8 cm,keepaspectratio=true]{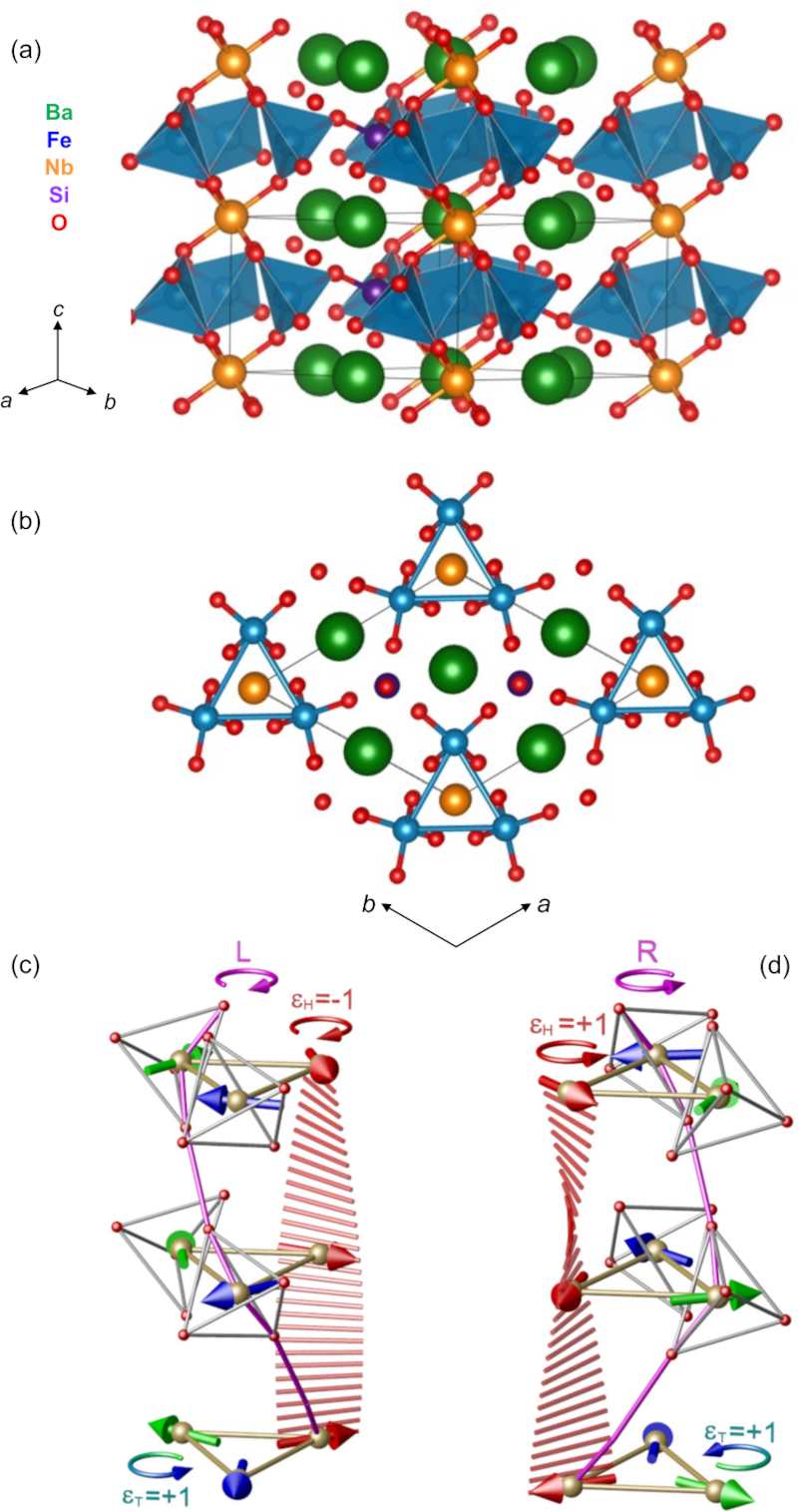}
 \caption{ (a) Crystal structure of Ba$_{3}$NbFe$_{3}$Si$_{2}$O$_{14}$ showing FeO$_{4}$ tetrahedra in blue and Ba, Nb, Si and O ions in green, orange, purple and red, respectively; (b) shows the structure viewed perpendicular to the layers (down [001]) highlighting the triangles of Fe$^{3+}$ cations in blue. Panels (c) and (d) show the left-handed magnetic helix for the left-handed enantiomorph and the right-handed magnetic helix for the right-handed enantiomorph, respectively; both magnetic helices have the same triangular chirality ($\epsilon_T=+1$ as shown in green), whilst their magnetic helices are of opposite handedness ($\epsilon_H=+1$ or -1 as shown in red),
 reproduced from reference~\cite{Qureshi2020}.}
 \label{fig:langasite}
\end{figure}

The arrangement of Fe$^{3+}$ ions in triangular units gives rise to the frustration of the antiferromagnetic Fe–O–Fe superexchange interactions and results in Fe$^{3+}$ moments within a triangle arranged at $120^{\circ}$ to one another. 
This “triangular chirality” is an example of vector chirality (see above), and analysis of results from inelastic neutron scattering experiments indicate that this vector chirality persists up to high temperatures, well above $T_{\rm{N}}$~\cite{Stock2011}.
Symmetric exchange interactions between triangles in successive layers result in the helical 3D magnetic order observed below $T_{\rm{N}}$, with crystals of opposite handedness switching the relative magnitudes of these interlayer exchange interactions and resulting in magnetic structures of opposite helices~\cite{Qureshi2020, Stock2011}.
Qureshi {\it{et al.}}~\cite{Qureshi2020} showed this very elegantly by carrying out scattering experiments on two samples of opposite crystal chirality to show that both enantiomorphs had the same vector chirality within the Fe triangles but opposite magnetic chiralities overall due to the opposite sense of the magnetic helices (Figure ~\ref{fig:langasite}).
Ba$_{3}$NbFe$_{3}$Si$_{2}$O$_{14}$ provides an exciting contrast to systems such as CsCuCl$_{3}$ (see Section ~\ref{CsCuCl3}) because the helical magnetic structure of Ba$_{3}$NbFe$_{3}$Si$_{2}$O$_{14}$ is thought to arise from frustrated symmetric (Heisenberg) exchange interactions, without the need for antisymmetric exchange (such as the Dzyaloshinskii-Moriya interaction)~\cite{Qureshi2020, marty2008single, Stock2011}.
In summary for Ba$_{3}$NbFe$_{3}$Si$_{2}$O$_{14}$:
\begin{itemize}
\item the magnetic chirality is fixed by the crystal chirality.
\item Magnetic chirality results from competing symmetric exchange interactions.
\end{itemize}

\subsubsection{\texorpdfstring{MnSb\textsubscript{2}O\textsubscript{6}}{MnSb2O6}}
\label{sec:MnSb2O6}

MnSb$_{2}$O$_{6}$ also crystallizes in a chiral but non-enantiomorphic structure of $P321$ symmetry. 
The structure is built from layers of (distorted) MnO$_{6}$ and SbO$_{6}$ octahedra, with Mn$^{2+}$ ions ordered into alternate layers to give distorted triangular networks of three MnO$_{6}$ octahedra around either an Sb$^{5+}$ site or a vacancy~\cite{REIMERS1989}. 
On cooling, it undergoes an antiferromagnetic phase transition at 12.5 K (with approximately 3D-Heisenberg-like behavior)~\cite{REIMERS1989} and Johnson {\it{et al.}} used single crystal neutron diffraction to investigate its magnetic ordering, characterized by co-rotating cycloids of Mn$^{2+}$ moments~\cite{johnson2013mnsb}.
This contrasts with Ba$_{3}$NbFe$_{3}$Si$_{2}$O$_{14}$ discussed above in that rather than a helical structure observed for Ba$_{3}$NbFe$_{3}$Si$_{2}$O$_{14}$, the spin plane has been rotated by $90^{\circ}$ to give a cycloidal spin structure, probably favored by out-of-plane anisotropy in MnSb$_{2}$O$_{6}$~\cite{PhysRevLett.111.017202}.
There’s some ambiguity as to the exact orientation of the planes in which the spins rotate without an applied magnetic field,~\cite{Kinoshita2016, chan2022neutron}, but it is close to the (110) plane~\cite{PhysRevLett.111.017202}.
This cycloidal spin structure again results from several symmetric exchange interactions (and anisotropy). 
The relative magnitudes of interlayer exchange interactions are directly related to the crystal structure's handedness, which determines the magnetic domain configuration~\cite{PhysRevLett.111.017202}.
The Dzyaloshinskii-Moriya interaction and cycloidal magnetic structure result in a polarisation perpendicular to the [001] axis~\cite{PhysRevLett.111.017202}. It seems that an applied magnetic field can tilt the plane of the spin cycloid (until a spin flop transition is reached at higher fields to give a helical spin structure), which can change the sign of the polarisation~\cite{Kinoshita2016, chan2022neutron}.
The opportunity to control electrical polarisation by applying a magnetic field in this magnetoelectric material is particularly exciting. 
In summary for MnSb$_{2}$O$_{6}$:
\begin{itemize}
\item The magnetic domain configuration of MnSb$_{2}$O$_{6}$ is directly related to the crystal chirality.
\item The cycloidal spin structure alone is achiral, resulting in an electric polarization.
\item An applied magnetic field can tilt the spin cycloid, switching the polarization and the ferroelectric domain configuration.
\end{itemize}

\subsubsection{\texorpdfstring{Cr\textsubscript{1/3}NbS\textsubscript{2}}{Cr1/3NbS2}}

\begin{figure}
\centering
\includegraphics[width=8.5cm,keepaspectratio=true]{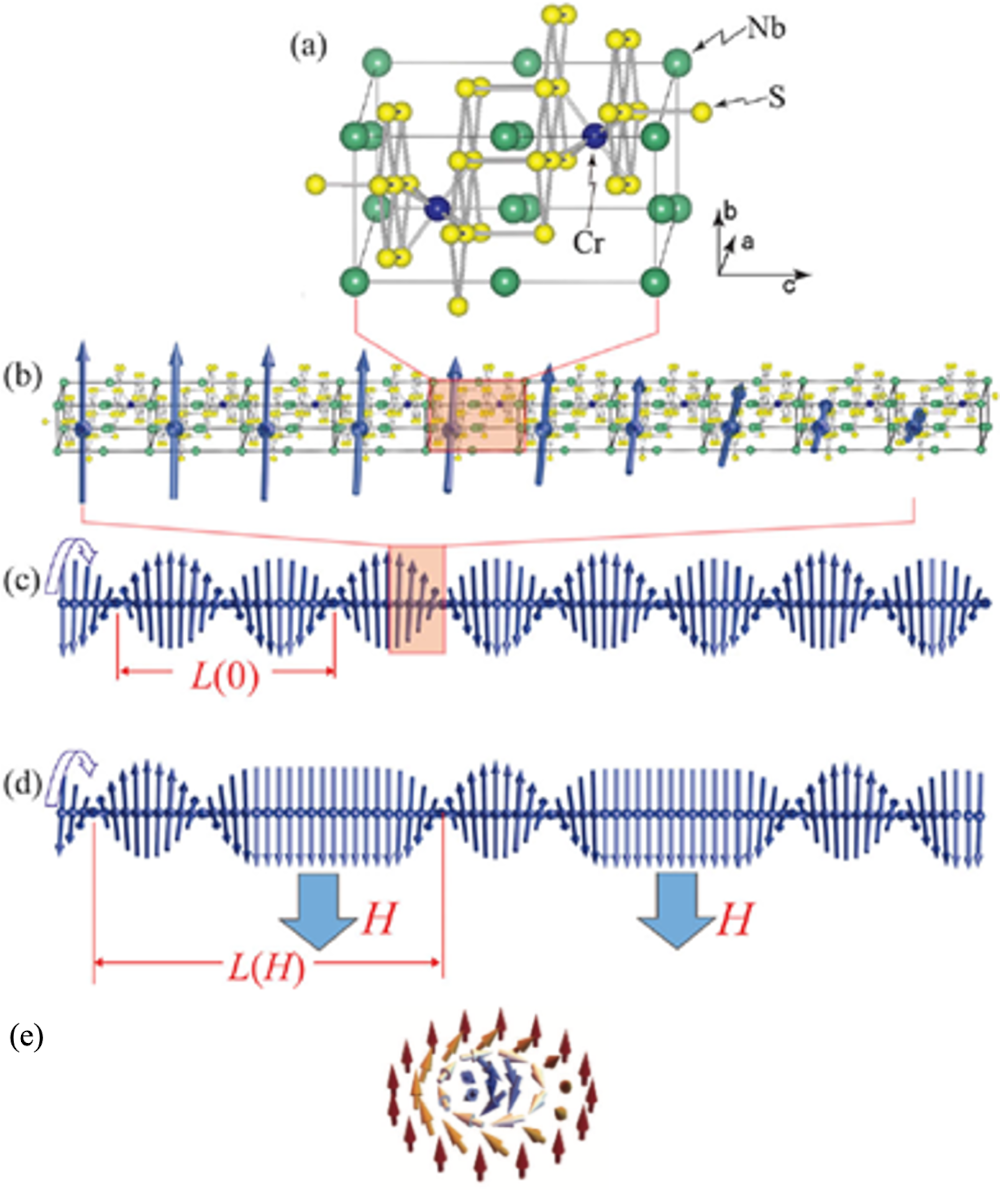}
 \caption{ (a) Crystal structure of Cr$_{1/3}$NbS$_{2}$ showing Cr, Nb, and S ions in blue, green, and yellow, respectively; (b) shows a part of a left-handed chiral helix observed as the magnetic ground state in zero applied field, and (c) shows the whole left-handed spin helix; (d) shows the chiral soliton lattice in applied fields perpendicular to the helical axis (reproduced from reference ~\cite{togawa2012chiral} with kind permission from APS); (e) shows a Bloch skyrmion as observed in MnSi (reproduced from reference ~\cite{Tokura2021} with kind permission from ACS).}
 \label{fig:CrNb3S6}
\end{figure}

The metallic ferromagnet Cr$_{1/3}$NbS$_{2}$ (also referred to as CrNb$_{3}$S$_{6}$) can be described as an intercalation product of the 2H disulfide NbS$_{2}$, with Cr$_{1/3}$NbS$_{2}$ adoping a hexagonal structure composed of sheets of edge-linked NbS$_{6}$ trigonal prisms, with Cr$^{3+}$ ions in octahedral sites between the layers. 
Depending on the precise stoichiometry and ordering of Cr$^{3+}$ ions over these sites, the structure is described by either $P6_{3}22$ or $P6_3$ symmetry (i.e., Sohncke but non-enantiomorphic space groups)~\cite{HULLIGER1970, Dyadkin2015}.
A range of magnetic ordering temperatures have been reported ($\sim$80 K $\leq$ $T_{\rm{C}}$ $\leq$ $\sim$130 K)~\cite{Dyadkin2015, Miyadai1983, KOUSAKA2009} and neutron scattering work indicated that in zero-field, the ground state is a chiral, helical magnetic structure, with Cr$^{3+}$ spins in the $ab$ plane (i.e., within the layer plane). 
The spin helices propagate along the $c$ axis (the direction perpendicular to the layers) with periodicity much longer than the unit cell dimensions (similar to the helical magnetic state observed in zero-field for MnSi discussed below). 
This helical spin structure results from several factors: the crystal field at the Cr$^{3+}$ sites gives strong uniaxial anisotropy favoring spins in the $ab$ plane, whilst intralayer symmetric ferromagnetic exchange and DMI give the helical spin structure~\cite{Miyadai1983}. 
As for MnSi, this combination of interactions means that Cr$_{1/3}$NbS$_{2}$ has a complex phase diagram~\cite{Tsuruta2016}. In addition to the chiral helical magnetic ground state, an applied field along the helix propagation direction can give a helical conical spin state with spins tilting towards the field direction before a forced ferromagnetic state is reached at higher fields. 
On the other hand, applying a magnetic field perpendicular to the helical axis (i.e., within the plane) gives a chiral soliton lattice (Figure ~\ref{fig:CrNb3S6}), with kinks separating ferromagnetic regions. 
The lengths of the ferromagnetic regions grow (as does the period of the chiral soliton lattice) with increasing applied field strength until a forced ferromagnetic state is reached above some critical field~\cite{togawa2012chiral, Yonemura2017}.
In summary for Cr$_{1/3}$NbS$_{2}$:
\begin{itemize}
\item multiple factors give rise to the helical magnetic states observed for Cr$_{1/3}$NbS$_{2}$ including magnetic anisotropy, symmetric exchange interactions, and antisymmetric DMI.
\item The long periodicity of the spin helices compared with the periodicity of the crystal lattice hints at the weaker coupling between magnetic and crystal structures and the potential for field control of the magnetic order.
\end{itemize}

\subsubsection{\texorpdfstring{MnSi}{MnSi}}

As noted in Section ~\ref{order-parameter}, intermetallic MnSi crystallizes with the B20 structure described by the Sohncke (but non-enantiomorphic) space group $P2_{1}3$,~\cite{Boren1934} with the absolute structure determined using resonant X-ray scattering~\cite{Dyadkin2016}.
It orders magnetically below 29 K with a helical magnetic ground state in zero-applied magnetic field with periodicity $\lambda$ much larger than the lattice dimensions ($\lambda \approx 190$ {\AA}; $a \approx 4.56$ {\AA}), similar to observations of Cr$_{1/3}$NbS$_{2}$ described above~\cite{Miyadai1983}.
The chiral helical state in MnSi is stabilized by ferromagnetic exchange and (weaker) DMI (allowed in this non-centrosymmetric crystal structure). 
The crystal field (anisotropic exchange) fixes the propagation vector of the helix along the [111] directions of the unit cell but the much longer periodicity compared with the unit cell dimensions reflects the relatively weak coupling between the magnetic and atomic structures~\cite{muhlbauer2009}.
The sense of the magnetic chirality is determined by the crystal chirality: left-handed crystals give left-handed magnetic helices~\cite{grigoriev2010interplay}.
The relative strengths of the symmetric ferromagnetic exchange and antisymmetric DMI change as a function of the applied magnetic field, giving a fascinating phase diagram~\cite{LEBECH1995, Bauer2012}.
A conical magnetic structure is observed at low temperatures with applied field along [111],~\cite{Grigoriev2006} before a ferromagnetic state emerges at higher fields. In a small region of the phase diagram in low applied fields and just below $T_{\rm{C}}$, the “A phase” was observed~\cite{LEBECH1995, Bauer2012}.
Neutron scattering experiments showed that this A phase consists of topologically-protected magnetic vortices known as skyrmions (Figure ~\ref{fig:CrNb3S6} e).~\cite{muhlbauer2009} 
Proposed by mathematician Tony Skyrme,~\cite{Skyrme1961, Skyrme1962} these skyrmion particles are relevant to various fields of physics~\cite{Tokura2021}, but magnetic skyrmions were predicted by Bogdanov and Hubert to be metastable phases in acentric materials, stabilized by DMI~\cite{BOGDANOV1994}. 
Mechanisms that give rise to non-collinear magnetic structures are necessary for magnetic skyrmions, and therefore, much research has focused on chiral magnets with non-collinear magnetic structures stabilized by competing symmetric exchange and DMI. 
However, a wide variety of skyrmion hosts are known, including those with achiral crystal structures, including ~\cite{Tokura2021}:
\begin{itemize}
\item the spinel GaV$_{4}$S$_{8}$ with achiral, but non-centrosymmetric crystal structure which hosts N\'eel skyrmions stabilised by DMI~\cite{Kezsmarki2015};
\item centrosymmetric metallic magnet Gd$_{2}$PdSi$_{3}$ which hosts Bloch skyrmions that result from frustrated magnetic exchange interactions (and not DMI)~\cite{Kurumaji2019};
\item centrosymmetric Gd$_{3}$Ru$_{4}$Al$_{12}$ in which Bloch skyrmions are stabilized by a combination of magnetic frustration and RKKY interactions~\cite{Hirschberger2019}.
\end{itemize}
The observation of these complex non-collinear magnetic textures in materials with a range of crystal symmetries (both centric and acentric) highlights the possibility of achiral crystal structures hosting chiral magnetic structures that result from combinations of symmetric exchange with antisymmetric DMI, frustration or anisotropies, as illustrated by Mn$_{3}$Sn and CaMn$_{7}$O$_{12}$ discussed below.

\subsubsection{\texorpdfstring{Mn\textsubscript{3}Sn}{Mn3Sn}}

The metallic half-Heusler Mn$_{3}$Sn crystallizes in a centrosymmetric structure of $P6_{3}/mmm$ symmetry with Mn sites arranged in a Kagome-type lattice in the $ab$ plane, with Sn sites within the hexagons. 
These Kagome layers are stacked along the $c$ axis, with triangles stacked staggered to give twisted triangular tubes. Below $T_{\rm{N}} \approx 420$ K Mn$_{3}$Sn orders magnetically with Mn spins in-plane in an almost antiferromagnetic arrangement but with a small in-plane ferromagnetic component due to a slight canting of spins~\cite{KREN1970, TOMIYOSHI1983}. 
The spin arrangement is referred to as an “inverse triangle” configuration, which results from exchange interactions (which are geometrically frustrated as a result of the Kagome lattice), as well as the magnetic anisotropy of the Mn sites and the DMI~\cite{Tomiyoshi1982}.
This non-collinear (but co-planar) spin arrangement has vector chirality (i.e., opposite signs for “inverse” and “normal” triangle configurations), and these configurations are non-degenerate due to spin-orbit coupling~\cite{Pradhan2023}. 
This has consequences for the anomalous Hall effect (AHE) observed in Mn$_{3}$Sn, with the sign of the AHE reversed by a small applied magnetic field~\cite{Nakatsuji2015}.

\subsubsection{\texorpdfstring{CaMn\textsubscript{7}O\textsubscript{12}}{CaMn7O12}}
The quadruple perovskite CaMn$_{7}$O$_{12}$ undergoes a series of phase transitions on cooling. 
The magnetic phases observed illustrate the delicate balance of interactions that can give rise to chiral magnetic order~\cite{Johnson2016}.
Starting from the $R\bar{3}$ phase below 400 K, charge ordering of Mn$^{3+}$ and Mn$^{4+}$ ions over the perovskite $B$ sites occurs on cooling before orbital ordering is observed below 250 K giving an incommensurate structural modulation~\cite{Slawinski2009} associated with an orbital density wave~\cite{Perks2012}.
Below $T_{\rm{N_1}} = 90$ K, CaMn$_{7}$O$_{12}$ orders antiferromagnetically and this magnetic phase transition induces a significant ferroelectric polarisation~\cite{Zhang2011}. 
Below $T_{\rm{N_1}}$ the Mn spins order in a chiral, in-plane helical structure~\cite{johnson2012} with periodicity related to the structural incommensurate modulation resulting from orbital ordering~\cite{Johnson2016, Slawinski2010}.
Because the nearest-neighbor magnetic exchange interactions are via Mn$^{3+}$ 3d orbitals, these interactions are modulated by the orbital density wave which affects the spin helix giving regions with more ferromagnetic interactions (and reduced helicity) and regions with more antiferromagnetic interactions (and increased helicity),~\cite{Johnson2016} and so the helix results from competition between these symmetric exchange interactions~\cite{Perks2012}.

It’s striking that this chiral magnetic order emerges from an achiral crystal structure but also that this magnetic ordering breaks inversion symmetry and gives rise to a large electric polarization. 
The polarization is perpendicular to the spin rotation plane, and the magnetoelectric coupling is thought to arise from ferroaxial coupling between the magnetic helix and the global rotation present in the crystal structure~\cite{Perks2012} (rotations of parts of a crystal structure concerning the rest, allowed by certain ferroaxial crystal classes)~\cite{Hlinka2016}.
Spin-orbit interactions are needed for this coupling between the electrical polarisation and the magnetic helicity, i.e., via the DMI, by which the helix's non-collinear spin arrangement stabilises structural distortions. 
In CaMn$_{7}$O$_{12}$, these distortions result in a local polarisation along $c$ (perpendicular to the spin plane), and the global chirality of the magnetic structure means that these local polarisations sum to give a net polarisation, with direction determined by the magnetic chirality~\cite{Perks2012}.

\subsubsection{\texorpdfstring{BaCoSiO\textsubscript{4}}{BaCoSiO4}}

\begin{figure}
\centering
\includegraphics[width=8.5cm,keepaspectratio=true]{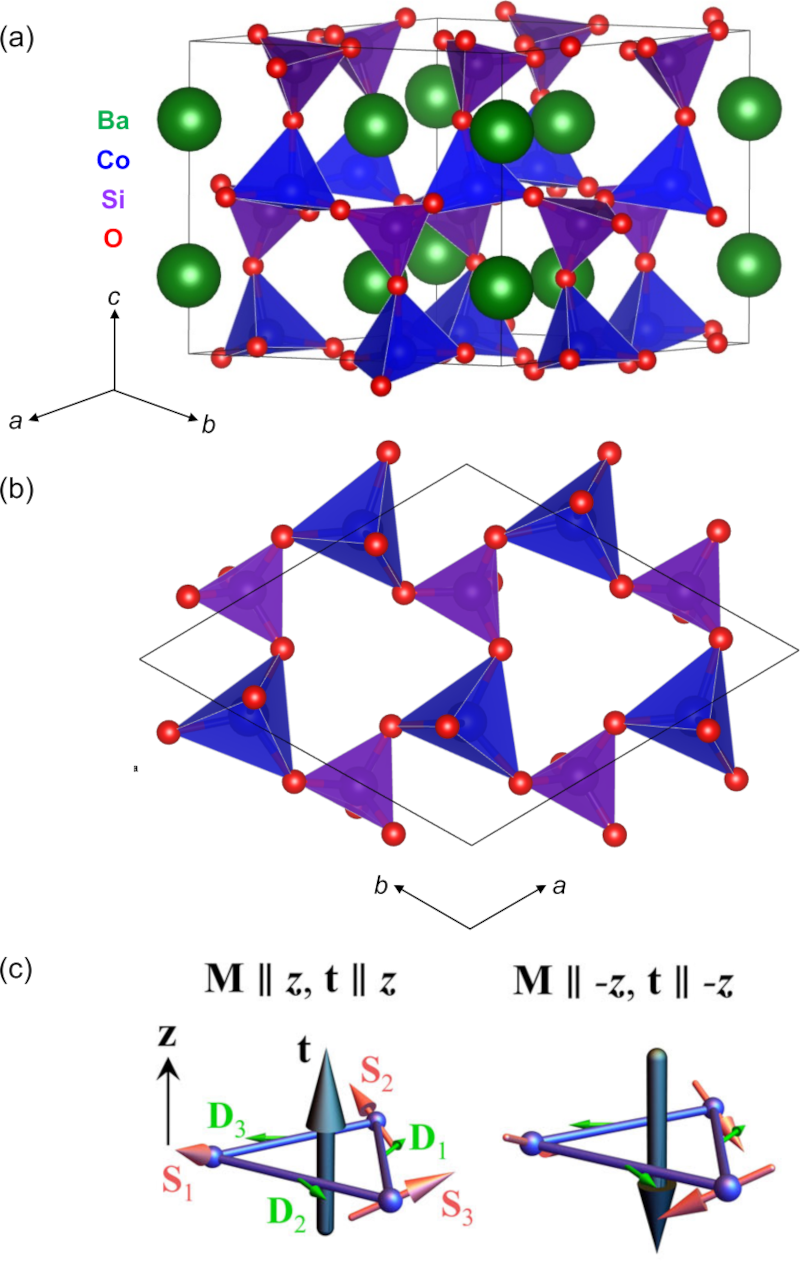}
 \caption{(a) Crystal structure of BaCoSiO$_{4}$ and (b) showing puckered rings of corner-linked SiO$_{4}$ and CoO$_{4}$ tetrahedra, with Ba, Co, Si and O sites shown in green, blue, purple and red, respectively. Panel (c) shows the arrangements of three Co$^{2+}$ arranged in a triangle (spins shown by pink arrows) resulting from frustrated Heisenberg exchange, magnetic anisotropy, and DMI (DM vectors shown by green arrows), giving a toroidal moment (large black arrows) parallel to the magnetization of each triangle (reproduced from reference ~\cite{ding2021} with kind permission from Nature publishing group).}
 \label{fig:BaCoSiO4}
\end{figure}

The stuffed tridymites often adopt chiral and polar crystal structures and, with their compositional flexibility, offer a promising field in which to explore magnetic chirality~\cite{Bhim2021}.
The chiral antiferromagnet BaCoSiO$_{4}$ shows coupling between several ferroic orders and helps illustrate the complex coupling between structural chirality, magnetic chirality, polarity, and ferrotoroidicity~\cite{Xu2022, ding2021}. 
The stuffed tridymite structure is built from corner-linked $B$O$_{4}$ and $B'$O$_{4}$ tetrahedra, with larger $A$ cations in interstitial sites. 
Slight rotations of these tetrahedra give crystal structures of opposite handedness of $P6_{3}22$ symmetry, while ordering of $B$ and $B^\prime$ cations can give a polar axis, giving chiral and polar structures of $P6_3$ symmetry, as observed for BaCoSiO$_{4}$ (Figure ~\ref{fig:BaCoSiO4}). 
Interestingly, $P6_3$ symmetry is of crystal class 6, which is pyroaxial~\cite{Hlinka2016} (allowing ferrorotational order), and so, as discussed for CaMn$_{7}$O$_{12}$ above, this symmetry allows coupling between the structural chirality, magnetic chirality, and electric polarisation via the DMI~\cite{Xu2022}. 
BaCoSiO$_{4}$ orders magnetically below $T_{\rm{N_1}} \approx 3.2$ K with a chiral, triangular arrangement of Co$^{2+}$ spins with a slight canting of spins towards the (polar) $c$ axis. 
This arrangement results from frustrated antiferromagnetic interactions within the triangular units and strong easy-plane magnetic anisotropy, with ferromagnetic canting due to the DMI~\cite{ding2021}.
This ferromagnetic canting gives these triangular units ferrotoroidal moments, with the direction of the toroidal moment linked to the scalar chirality within each triangle (Figure ~\ref{fig:BaCoSiO4}c). 
The antiferromagnetic exchange between triangular units is frustrated, so the ground state is ferritoroidal, but a small magnetic field applied along $c$ can induce a transition to a ferrotoroidal phase~\cite{Xu2022, ding2021}.

\subsubsection {Altermagnetism}
While magnetochirality is a well-recognized effect with widespread applications, we aim to spotlight a novel form of magnetism that holds the potential to reshape our understanding of magnetic properties in chiral materials: altermagnetism~\cite{vsmejkal2022emerging}. 

These new altermagnetic materials share a lack of the nonrelativistic Kramer degeneracy at a general point in the Brillouin zone, finite anomalous Hall effect, and finite magnetooptical effect similar to ferromagnets. 
Typically, spin degeneracy is lifted by one of two means. The first is breaking inversion symmetry in relativistic spin-orbit coupling, which introduces momentum-dependent splitting of the band structure by spin. 
The second is breaking time-reversal symmetry, which results in moment-independent spin splitting. 
This would seemingly indicate that collinear antiferromagnets must have spin degenerate non-relativistic bands. 
Spin degeneracy can be traced back to crystallography, mainly when nonrelativistic spin groups are employed instead of solely relativistic spin-orbit coupling symmetries. 
This approach involves two sequential transformations: the first targets the crystal structure, while the second manipulates the spin space, offering a novel perspective on the underlying mechanisms of these magnetic materials~\cite{vsmejkal2022emerging}. 
In this scheme, the spin sub-lattices of conventional antiferromagnets are connected through a spin rotation and a spatial translation, which preserves time-reversal symmetry, causing the band structure to have spin degeneracy. 
Nonetheless, time reversal symmetry can still be broken when the spin sub-lattices are connected by a spatial translation combined with a spatial rotation in addition to the spin rotation. 
The broken time-reversal symmetry results in spin splitting in the band structure, similar to ferromagnets, while there is zero net magnetization, which defines the new altermagnetic phase.

Given that sublattices units can be connected by rotation and translation, it should be expected that some antiferromagnetic enantiomorphic chiral crystals should also manifest altermagnetism, as they can host the primary magnetic anticipated groups to show this effect~\cite{turek2022altermagnetism, cheong2024altermagnetism}. 
Though this effect has not (yet) been reported in chiral structures belonging to one of the enantiomorphic space groups, it has already been reported to be the case in the non-enantiomorphic Sohncke space groups~\cite{chang2018topological, sanchez2019topological, vsmejkal2020crystal}, hence, this is calling for further studies.

\subsection{Topological Chiral Materials}

\begin{figure}[htb!]
\centering
\includegraphics[width=8.5cm,keepaspectratio=true]{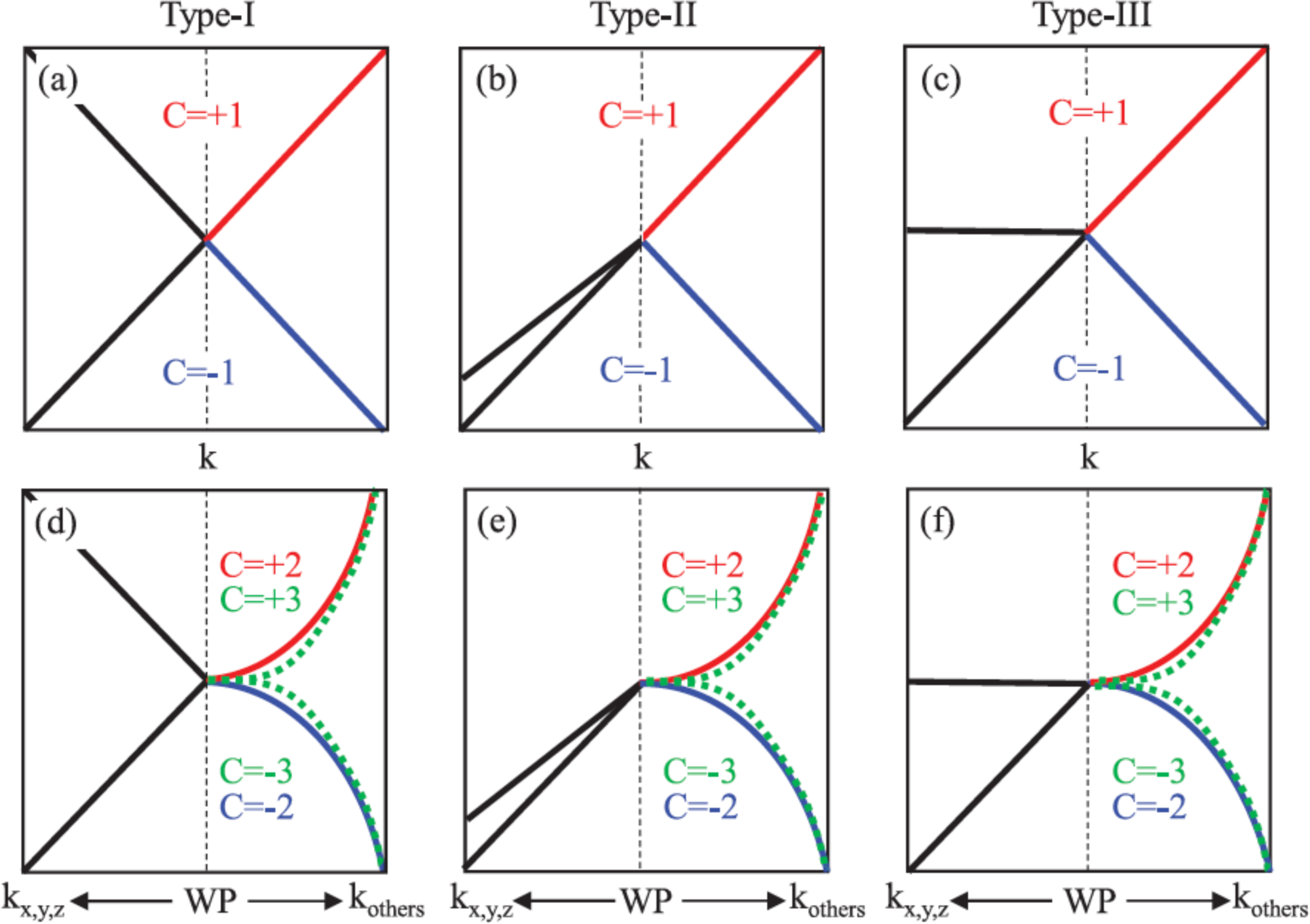}
 \caption{Three types of unconventional Weyl phonons (WPs) localized at HSLs. a–c) Phononic dispersions along the kx(y, z)-WP-kother path for Type-I,
and Type-II and Type-III charge-one WP. d–f) Phononic dispersions along the kx(y, z)-WP-k other path for Type-I, Type-II, and Type-III CTWP and charge-three
WP, where the solid red-blue and green-dashed lines denote the phononic dispersions of charge-three WP. Taken from Ref.~\cite{yang2023unconventional}.}
\label{fig:WPClassificationPhonons}
\end{figure}

The intricate relationship between structural chirality and spin has garnered significant attention recently, as evidenced by several notable publications~\cite{schroter2019, chang2018topological, adhikari2023interplay, wieder2022topological, lv2021experimental}. 
In particular, the work of Chang {\it{et al.}} delves into the intriguing connection between these two facets of condensed matter physics~\cite{chang2018topological}.
They have elucidated how Kramers–Weyl fermions, a fascinating and universal topological electronic property, are inherent to non-magnetic chiral crystals characterized by spin-orbit coupling. 
This discovery is particularly intriguing because these Kramers–Weyl fermions are not merely coincidental but are intricately tied to structural chirality, lattice translation symmetries, and time-reversal symmetry. 
In particular, to realize Kramers-Weyl points, two criteria should be met: one breaks inversion symmetry, $\mathcal{P}$, but respects time-reversal symmetry, $\mathcal{T}$, whereas the other is the removal of double degeneracy at nontime reversal invariant points (TRIMs). Figure~\ref{fig:WPClassificationPhonons} presents a simple description of Weyl point classification in phonons, described in Ref.~\cite{yang2023unconventional}.
Therefore, a chiral lattice becomes a prime candidate for realizing Kramers-Weyl fermions as it inherently lacks inversion, mirror, or other rotoinversion symmetries, giving it a well-defined ``handedness''. 

Recognizing these interrelationships opens up exciting avenues for further exploration and understanding of chiral materials, their topological properties, and the intriguing role of chirality in governing electronic behavior. 
For example, it has a unique spin texture, giant helicoid Fermi arcs, large topologically nontrivial energy windows, quantized circular photogalvanic effects, etc~\cite{hasan2021}. 
The emergence of these topological states represents a unique scenario in which traditional frameworks like the Landau-Ginzburg-Wilson theory fail to explain the transitions leading to these states. 
Notably, these are conditions where no symmetry is spontaneously broken. 
Instead, understanding these states relies on a concept known as topological invariance, commonly referred to as the Chern number~\cite{thouless1982quantized, simon1983holonomy}. 
This Chern number serves as a global parameter that encapsulates the topological essence of the system and is determined by the structure of the manifold that encloses these states. 
This approach diverges from conventional theories and offers a more nuanced understanding of the topological properties of the system. 
In particular, Chang {\it et al.} suggest several materials (where we only report the enantiomorphs) that can have Kramers–Weyl fermions: TlBO$_2$ ($P4_1$ space group 76),
Sr$_2$As$_2$O$_7$ ($P4_3$ space group 78),
Ag$_3$SbO$_4$ ($P4_122$ space group 91),
MgAs$_4$ ($P4_12_12$ space group 92),
H$_4$Ca$_2$AsF$_{13}$  ($P4_322$ space group 95),
m-Cu$_2$S ($P4_32_12$ space group 96),
DyAl$_3$Cl$_{12}$ ($P3_112$ space group 151),
IrGe$_4$ ($P3_121$ space group 152),
SrIr$_2$P$_2$ ($P3_221$ space group 154),
$\alpha$-In$_2$Se$_3$ ($P6_1$ space group 169),
BaN$_2$O$_4 \cdot $H$_2$O ($P6_5$ space group 170),
Hf$_5$Ir$_3$ ($P6_122$ space group 178),
Na$_3$B$_4$O$_7$Br ($P6_522$ space group 179),
NbGe$_2$  ($P6_222$ space group 180),
WAl$_2$  ($P6_422$ space group 181),
Li$_2$Pd$_3$B ($P4_332$ space group 212),
Mg$_3$Ru$_2$ ($P4_132$ space group 213). 
From this list, we find semi-metals and small-gap insulators; some can also host unconventional four-fold-degenerate chiral fermions at TRIMs and may thus host the unconventional spin-3/2 chiral fermion~\cite{bradlyn2016beyond} or even quantized photocurrent. 
Other materials also reported recently are, e.g. Li$_2$Pt$_3$B ($P4_332$ space group 212), which is a  s-wave superconductor with an antisymmetric spin-orbit coupling induced by the broken inversion symmetry that leads to the spin-splitting of the Fermi surface and gives to unique superconducting properties, such as the parity mixing of Cooper pairs; 
SrSi$_2$ is a nonmagnetic double Weyl semimetal with a Circular Photogalvanic Effect reported~\cite{sadhukhan2021electronic}.

The low numbers of Kramers-Weyl fermion hosts reported for some chiral space groups (indeed, no reported structures for several chiral symmetries!) highlights the gaps in the landscape of chiral materials and motivates materials design and discovery (see Section ~\ref{databases}).
According to Ref.~\cite{yu2022encyclopedia}, all the enantiomorphic groups should have materials with emergent particles. 
This last reference provides a comprehensive classification of emergent particles in time-reversal-invariant systems. 
The paper focuses on spinful and spinless particles in three-dimensional (3D) systems and categorizes them based on the 230 space groups. 
For example, they report that most enantiomorphs (magnetic and nonmagnetic) have topological properties such as Charge $-2$, $-3$, $-4$ Weyl points, can have nodal surfaces, and have Dirac points. 
Here we want to stress that the lack of known and reported chiral materials hinders the exploration of the topological materials in enantiomorphic groups as the list of enantiomorphic groups with representative compounds for emergent particles is incomplete. 

On the other hand, we encounter a fascinating phenomenon in condensed matter physics: Emerging particles or excitations owe their existence to degenerate band crossings within the dispersion relation. 
These band crossings can involve electrons, phonons, or magnons and are safeguarded by specific symmetries inherent to the system. 
Topological materials can typically be categorized based on the type of band crossing they exhibit. 
The first category consists of materials with zero-dimensional (0D) band crossings, commonly called nodes or nodal points. 
These include Dirac Semimetals (DSMs) and Weyl Semimetals (WSMs), where either doubly or singly degenerate bands intersect at discrete points near the Fermi level ($E_F$). 
The second category features materials with one-dimensional (1D) band crossings, which occur along lines in momentum space.
Such materials are generally termed Topological Nodal-Line Semimetals (TNLSMs) and exhibit four or twofold band crossings along these lines. 
Finally, the third category encompasses materials with two-dimensional (2D) band crossings preserved on a surface within the 3D Brillouin Zone (BZ). 
These materials are topological nodal-surface semiconductors and represent a more complex form of band crossing.
An excellent discussion of the differences between these groups can be found in Ref.~\cite{lv2021experimental}.
Weyl particles serve as a pioneering example of topological materials. 
In this context, these particles emerge due to the convergence of two distinct bands in the material's energy dispersion, one for each spin channel. 
What distinguishes Weyl particles is their unique status as symmetry-protected entities arising from the crystal's inherent symmetries. 
This symmetrical protection bestows these particles a distinctive topological charge, a quantized Berry flux residing in momentum space. 
Within Weyl particles, this topological charge is quantified by an integer known as the Chern number, as introduced before. 
This numerical value plays a pivotal role, imparting a sense of handedness to the electronic wavefunction associated with these particles and capturing the topology of the filled electron bands. 
In essence, the Chern number encapsulates the topological properties of Weyl particles, dictating their behavior and providing an intrinsic link between their emergence and the underlying symmetries of the crystal lattice. 
The Chern number generally identifies the topological invariance of quantum Hall insulators. 
This connection between topology and symmetry is a profound revelation in condensed matter physics, offering a deeper understanding of how exotic particles like Weyl fermions come into existence and how they influence the electronic properties of materials.

Given the structural attributes of the chiral materials under discussion, it is evident that their unique topological properties and emergent particles extend to other excitations, such as in phonon band structures. 
For instance, KMgBO$_3$ has been reported to feature multifold and multidimensional topological phonons and charge-four Weyl phonons~\cite{sreeparvathy2022coexistence}. 
Similarly, TlBO$_2$ (space group 76) exhibits multi-Weyl points in its nonmagnetic crystals~\cite{wang2022single}. 
In the case of $\alpha$-TeO$_2$ (space group 92), non-symmorphic twist symmetries underlie phonon band anti-crossings and topological behaviors~\cite{juneja2021materials}. 
CsBe$_2$F$_5$ (space group 213) uniquely hosts Type-I and Type-II unconventional charge-two Weyl points with varying Chern numbers~\cite{yang2023unconventional}. 
MgAs$_4$ ($P4_12_12$ space group 92) features topological Weyl points arising from twisting phonon symmetries that combine rotational and translational elements~\cite{juneja2022phonons}. 
The coexistence of charge-2 Weyl and charge-2 Dirac points characterizes BaPt$_2$S$_3$ ($P4_12_12$ space group 92)~\cite{wu2023paired}. 
In contrast, K$_2$Mg$_2$O$_3$ ($P4_32_12$ space group 96) and Nb$_3$Al$_2$N ($P4_132$ space group 213) host isolated Weyl phonons with two or four Chern numbers in their acoustic branches~\cite{liu2022beyond}. 
Moreover, there are instances where topological phonons coexist with other topological particles, such as the reported coexistence between charge-2 Dirac points and Weyl phonons in enantiomorph space groups 92, 94, and 96, specifically in Na$_2$Zn$_2$O$_3$ ($P4_32_12$ space group 96)~\cite{liu2022coexistence}.
This extensive work underscores chiral materials' rich topological landscape regarding emergent particles and complex phonon band structures.

    \subsection{Electric properties}    
\label{sec:Gyroelectricity}
While the change of optical activity of magnets (or diamagnetic substances)  
under an applied magnetic field was discovered by Faraday in 1845 (the Faraday effect, see Section~\ref{magnetism}),
the analysis of the equivalent concept for the electric case, i.e. the change of optical activity under an applied electric field, came more than a century later~\cite{aizu1964, zheludev1976, stasyuk1983, kobayashi1983, koralewski1984, vlokh1987, uesu1990, gunning1997, vlokh2001}.
The (linear) electrogyration tensor $\eta$ can be defined as the change of optical activity under an applied electric field $\mathbf{E}$:
\begin{equation}
\label{eq:electrogyration-tensor}
    \gamma_{ijk} = \left .
    \frac{\partial g_{ij}}{\partial E_k}
    \right \vert_{\mathbf{E}=0},
\end{equation}
where $i$, $j$ and $k$ refer to the Cartesian directions and $g$ is the gyration or optical activity tensor.
Hence, as for the Faraday effect, a crystal not optically active can show electrogyration as an applied electric field can induce a non-zero optical activity.
All crystals can exhibit linear electrogyration except those with symmetry $m3m$, $\bar{4}3m$ and $432$~\cite{vlokh1987, buckingham1971}. 
Of course, higher-order electrogyration responses (quadratic electrogyration) can exist if one treats higher derivatives of the optical activity tensor.

\begin{figure*}[htb!]
\centering
\includegraphics[width=17cm,keepaspectratio=true]{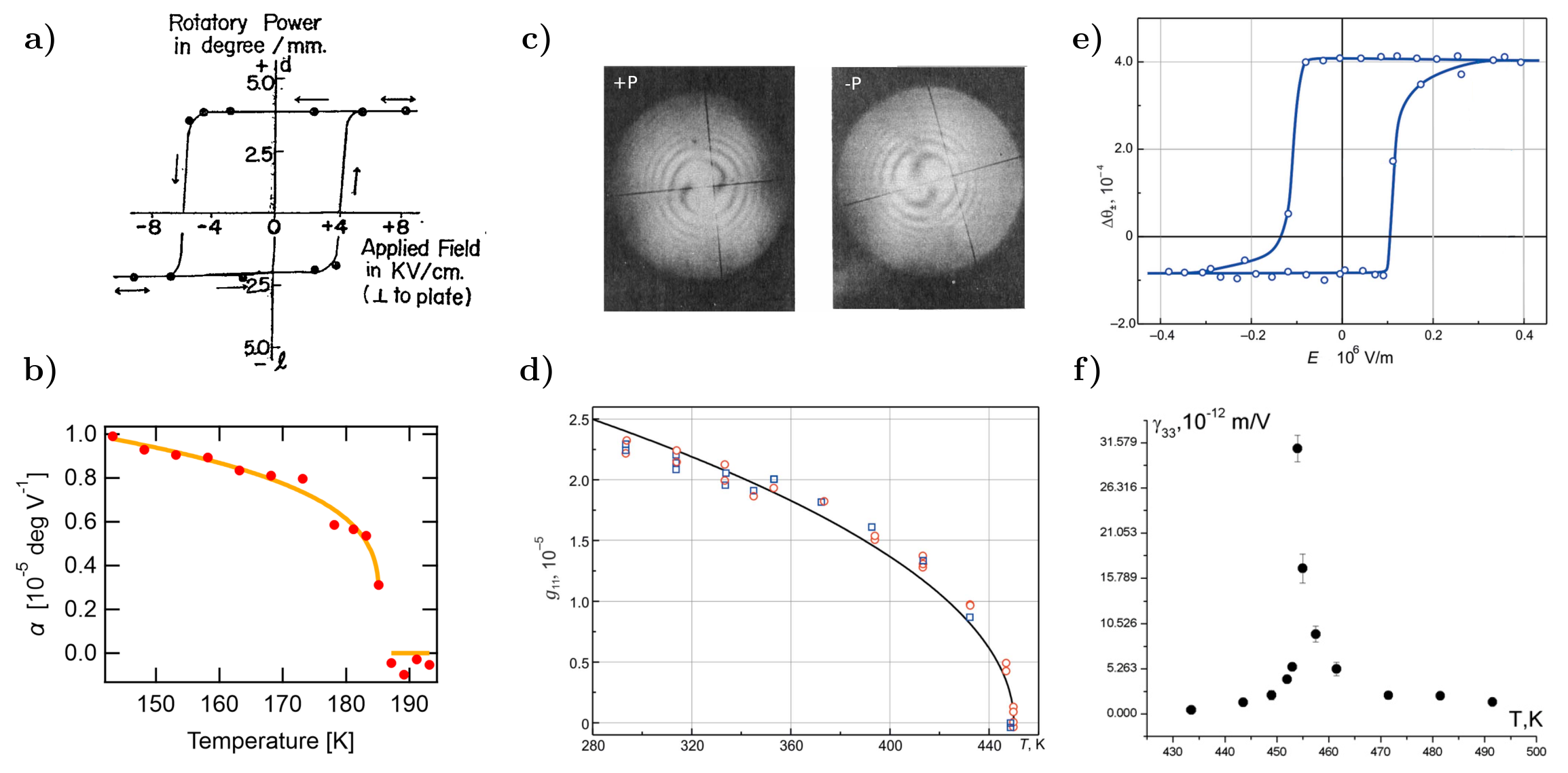}
 \caption{a) Hysteresis loop of the optical activity rotation versus electric field as measured in LiH$_3$(SeO$_4$)$_2$, from~\cite{futana1962}. b) Temperature evolution of the absolute value of the electrogyration coefficient of a single domain of RbFe(MoO$_4$)$_2$, from~\cite{hayashida2021}. c) Airy's spirals of the ferroelectric Pb$_5$Ge$_3$O$_{11}$ crystal for the two polarized domains ($+P$ and $-P$), adapted from~\cite{iwasaki2011}. d) Temperature dependence of the absolute value of the gyration component of Pb$_5$Ge$_3$O$_{11}$ as measured experimentally (red circles and blue squares), the black curve shows the theoretical evaluation, from~\cite{shopa2021}. e) Hysteresis flip of the optical activity angle versus electric of Pb$_5$Ge$_3$O$_{11}$, adapted from~\cite{shopa2021}. f) Temperature dependence of the electrogyration coefficient of Pb$_5$Ge$_3$O$_{11}$:Cr crystal, from~\cite{adamenko2008}. }
 \label{fig:electrogyration}
\end{figure*}

The special case where the optical activity is reversed by an applied electric field is called gyroelectricity~\cite{aizu1964, kizel1975} or ferrogyrotropy~\cite{wadhawan1979, wadhawan1982}.
There is also the special case of  hypergyroelectricity~\cite{aizu1964, vlokh1987}), which is defined as crystals where it is the electrogyration (vs the optical activity reversal for gyroelectrics) that can be reversed by an electric field.
There is an intimate link between gyroelectricity and ferroelectricity, as reversing the optical activity under an electric field is done by reversing the polarization.
Some of the first observations of gyroelectricity were made on the Ca$_2$Sr(C$_2$H$_5$CO$_2$)$_6$~\cite{kobayashi1962} and  LiH$_3$(SeO$_4$)$_2$~\cite{futana1962} ferroelectric crystals in 1962. 
In these measurements, a hysteresis loop is observed when plotting the optical activity versus the applied electric field (see Figure~\ref{fig:electrogyration} a for the case of LiH$_3$(SeO$_4$)$_2$).
Having such a switchable optical activity under an applied electric field with a hysteresis loop shape raised the question of whether the optical activity can form or be associated with a new ferroic order~\cite{wadhawan1979}. 
However, as the optical activity tensor is mediated by the ferroelectric order, it has been defined as an implicit form of ferroicity, i.e., it can only be affected through an accompanying explicitly ferroic order~\cite{wadhawan1979}.
Hence, in a gyroelectric crystal, the optical activity tensor evolution with temperature follows the ferroelectric polarization, i.e. evolves as a square root function below $T_C$ (see, e.g. Figure~\ref{fig:electrogyration} b for RbFe(MoO$_4$)$_2$ or Figure~\ref{fig:electrogyration} d for Pb$_5$Ge$_3$O$_{11}$).
Then, several crystals were reported for their electrogyration and/or their gyroelectric response, such as the boracites Co$_3$B$_7$O$_{13}$I and Cu$_3$B$_7$O$_{13}$Cl~\cite{takahashi1992}, the sillenites Bi$_{12}$SiO$_{20}$ and  Bi$_{12}$GeO$_{20}$ ~\cite{deliolanis2004, deliolanis2006}, Sn$_2$P$_2$S$_6$~\cite{vlokh2008}, Stolzite PbMoO$_4$ and  PbWO$_4$~\cite{novikov2017} or more recently in the ferroaxial phase transitions reported in NiTiO$_3$~\cite{hayashida2020b} and RbFe(MoO$_4$)$_2$~\cite{hayashida2021} (see Figure~\ref{fig:electrogyration} b).
Among them, one of the most studied crystals for its large gyroelectric and electrogyration response is the aforementioned lead germanate Pb$_5$Ge$_3$O$_{11}$ 
(see Sections ~\ref{PGO} and ~\ref{order-parameter} and Figure~\ref{fig:electrogyration}).
Figure~\ref{fig:electrogyration} c shows the Airy's spirals of the two ferroelectric domains of Pb$_5$Ge$_3$O$_{11}$ where we can observe the change of the rotation direction in the two cases.
Figure~\ref{fig:electrogyration} d shows the evolution of the temperature dependence of the gyration amplitude of Pb$_5$Ge$_3$O$_{11}$ which becomes non-zero at the ferroelectric phase transition and evolves as a square root function, i.e., like the polarization of the material.
Figure~\ref{fig:electrogyration} e shows how the optical activity angle evolves under an applied electric field in Pb$_5$Ge$_3$O$_{11}$.
Here again, we can see that the optical activity exhibits a hysteresis shape as it follows the switching of ferroelectric polarisation under an applied electric field.
Finally, Figure~\ref{fig:electrogyration} f shows the evolution of the electrogyration coefficient of doped Pb$_5$Ge$_3$O$_{11}$:Cr crystal when varying the temperature.
Here again, at the critical temperature where the ferroelectric phase transition appears, we can observe the divergence of the electrogyroelectric response, with very large values compared with crystals that do not have such a phase transition.
Recently, M. Fava {\it{et al.}}~\cite{fava2023ferroelectricity}, have shown from DFT calculations that the chirality of the ferroelectric phase 
comes from a polar soft mode that induces 
the phase transition. 
This offers a microscopic understanding of why the optical activity follows the ferroelectric polarization in a one-to-one way, given that the polar distortion is also chiral.

We can see that electrogyration and gyroelectricity were topics put aside in the past 10 years. Still, they could be brought back to the forefront with the increase in interest in chiral crystals and the recent advance in photonic applications.

\subsection{Strain}

The study of elastic properties in chiral materials is an emerging field that has garnered increasing attention over the years. 
One of the pioneering works in this area can be traced back to the study by Ben {\it{et al.}}~\cite{ben2001chromatin}, which delved into the elastic constants of the 30 nm chromatin fiber. 
The study examined the fiber's DNA elastic properties and geometric attributes and made significant inferences about how the inherent chirality of the chromatin fiber influences its mechanical characteristics. 

In the context of structural chiral materials, the work by Huang {\it{et al.}}~\cite{huang2020}, which delves into the atomic-scale sensitivity of elastic properties in chiral crystals, stands out in the field.
The paper is particularly noteworthy for exploring the Dresselhaus effect in ferri-chiral crystals and specifically identifying NaCu$_5$S$_3$ as a material that exhibits this unique behavior. 
The study further discusses the potential for switching chirality, affecting the Dresselhaus spin splitting. 
This opens up new avenues for understanding the intricate relationship between chirality and spin in periodic crystals, offering a compelling alternative to the conventional Rashba spin textures. 

On the other hand, the interplay between chirality and optical activity as influenced by strain remains a relatively unexplored area in materials science, however a good theoretical description of the phenomena is provided in ref.~\cite{wadhawan2000introduction}. 
While some studies have touched upon this subject, as seen in materials like NaCu$_5$S$_3$~\cite{zhang2021show} or  Co$_3$Ni$_3$Ga$_8$~\cite{singh2022cobalt}, they offer limited insights into how strain specifically alters the optical activity. 
One of the seminal studies in this field centers on dicalcium strontium propionate, Ca$_2$Sr(C$_3$H$_5$CO$_2$)$_6$~\cite{wadhawan1979, sawada1977ferroelasticity,Glazer_1981}. 
This unique material is both ferroelastic and ferroelectric, exhibiting gyrotropic behavior under the influence of uniaxial stress. 
The authors of these works provide an in-depth analysis of how mechanical strain impacts the optical properties of dicalcium strontium propionate (DSP) across its various phases and domains.
The material undergoes a subtle lattice distortion upon applying strain, transitioning from a high-symmetry phase with $m3m$ symmetry to a low-symmetry phase characterized by $422$ symmetry. This phase transition occurs at a critical temperature of 281.7 K, leading to six distinct gyrotropic states, each defined by a unique gyration tensor. 
This structural phase transition can be generalized regarding macroscopic symmetry breaking, as discussed in Ref.~\cite{erb2020vector}.

The elastic properties of other chiral materials have also been studied, including
TeO$_2$~\cite{thomas1988crystal}, where $\alpha$-TeO$_2$ is noted for its acoustic-optic properties, which result from its unusual elastic behavior and high birefringence;  (C$_5$H$_{11}$NH$_3$)$_2$ZnCl$_4$~\cite{srinivasan2004structural}, where strain can break the symmetry of this electro-toroid system and the system becomes spontaneously optical active; 
Cu$_2$OSeO$_3$~\cite{zhang2017direct, nomura2019phonon}, where the dichroism extinction effect measured in this material which shows helical, conical, and skyrmion phases, and the sound velocity is affected by direction of an applied magnetic field CrNb$_3$S$_6$~\cite{kishine2020}, where the hybridization of the rotational and translational vibrational modes provides a unique response related to elastic response. 
(The effect on the optical activity of CrNb$_3$S$_6$ is not yet reported.)

Interestingly, the direction and magnitude of the applied stress can induce shifts in both the ferroelastic and gyrotropic states across different domains within the material~\cite{wadhawan2000introduction}. 
This makes DSP one of the rare instances where external forces can selectively stabilize certain domains over others~\cite{wadhawan1979}. 
Transitions between states in different domains are triggered when the applied force creates a sufficient disparity in the stored free enthalpy between the domains, overcoming the energy barrier that separates them. 
This manifests as a change in sign in the optical activity tensor.
Furthermore, although the activity tensor remains invariant under time inversion, its behavior can be modulated in magnetism through coupling with the magnetoelectric effect.

\section{Chiral phonons}\label{sec:chiral-phonons}

In an attempt to respond to the question of whether "if electrons can carry chirality at valleys in hexagonal 2D arrays, might photons do the same?",
Zhang and Niu~\cite{zhang2015} put the notion of phonon chirality in the spotlight in 2014 from a theoretical perspective.
The experimental validation through an optical pump-probe technique and infrared circular dichroism came a few years later~\cite{zhu2018}, which has pushed the interest in chiral phonons to the forefront~\cite{chen2019, romao2019, kishine2020, ptok2021, suri2021, juraschek2022, ishito2022, chen2022chiral, xiong2022b, yao2022, tsunetsugu2023, geilhufe2023, ueda2023}.
Indeed, observing and manipulating chiral phonons has opened new research directions such as the phonon Hall effect, phonon Berry curvature~\cite{qin2012,saito2019,park2020,saparov2022}, phonon effective magnetic field and induced orbital or spin magnetization~\cite{geilhufe2023, xiong2022, juraschek2019,juraschek2022, fransson2023}, current induced by chiral phonons~\cite{yao2022}, axial thermal expansion~\cite{romao2019} or even using chiral phonons as dark matter detector~\cite{romao2023}, all of them extending the field of phononic applications to new horizons~\cite{maldovan2013}.

\begin{figure*}
\centering
\includegraphics[width=18cm,keepaspectratio=true]{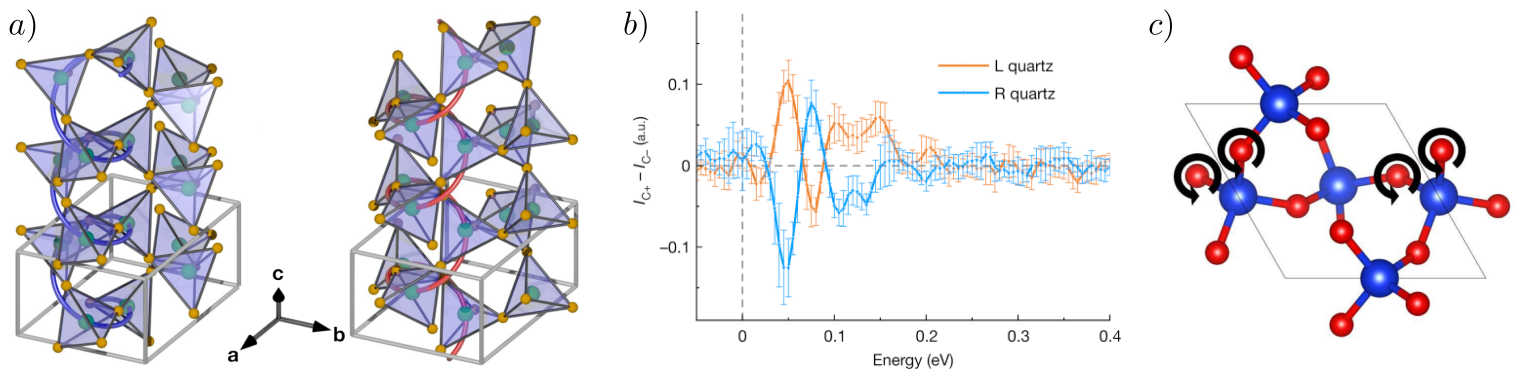}
 \caption{a) Schematic picture of left ($P3_221$ space group) and right ($P3_121$ space group) handed quartz crystal structures. The tetrahedra represent the SiO$_4$ units. b) Circular dichroism of the left and the right-handed quartz as extracted from the resonant X-ray scattering at the reciprocal lattice vector q=(-0.25, 0, 0.32). c) Schematic picture of the oxygen chiral motions of the phonon mode probed at  q=(-0.25, 0, 0.32). Red balls are for the O, and blue balls for the Si. The three figures are adapted from~\cite{ueda2023}.}
 \label{fig:phonon-quartz}
\end{figure*}

To qualify the phonon chirality, Zhang and Niu used the pseudo angular momentum (PAM) of the eigendisplacements associated with this phonons~\cite {streib2021}.
In this way, they split the phonon eigenvectors into right and left-handed sublattices and could define a circular polarization of the phonons.
The direct use of PAM was barely studied except when it interacts with the magnetic properties.
Another interest for PAM is observing, characterizing, and exciting chiral phonons through polarized light. 
This was exemplified by Ishito {\it{et al.}} in their polarized Raman scattering experiments to identify chiral phonons in cinnabar ($\alpha$-HgS)~\cite{ishito2022}.
They demonstrated the possibility of probing the chirality of the crystal domains through non-contact and non-destructive measurements. They showed that the conservation law of the pseudo-angular momentum between light and phonons allows momentum transfer, hence ideal for combined opto-phononic technological applications~\cite{juraschek2020}.
Clear experimental measurement of chiral phonons has recently been made in quartz through resonant inelastic X-ray scattering with circularly polarized X-rays~\cite{ueda2023}.
By construction, the circularly polarized X-rays are chiral such that they can interact with chiral phonons of the reciprocal space and, hence, dispersion of the chiral phonon modes can be obtained, see Figure~\ref{fig:phonon-quartz}.
The authors also performed DFT calculations and showed that the charge quadrupole of the O 2p orbitals in the chiral phonon eigenvector is the source of the dichroic X-ray signal detected experimentally.

In their article ``Orbital magnetic moments of phonons''~\cite{juraschek2019}, Juraschek and Spaldin made a systematic DFT calculation of the induced orbital magnetization coming from PAM in ionic crystals. 
Because chiral phonons will carry PAM by definition, they will naturally carry an orbital magnetic moment.
In ionic crystals, the rotation of charged ions will induce magnetic moments through their gyromagnetic ratio. 
Hence, defining a phonon magneton as a unit measure of this effect is possible. 
Juraschek and Spaldin found that the order of magnitude of this phonon magneton is about 10$^{-4} \mu_B$.
This effect could be observed experimentally by ultrafast laser excitations of phonons.

\begin{figure}
\centering
\includegraphics[width=8.5cm,keepaspectratio=true]{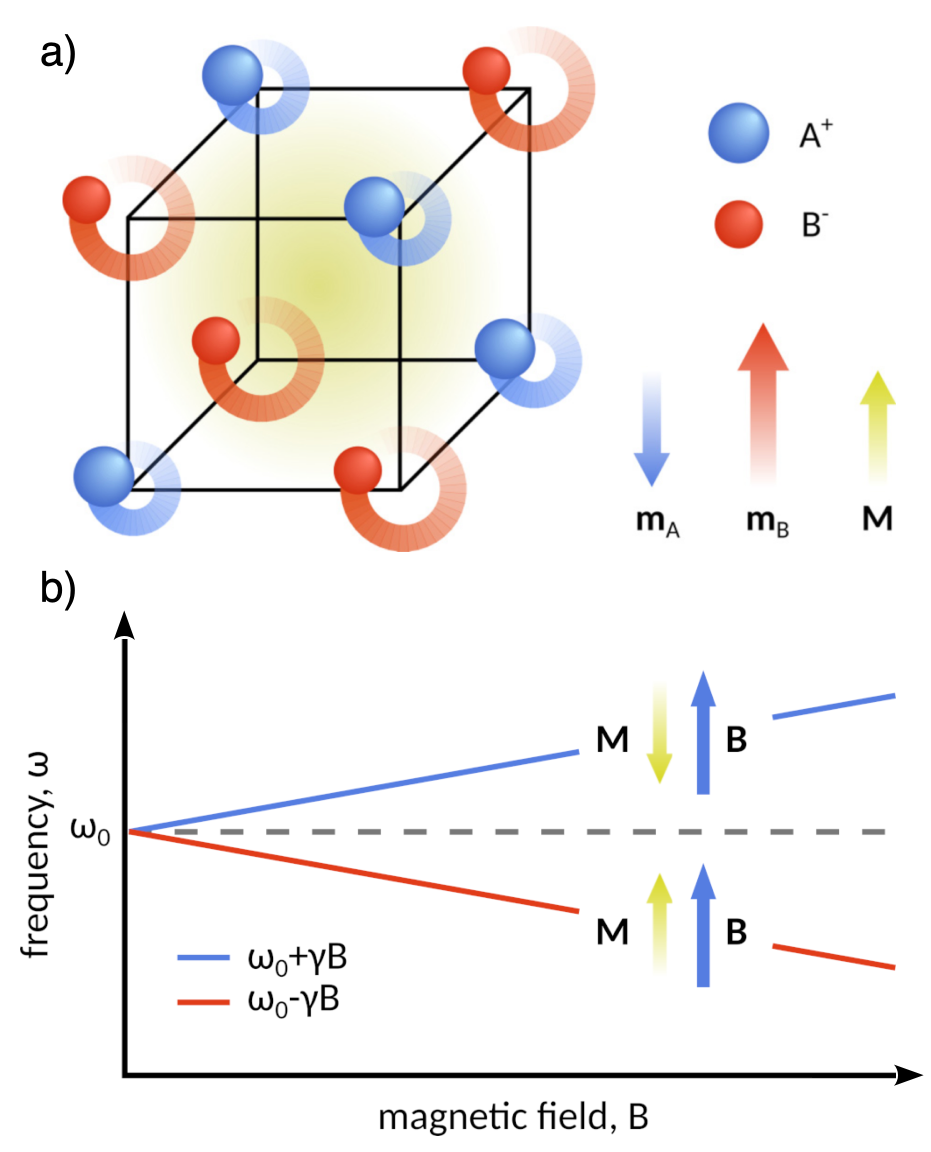}
 \caption{a) Schematic picture of local circular motions of ions as present in chiral phonons, here for a hypothetical diatomic ionic crystal (e.g., PbTe~\cite{baydin2022}). The circular motions of the ions create local magnetic moments of different amplitudes $m_A$ and $m_B$ such that the total magnetic moment $M$ is non-zero. b) Schematic picture showing the phonon Zeeman splitting of a degenerate mode. The phonon frequency ($\omega$) is plotted concerning the magnetic field amplitude ($B$) where the phonon eigenvector with opposite magnetic moment to the field is shifted upward ($\omega_0+\gamma B$). In contrast, the one with moment aligned with the field is shifted downward ($\omega_0-\gamma B$) where $\gamma$ would represent the gyromagnetic ratio of the phonon.  Adapted from ~\cite{juraschek2017}.}
 \label{fig:phonon-magnetic-moment}
\end{figure}

The effect of chiral phonons on magnetism can be seen as an effective magnetic field on the system, also described as a phonon analog of the inverse Faraday effect in optics~\cite{vanderziel1965, popova2011, juraschek2020, juraschek2022, xiong2022}.
Xiong {\it{et al.}} ~\cite{xiong2022} used a classical point charge model (using the Born effective charges of the atoms) and a fundamental magnetostatic Biot and Savart law to show that the circular motion of charged atoms of chiral phonons will induce a local effective magnetic field.
They show that in systems with time-reversal symmetry, chiral phonons' total induced effective field is zero but can be observed through temperature gradients.
When applied to bulk tellurium, their model can give a total effective magnetic field of 0.01 Tesla at room temperature.
They also show that this effective magnetic field from chiral phonons can renormalize the Curie temperature of ferromagnetic substances with Heisenberg model interactions.
While this effective magnetic field of chiral phonons was reported to be relatively small, recently, Juraschek, Nueman, and Narang~\cite{juraschek2022} have estimated from first-principles calculations and microscopic models that the effective magnetic field can be larger than 100 Tesla in rare earth triclhorides under the conditions of ultrafast laser excitation experiments.
This can also explain the large phonon splitting between left and right circularly polarized eigenvectors under applied magnetic field reported by Schaack in the paramagnetic CeF$_3$~\cite{schaack1976} or more recently in the Dirac semimetal Cd$_3$As$_2$~\cite{cheng2020} where a large phonon Zeeman effect was observed (with an effective phonon moment of 2.7$\mu_B$).
If such a giant magnetic effective field induced by ultrashort terahertz pulses is confirmed experimentally, this will put the field of phonomagnetism at a new scale with a high impact in materials science~\cite{nova2017,afanasiev2021,kimel2022}.
Such an effect being possible in non-magnetic insulators opens the way for inducing dynamical magnetoelectricity in non-magnetic non-polar materials through chiral phonons is now also at its infancy~\cite{juraschek2017, geilhufe2021, baydin2022, geilhufe2023, bossini2023}.

Another effect that has been raised by Geilhufe~\cite{geilhufe2022} is that the circular motion of ions introduces inertial effects on electrons, which can be seen as effective Zeeman-like splitting of the phonon frequencies. 

The transfer of angular momentum between phonons, electrons, and magnetism (spin and/or orbital)
relies on the electron-phonon coupling. However, its exact microscopic mechanism was not clearly defined until recently~\cite{hamada2020}. 
This transfer of angular momentum to magnetism is intimately linked to the Einstein/de Haas or Barnett effect, which highlights the fact that changing the magnetization of a material induces a mechanical rotation of the sample and the other way around, respectively~\cite{barnett1935, zhang2016}.
Recently, Mentink, Katsnelson, and Lemeshko~\cite{mentink2019} have developed a microscopic theory of an ``angulon'' quasiparticle to unify these effects. Through this theory, they show that new high-frequency effects should be observed in electron spin resonance that could help understand the transfer of angular momentum appearing in ultrafast experiments.

The chiral phonons field has been particularly active within the past few years, where a first dedicated CECAM workshop was organized in July 2023~\cite{CECAM-chiral-phonons}. 
Hence, we can expect more regarding chiral phonons in both fundamental research (chiral bosons) and new technological applications (phononics).

To conclude this section, we mention that the concept of chirality can be associated with zone-center phonons, although the corresponding modes must have zero angular momentum. Different metrics should be adopted in that case, such as the Continuous Chirality Measure (CCM)~\cite{fecher2022} for example. Recently applied to the chiral eigendisplacements of Pb$_{5}$Ge$_{3}$O$_{11}$ at the $\Gamma$ point, the CCM has been found to increase concerning the phonon frequency~\cite{fava2023ferroelectricity} due to the significant presence of oxygen in the unit cell. 
This suggests the use of high-frequency electric fields as a way to realize the control of this chiral degree of freedom through a fast-switching mechanism~\cite{fava2023ferroelectricity}.
Moreover, a $\Delta\mathbf{L}_{ph}$ difference - which could be interpreted as a torque - can be induced by a nonchiral to chiral phase transition as also reported in Ref.~\cite{fava2023ferroelectricity}.


\section{Chiral order parameter and control of structural chirality}
\label{order-parameter}
An order parameter can be defined as a degree of freedom whose magnitude signifies the transition from one phase to another. Typically, these phases are distinguished by the presence or absence of specific symmetries. The existence of an order parameter is linked to the notion of (spontaneous) symmetry breaking (SSB), which aligns with the Landau theory of phase transitions
~\cite{toledano1987landau}, although several exceptions are known (e.g. topological~\cite{doi:10.1142/11016}, liquid-gas or reconstructive~\cite{dmitriev1996reconstructive} phase transitions, for instance).
The idea of a spontaneous order appearing below a critical temperature $T_{c}$ is thus generally associated with the breaking of symmetry as a result of an instability of some kind in the system: a magnetic order breaks the SO(3) rotation of the spin, a polarisation breaks the inversion along some mirror plane (while leaving more than one position unmoved under point group operations,  thus removing the inversion center) and so on. 
Hence, we can raise the following questions: Does the spontaneous emergence of a structural distortion from a high-symmetry nonchiral phase to a low-symmetry chiral phase produce an independent order parameter? And can we treat this distortion as an internal degree of freedom in a crystal?

\begin{figure}[h!]
\centering
\includegraphics[width=8cm,keepaspectratio=true]{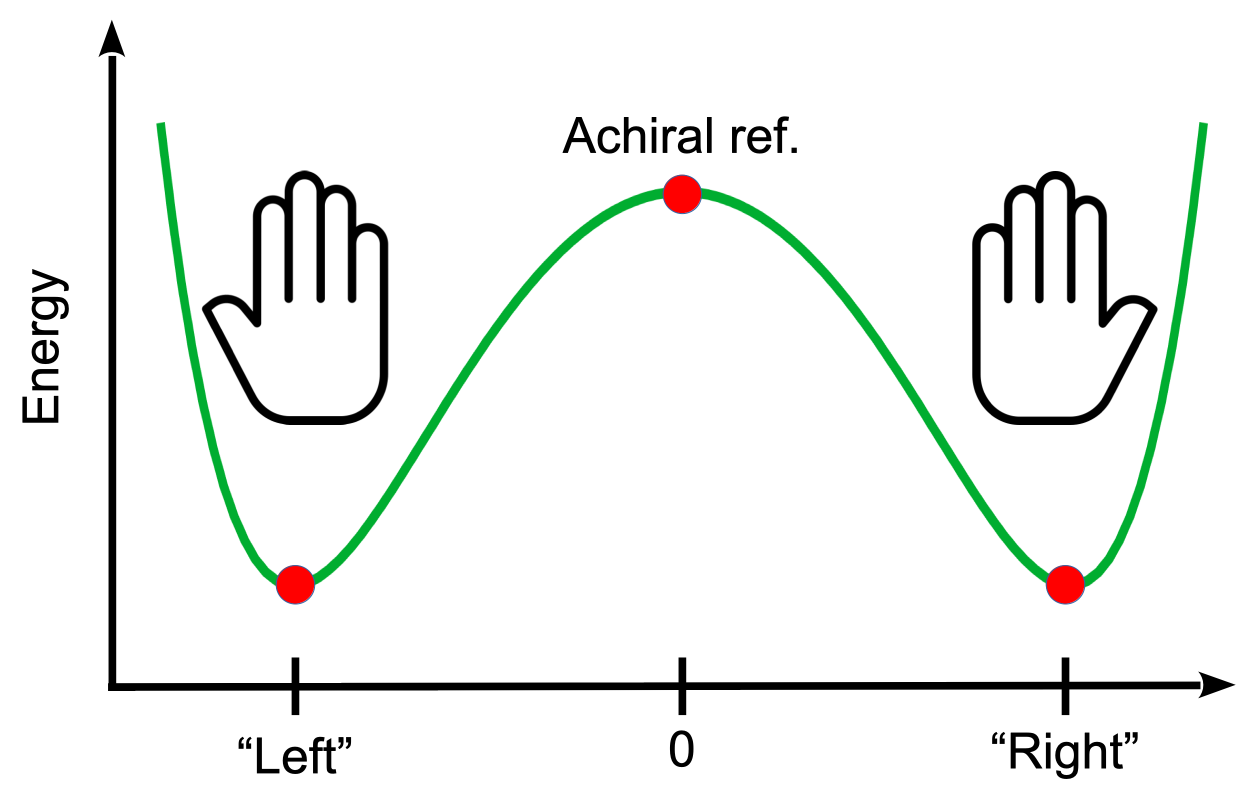}
 \caption{Schematic picture of a double well associated to a chiral structural instability coming from an achiral reference structure.}
 \label{fig:double_well_chirality}
\end{figure}

If a crystal possesses an unstable phonon that drives a transition from an achiral to a chiral phase, it is natural to adopt the amplitude of the symmetry-lowering 
mode as the order parameter.
In this case, the phase transition is displacive or order-disorder (i.e., not reconstructive) when a group-subgroup relation between the two phases exists, which also means it preserves the internal connectivity and does not break any bonds~\cite{dove1997}. 
It can also be linked to free energy double well, as illustrated in Figure~\ref{fig:double_well_chirality}, where both the achiral and chiral configurations are energetically feasible within the solid-state transition. 
In such scenarios, we can classify this structural phase transition as primarily chiral, where chirality is the primary order parameter.
As we have seen in Section~\ref{Sec:Examples_Struct_chiral}, a few examples of displacive chiral phase transitions are known (e.g. K$_{3}$NiO$_{2}$~\cite{favaKNO}).
While the distortion originating from the unstable mode drives the system to be chiral, there exists ambiguity regarding its classification as the chiral order parameter. 
Despite the distortion embodying chirality is coming from the soft phonon mode, determining it as the definitive chiral order parameter requires a more profound investigation. 
Additionally, a close examination is essential to clarify whether the magnitude of this distortion inherently captures chirality to the degree that justifies its acknowledgment as the chiral order parameter.

This section delves into the current literature surrounding the definition of an order parameter for chirality in periodic crystals and the mechanisms by which external stimuli can trigger chirality flipping. 
Additionally, we investigate the case of coupled order parameters to chirality, in which enantiomorphic switching could occur by flipping those other order parameters.
Controlling the associated structurally
chiral domains by external means are enormously appealing as this would open materials science to new functionalities based on chirality.
Defining an order parameter for chirality, quantifying it (see Section~\ref{Sec:Quantifying chirality}), and having an associated conjugate field would also have substantial consequences in the fundamentals of solid state physics as chirality would enter as a new ferroic order aside to ferroelectricity, ferromagnetism, ferroelasticity, and ferrotoroidicity~\cite{Gnewuch2019}. 

\subsection{Definition of a chiral order parameter}
\label{toroidal}

\begin{figure}
\centering
\includegraphics[width=8.5cm,keepaspectratio=true]{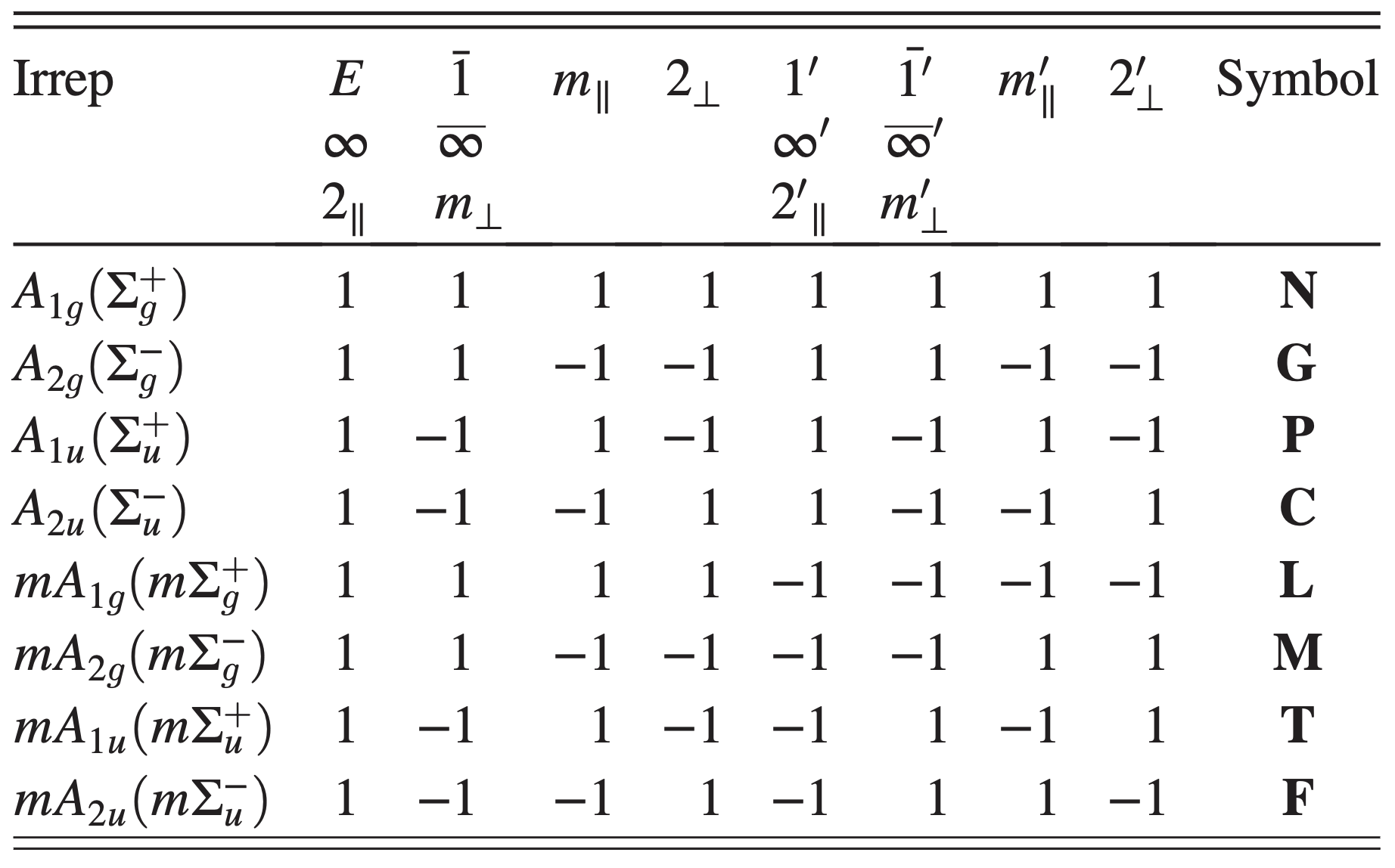}
 \caption{Character table of the point group $\infty$/mm1' point group from  Ref.~\cite{hlinka2014}. 
Here $\mathbf{N}$, $\mathbf{G}$, $\mathbf{P}$ and $\mathbf{C}$ represent the time-even nematic, axial, polar and chiral orders respectively. At the same time, $\mathbf{L}$, $\mathbf{M}$, $\mathbf{T}$ and $\mathbf{F}$ indicate the antiferromagnetic, magnetic, magneto-toroidal and time-odd chiral representations.}
\label{fig:hlinka_table}
\end{figure}

\begin{figure}
\centering
\includegraphics[width=8.5cm,keepaspectratio=true]{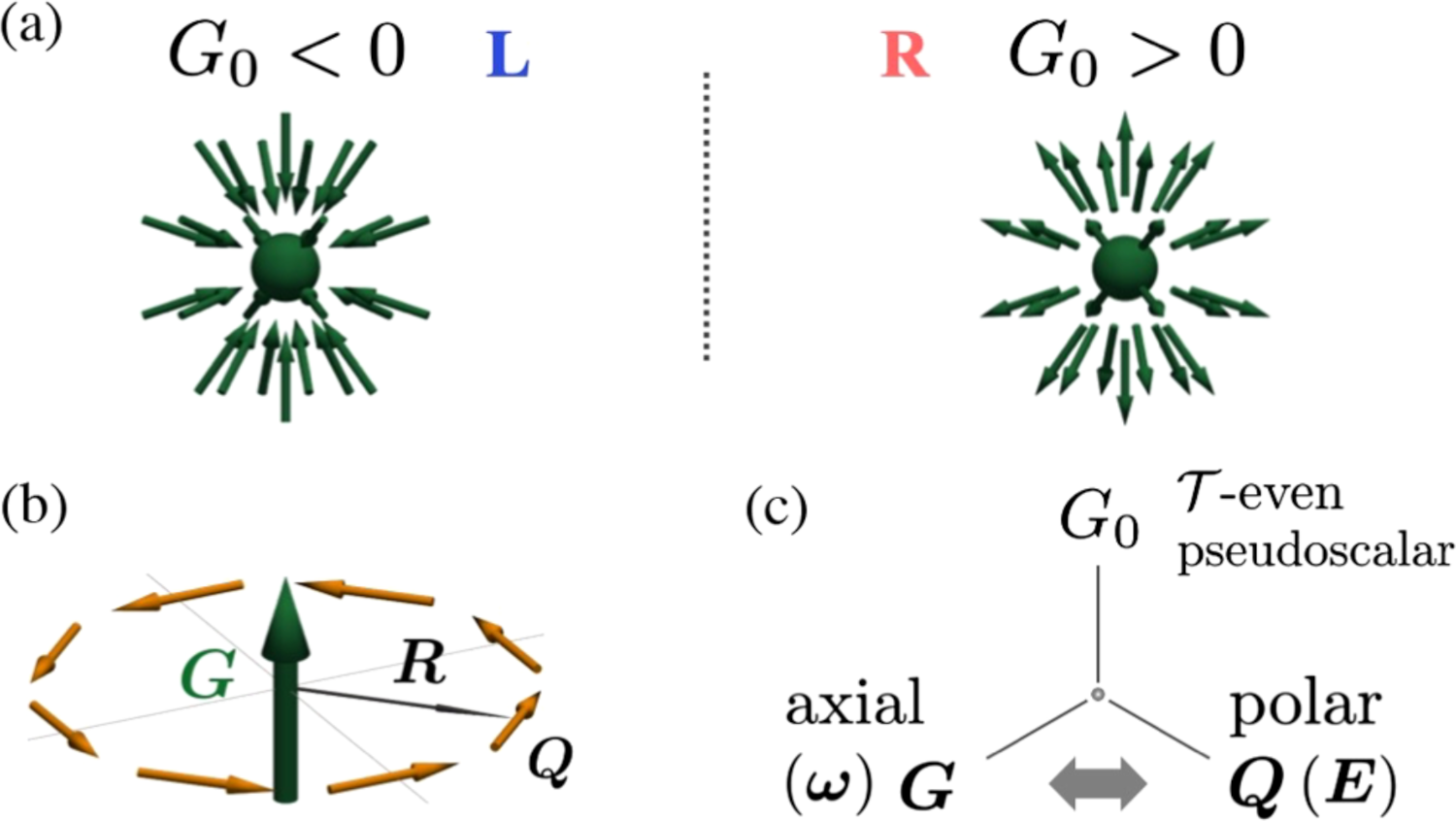}
 \caption{(a) Schematic picture of the left (L) and right (R) handed pseudoscalar electrotoroidal monopole $G_0$ where the green arrows represent the flux of  electrotoroidal dipoles. 
 (b) ``Classical'' representation of the toroidal moment as a vortex-like configuration of dipoles $Q$. 
 The axial-polar-chiral (pseudoscalar) interaction is represented in (c), where $\mathbf{Q}$ is the electrotroidal dipole, $\mathbf{E}$ is the electric field and $\mathbf{\omega}$ is an axial deformation of either electromagnetic or mechanical origin. Reproduced from Ref.~\cite{oiwa2022}}
\label{fig:electrotroidal-monopole-G0}
\end{figure}

A crystal is chiral if its space group only contains proper operations~\cite{fecher2022}.
Hence a straightforward definition of a chiral order parameter, associated to a nonchiral phase, would be that of a degree of freedom - for instance a phonon or symmetry adapted mode - which spontaneously breaks some improper (e.g. mirror) operations, thus inducing chirality in the system.  
Some additional restriction requires having the chiral order to belong to a non-invariant irreducible representation of a nonchiral space group, and the most logic choice is represented by the time-reversal even pseudoscalar representation. It has in fact been proven that every chiral space group possess the so called $G_0$ monopole~\cite{Kishine2022}, which is in fact a pseudoscalar property. One can associate the sign of $G_0$ to a specific handedness following a convention as in Fig.~\ref{fig:electrotroidal-monopole-G0} from Ref.~\cite{oiwa2022}.
Naturally, many distinct properties may fulfill the requirement of being pseudoscalars and in several physical contexts. 
As explained in Appendix~\ref{appendix_multipole}
polar and axial components may be extracted from the multipole expansion of a generic observable operator. While neither is necessarily by itself chiral, assuming for example a polar vortex $\mathbf{P}(\mathbf{r})$ such that $\mathbf{\nabla}\times\mathbf{P}\neq$0, the dot product $\mathbf{P}(\mathbf{r})\cdot(\mathbf{\nabla}\times\mathbf{P}(\mathbf{r}))$ represents a local three-dimensional formulation of $G_0$, albeit not a unique one.
While the specific formulation of a chiral order parameter depends on the problem under consideration and may change upon the context (with the sole requirement of being zero in space groups possessing improper symmetry operations), the possibility of a spontaneous transition from a nonchiral reference phase to a chiral state has been elucidated in several recent works~\cite{hlinka2014,erb2020vector,chiroaxial_Erb_Hlinka}. In particular and with reference to the work of Hlinka~\cite{hlinka2014}, it is easy to spot that the chiral representation can be obtained as the product between the polar and the axial representations as also highlighted in the character table~\ref{fig:hlinka_table}.
This detail is crucial to define a so called enantioselective effect, namely an interaction between a chiral material and an external parameter cabable of removing the degeneracy between states equidistant from a nonchiral reference but with opposite handedness.

\subsubsection{Enantioselectivity mediated by physical effects}

After defining the electric toroidal monopole as a chiral order parameter, Kishine {\it{et al.}}~\cite{Kishine2022} try to define an associated conjugate field.
According to them, this conjugate field should be able to favor one or the other enantiomorphic phase through flipping the sign of $G_0$, i.e., it should have the same symmetry transformation as $G_0$ (the field times $G_0$ should be an invariant of the Hamiltonian).
One proposed example of such a field  is the so-called Lipkin ``zilch'' $\rho_{\chi}$~\cite{10.1063/1.1704165,PhysRevLett.104.163901,Proskurin_2017,Kishine2022}:

\begin{equation}\label{zilch_eq}
    \rho_{\chi} = \frac{\epsilon_{0}}{2}\mathbf{E}\cdot(\mathbf{\nabla}\times\mathbf{E}) + \frac{1}{2\mu_{0}}\mathbf{B}\cdot(\mathbf{\nabla}\times\mathbf{B}),
\end{equation}

\noindent which can be a measure the chirality of an electromagnetic field~\cite{PhysRevLett.104.163901} (e.g. a circularly polarized wave) and where $\mathbf{E}$ and $\mathbf{B}$ are the electric and magnetic fields, respectively. 
This spacial-dependent zilch field is a conserved property of the Hamiltonian~\cite{10.1063/1.1704165,PhysRevA.87.043843}. 
Such an object may couple with the chirality since the product between a vector (namely, the electric field) and a pseudovector (that is, $\mathbf{\nabla}\times\mathbf{E}$) must contain the pseudoscalar irreducible representation.
The electromagnetic force associated with circularly polarized waves contains a chiral contribution and produces a torque that can be used to trigger enantiomeric separation in, e.g., plasmonic nanoparticles~\cite{Rukhlenko2016,mun2020} or  molecules~\cite{Genet2022}. 
While it has recently been proposed in the context of crystal enantioselectivity~\cite{oiwa2022,favaKNO}, to the best of our current knowledge, no experiment in this sense has been performed.

If we define $\chi$ as the pseudoscalar gyrotropic order parameter, the following interaction:

\begin{equation}\label{chiral_switch_n}
    V = h_{ij}\chi\cdot{P}_i\cdot{G}_j 
    -\mathbf{P}\cdot\mathbf{E} -  \mathbf{G}_i\cdot(\mathbf{\nabla}\times\mathbf{E})_i
\end{equation}

\noindent can be constructed where $\mathbf{G}_i$ is the axial electrotoroidal moment~\cite{dubovik1983toroidal, dubovik1990toroid} and h$_{ij}$ a material-dependent coupling.
If the polar and toroidal ``auxiliary'' orders are non-spontaneous, we can recast the previous expression as $V =\chi{h'_{ij}}\cdot{E}_i\cdot{(\mathbf{\nabla}\times\mathbf{E})}_j$. Naturally, enantioselectivity comes with the additional energy cost of creating the auxiliary orders from the same electromagnetic perturbation. 
Finally, along with the zilch, we ought to mention that electromechanical interactions may also couple with a chiral order. An axial configuration of dipoles 
can also be triggered 
by a rotational lattice deformation ($\mathbf{\nabla}\times\mathbf{u}$, where $\mathbf{u}$ is an ion displacement vector). Thus, such rotational distortion applied simultaneously with an electric field could be used to directly affect the chirality as much as the aforementioned zilch and as suggested by refs.~\cite{oiwa2022,favaKNO}.
This would also mean that an electric field could induce a rotation of the crystal and, the other way around, a rotation of the crystal could induce an electric polarization in chiral compounds (rotoelectricity~\cite{gopalan2011}, analogous to the Barnett and Einstein-de-Haas effects in magnets~\cite{barnett1935}).
Interestingly, the character table of the parent phase of enantiomorphic compounds shows that tensors behaving as $[\mathbf{V}^2]\times\mathbf{V}$ ($\mathbf{V}$ being a vector) - such as the strain gradient $\mathbf{\nabla}\epsilon_{ij}$ - may contain the pseudoscalar IRREP, which would allow them to directly interact with $\chi$ without the need of an intermediary electromagnetic field~\cite{favaKNO}. This hypothetical phenomenon, dubbed ``flexochirality'', remains completely unexplored.

\subsection{Chirality and coupled order parameters}
\label{Sec:coupledOP}
Although the application of a conjugate field to flip structural chirality has not (yet) been reported,
the coupling of chirality with other order parameters that can be tuned by external fields can be considered.
Such a mechanism is well known in, e.g., ferroelectric crystals, such as improper, hybrid improper, or pseudoproper  ferroelectricity~\cite{levanyuk1974, bousquet2008, cano2010, benedek2011},  where the polarization couples with another non-polar order parameter that will be tunable by an electric field in return or, vice versa, the polarization is induced by the non-polar order parameter (particularly sought after in magnetoelectrics to tune the magnetic order parameter with an electric field~\cite{bousquet2016}).
These possible couplings can be determined from symmetry analysis and group theory~\cite{dresselhaus2007group,Campbell2006, stokes2007isotropy,https://doi.org/10.1107/S0021889803005946}.
Throughout the rest of this section, we will explore such mechanisms that may prove helpful concerning the switching and flipping of the chirality and show that, 
generally, an interaction between external parameters (namely, an electric field and/or lattice deformations) and the chirality can exist and could induce enantioselectivity if certain conditions are met.

\subsubsection{Coupling with other structural deformations}

Before going through the proper coupling of chirality with other order parameters, we would like to discuss the case where chirality is described by the product between an axial pseudovector $\mathbf{A}$ and a regular vector $\mathbf{V}$. 
While an irreducible representation solely associated with the chirality generally exists~\cite{hlinka2014}, it is also contained in the (generally reducible) product between a vector $\mathbf{V}$ and an axial $\mathbf{A}$ representation. 
Consequently, it is reasonable to consider such a product as an effective conjugate field 
for the chirality, although subject to competition with other phases as a consequence of its generally reducible character
(see also Sections~\ref{toroidal} and Appendix~\ref{appendix:polar-vortices}). 
We have discussed the example of 
Pb$_{5}$Ge$_{3}$O$_{11}$ (Section~\ref{PGO}) for which the handedness and 
optical activity can be flipped via an applied electric field  (i.e., gyroelectricity, see Section~\ref{sec:Gyroelectricity}) in which the polar domains are spontaneously chiral and optically active since the polar soft mode driving the ferroelectric transition drives also the chirality~\cite{fava2023ferroelectricity}.
In this case, the high symmetry phase already contains axiality $\mathbf{A}$ such that when the polar vector $\mathbf{P}$ develops in the crystal, the chirality is automatically induced through the presence of both $\mathbf{A}$ and $\mathbf{P}$ together.
Hence, chirality is flipped when the polarization is reversed because the axial vector is frozen into the structure. 
Remarkably, the low symmetry ferroelectric phase of Pb$_{5}$Ge$_{3}$O$_{11}$ is at the same time axial, polar, and chiral, which, in a way, makes an electric field a conjugate field for chirality in this compound.
This simple but essential fact is possibly the most efficient and unique way to control the global gyrotropic properties in chiral crystals. 
However, it requires the additional and spontaneous presence of a fixed-sign axiality.
We note that Fabini et al.~\cite{fabini2024} recently reported several possible chiral phases in the CsSnBr$_3$ perovskite crystal where chirality can be present with other structural distortions, like ferroelectric or antiferroelectric ones, and is hence a good case for looking at chirality coupling with other order parameters.

\begin{figure}
\centering
\includegraphics[width=8.5cm,keepaspectratio=true]{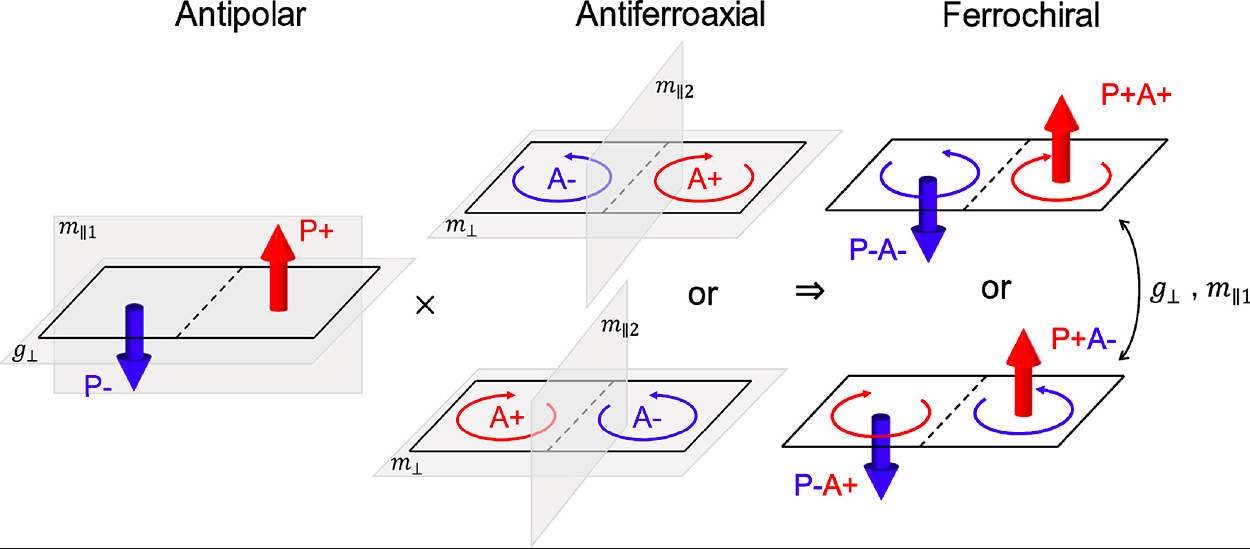}
  \caption{Illustration of chirality as a combination of antipolar and antiferroaxial orders. A($\pm$) represents a clockwise/counterclockwise axial order, while P($\pm$) is a positive/negative local polarization. The pairs ($-$P,$-$A;P,A) and ($-$P,A;P,$-$A) have opposite handedness. From Ref.~\cite{hayashida2021b}}.
  \label{fig:axial_polar}
\end{figure}

Another case of chirality driven by coupled orders has been reported in the afore-discussed Ba(TiO)Cu$_4$(PO$_4$)$_4$ crystal (see Section~\ref{Sec:Examples_Struct_chiral}).
With the help of x-ray diffraction, Hayashida {\it{et al.}}~\cite{hayashida2021b} report a second-order ferrochiral transition at 710 $^{\circ}$C, induced by the coupling between an antipolar and an antiferroaxial order as depicted in  Figure~\ref{fig:axial_polar}. 
Hence, strictly speaking, neither polar nor axial order exists in the high or low symmetry phases of Ba(TiO)Cu$_4$(PO$_4$)$_4$. 
The chosen order parameter, in this case, is the staggered rotation of the antipolar structural Cu$_{4}$O$_{12}$ units, and the phase transition appears to be a second-order one.

In the case of K$_3$NiO$_2$, CsCuCl$_3$ and MgTi$_2$O$_4$ crystals (see Section~\ref{Sec:Examples_Struct_chiral}), the chiral phase transition is coming from a zone boundary mode that does not involve a polar vector nor an axial distortion.
This raises a significant question: is it possible to flip the gyrotropic properties of such nonpolar nonaxial systems possessing a chiral order, including materials crystallizing into enantiomorphic pairs, by 
external means? 
The case of K$_3$NiO$_2$ has been scrutinized from DFT calculations by Fava {\it{et al.}}~\cite{favaKNO} where the chiral distortion can be seen as a helix of atomic displacements (see Figure~\ref{fig:K3NiO2}).
Nevertheless, possible couplings with polar distortions have been explored, where it appeared that the first coupling with the chiral distortion would be at the fourth order in an energy expansion and involve a Raman mode together with the polar one.
This would also mean that 
chirality could be dynamically tuned through non-linear phononic ultrafast laser excitation~\cite{afanasiev2021, juraschek2017b, buzzi2018}.
 
\subsubsection{Coupling with strain}

Another possible coupled order to chirality is the strain.
Chirality switching with strain has been demonstrated from DFT calculations in the NaCu$_{5}$S$_{3}$ crystal~\cite{PhysRevB.102.235127}.
NaCu$_{5}$S$_{3}$ crystallizes in the space groups $P6_{3}/mcm$ and $P6_{3}$22 (Sohncke non-enantiomorphic space group) in the high and low symmetry phases, respectively.
The high symmetry $P6_{3}/mcm$ phase is described to be antichiral as it contains compensated opposite chirality. In comparison, the low symmetry $P6_{3}$22 phase can be described \textit{ferrichiral} as the phase transition distortions partially uncompensated the perfect balance between the opposite chiralities. 
In this case, it has been demonstrated that a few percent of biaxial strain can tune and quench the barrier between the two ferrichiral domains, hence inducing a switching between the two chiral cases.
Another example of strain tuning chirality is the case of BiFeO$_3$ perovskite where it has been predicted numerically that (-110) epitaxial strain (of about 7\%) can induce a gyrotropic phase transition from non-chiral $Pnma$ space group to the non-enantiomorphic Sohncke $P2_{1}2_{1}2_{1}$ space group~\cite{prosandev2011}.
The associated distortions present in the $P2_{1}2_{1}2_{1}$ phase can be described by interpenetrated arrays of ferroelectric vortices and antivortices such that the resulting distortion is chiral (the one induced by the vortices).
However, the gyrotropic properties alone and the chirality flipping possibilities were not explored further in this case.

\subsubsection{Coupling with magnetism}
\label{subsec:coupling-order-magnetism}

 As discussed in Section~\ref{subsec:magnetic-chirality}, there can be an intimate link between magnetism and crystal chirality.
Indeed, if the magnetic order developing in the crystal is chiral, it will induce crystal chirality through the symmetry breaking of the magnetic order parameter and the spin-lattice coupling (e.g. langasite Ba$_3$NbFe$_3$Si$_2$O$_{14}$).
We have also seen that the other way around is proper, crystal chirality can be  imposed on  the magnetic order (e.g. Mn$_3$Sn).
In this section, we discuss the coupling between chirality and magnetism from the perspective of an order parameter.

In the light of the previous subsections, we can get chirality through axiality times a vector, which can be produced by combining a structural distortions and a magnetic order. 
For example, combining a structural polar vector and spin axiality can drive chirality as seen in Section~\ref{subsec:magnetic-chirality} for the case of Cu$_{3}$Nb$_{2}$O$_{8}$.
Indeed, in Cu$_{3}$Nb$_{2}$O$_{8}$~\cite{PhysRevLett.107.137205}, the onset of polarisation coincides with the appearance of a helicoidal magnetic ordering. 
Since the polarisation is oriented perpendicularly to the plane where the spins rotate, this results in chirality.

The other possibility is when a vector comes from the spin and axiality from the structure is at play, as seen in Section~\ref{subsec:magnetic-chirality} for the case of CaMn$_7$O$_{12}$.
The spin helical order $\sigma$ breaks the inversion symmetry. It induces a polarization $\mathbf{P}$ perpendicular to the helix rotation, similarly to Cu$_{3}$Nb$_{2}$O$_{8}$ but with the difference that an axial structural order $\mathbf{A}$ is also present in the crystal structure such that an overall coupling $\sigma\mathbf{P}\mathbf{A}$ is expected to be at play for the multiferroic properties~\cite{Perks2012}.

Of course, both structural and magnetic chiralities can coexist, as observed in materials such as in the Langasite Ba$_3$NbFe$_3$Si$_2$O$_{14}$ where the crystal structure is chiral and the spin structure is helicoidal (definitely chiral too). 
This situation has been recognized as an alternative to the cases where the magnetic chirality stems from the DMI and frustration, respectively~\cite{PhysRevLett.111.017202}.

\subsubsection{Coupling with electronic degrees of freedom}

A general requirement for a quantity to be chiral is to be parity-odd and time-reversal even, according to Kelvin and Barron~\cite{Barron_1986a,Barron1986,Kishine2022}.
Since chirality is associated with the absence of improper operations (not simply the lack of inversion symmetry), it is natural to seek time-reversal-even pseudoscalar quantities as chiral objects. 
By studying the Fermi liquid instabilities of spin-orbit-coupled metals, Fu~\cite{fu2015} identified that the pseudoscalar quantity  
$\sum_{\mathbf{k}}\hat{\mathbf{k}}\cdot\mathbf{s}(\mathbf{k})$, where $\mathbf{s}$ is the wavevector dependent spin and $\hat{\mathbf{k}}$ its normalized momentum, gives an isotropic gyrotropic liquid if a spin instability makes this quantity non-zero.

A gyrotropic order of electronic origin has been measured in the layered semimetal 1T-TiSe$_{2}$~\cite{Xu2020}. 
In particular, near the critical temperature ($\sim$ 174 K), a circularly polarised mid-infrared light can favor one chiral domain over the other. 
The proposed energy invariant ($\delta{F}$), coupling matter and electromagnetic chiralities, in this case takes the form $\delta{F} \sim \phi_{\text{gyro}}[(\mathbf{E}\times\partial_z\mathbf{E})\cdot\hat{z}]$ where $\phi_{\text{gyro}}$ is the gyrotropic order parameter. 
However, due to the system's metallic nature, a gyrotropic switchable order cannot be detected via bulk optical activity measurement due to light absorption. 
Thus, the circular photogalvanic current generated by (and parallel to) the incident light perpendicular to the sample surface was used to detect the chirality and its flipping in this metallic compound.
It is nevertheless important to realize that the measurements in Ref.~\cite{Xu2020} do not show any observable hysteresis upon thermal cycling, which means that the right-to-left/left-to-right permanent flip is probably not realized in the sample. 
Finally, while understanding the origin of the gyrotropic state requires further analysis, two possible mechanisms have been theorized: i) the modulation of a charge density wave state (CDW) and (ii) the realization of a Pomeranchuk instability in the $p$ channel~\cite{Xu2020}.
Indeed, and since this material is layered, once formed at $T_{\text{CDW}}$ = 198 K, the CDW direction can orient chirally for the propagation direction (e.g., perpendicular to each layer). 
The proposed gyrotropic invariant thus describes the interaction between the CDW's chirality and that of the incoming light.
On the other hand, a chiral order generated by an instability in the Fermi surface would be unrelated to the CDW and coexist with it~\cite{Xu2020}.

An equivalent to helical or helicoidal spin orders, but purely structural,  chiral phase transition from an achiral reference has been identified in the doped BiMn$_{7}$O$_{12}$~\cite{doi:10.1126/science.aay7356} compound. 
The chirality, in this case, takes the form $\mathbf{k}\cdot (\mathbf{u}_{i}\times\mathbf{u}_{j})$ where $\mathbf{k}$ 
is the modulation vector of the low-temperature phase and $\mathbf{u}_i$ is an atomic displacement, 
and can be identified as a helical configuration of electric dipoles, further stabilized by the competition between lone pairs and the orbital ordering, and with the mixing of $d$-orbitals along the direction of propagation~\cite{doi:10.1126/science.aay7356}.

\subsubsection{Additional sources of chirality inversion through lasers}

While circularly polarized femtosecond laser-induced enantioselectivity has been extensively used in  solutions~\cite{korede2023}, this has been recently achieved in a dry state of NaClO$_{3}$~\cite{WANG2023101323}. 
From a metastable $P2_{1}/a$ phase, circularly polarised light can achieve an enantiomeric excess of about 20 $\%$. 
More specifically, while heating can locally reduce the energy gap between achiral and chiral phases, a nonzero enantiomeric excess can only be created if circularly polarised light is employed.
It is found that the transition proceeds in two steps, starting from the $P2_{1}/a$ reference structure. 
More specifically, the sliding 
of the unit cell by $c$/4 to form transient octahedral units occurs. 
As the light momentum is transferred to this intermediate phase, it determines the transition to either the right or left-handed chiral structure via twisting of the octahedra. 
The force associated with this twisting is proposed to result from combining a photon scattering term and the circularly polarised light's spin angular momentum.
While the electromagnetic spin angular momentum alone cannot induce the chiral deformation, the occurrence of a spin-to-orbital conversion process~\cite{PhysRevLett.99.073901,Bliokh2015} generates an orbital angular momentum whose associated torque is likely to be responsible~\cite{WANG2023101323} for the observed structural enantioselectivity in NaClO$_{3}$.
It is worth mentioning that, when crystallized from an aqueous solution, Kondenpudi {\it{et al.}}~\cite{kondepudi1990} have reported that a quasi-full enantioselection can be achieved by simply stirring the solution.
This is an experimental demonstration that an axial macroscopic rotation can tune chirality during growth, and it would be interesting to test the same effect in a fully solid state case during the achiral to chiral phase transition~\cite{niinomi2014}.

Recent experiments performed on the previously discussed Ba(TiO)Cu$_4$(PO$_4$)$_4$ crystal (Section~\ref{Sec:Examples_Struct_chiral} and \ref{Sec:coupledOP}) support the idea that chiral domains can also be flipped by a laser but via local heating and through handedness-independent source~\cite{hayashida2022}. 
A laser wavelength associated with a peak in the material's absorption coefficient induces local and radial heating. 
If the power density is sufficiently high, chiral domains could be manipulated. 
Surprisingly, the switching process was observed to be independent from the polarization of the electromagnetic field and dominated by local heating. Domain reconstruction was explained in terms of domain boundary minimization.
It is found under laser scanning, where the laser hitting point is moved from one domain to the other at constant power, and laser cooling, where the hitting point does not change but decreases the power step-wise.

\section{Chirality measure and quantification}
\label{Sec:Quantifying chirality}

Whilst it is very tangible to consider chirality regarding the non-superimposability of mirror images ~\cite{Thompson1894}, this simplistic approach fails to capture many subtleties and consequences.

\begin{figure}
\centering
\includegraphics[width=\columnwidth,keepaspectratio=true]{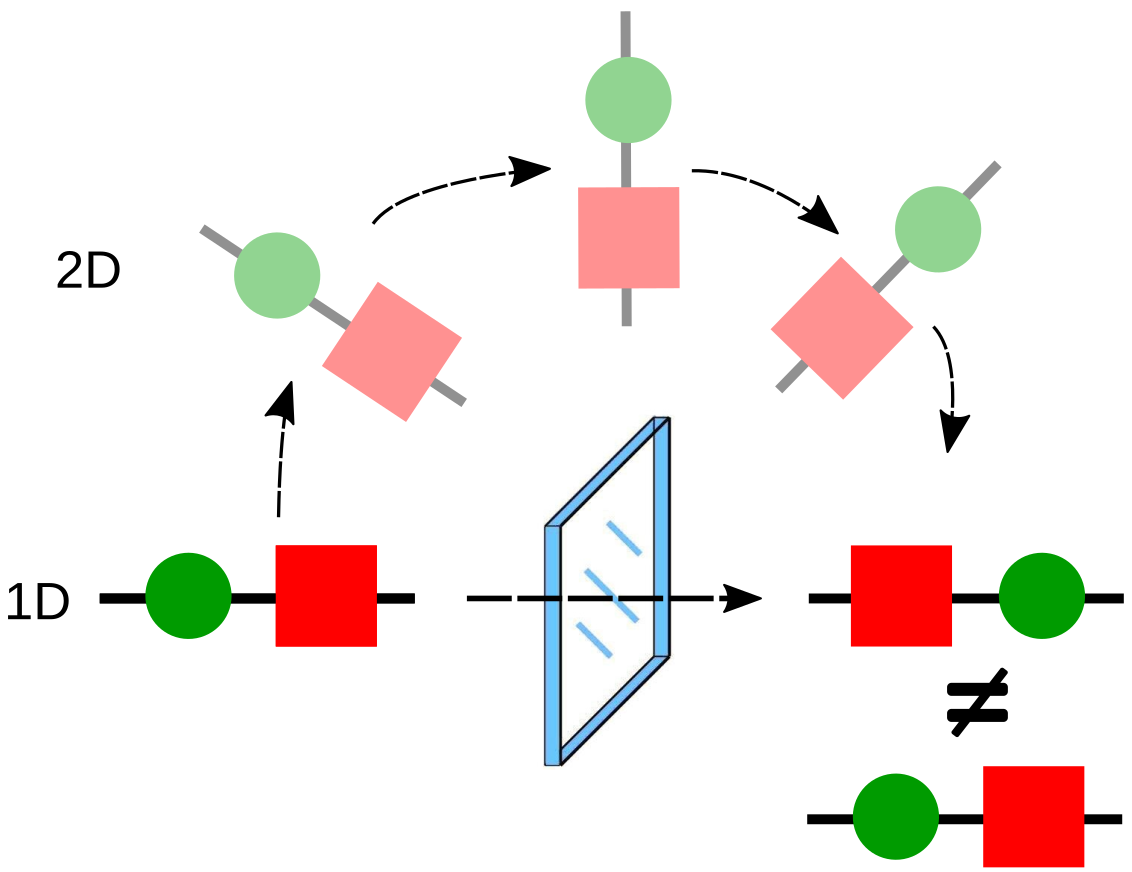}
 \caption{Schematic view of a one-dimensional asymmetric object. 
 In 1D the object on the left is chiral since we cannot superimpose it with its mirror image (on the right). 
 However, in 2D the additional (azimuthal) rotational symmetry allows the overlap of the object with its reflected image as depicted in the top transformation.}
 \label{fig:ch_dimensions}
\end{figure}

This section addresses concerns
regarding chirality and its measure or quantification.
Primarily, it is crucial to consider the dimensionality of the ambient space, as chirality may only occur in the lowest dimension in which the object under study is embedded. 
Chiral objects in a given dimension $n$ will lose their chirality when placed in an $(n+1)$-dimensional framework since the $n$-dimensional hyperplane accommodating them becomes a mirror symmetry hyperplane~\cite{Mezey-95}.
In Figure~\ref{fig:ch_dimensions}, we show a schematic picture of this fact where a one-dimensional chiral object becomes achiral when considered in two dimensions.
We will limit our discussion to three-dimensional spaces for this review to avoid confusion.

Moreover, a binary chiral vs achiral classification sometimes proves inadequate, prompting significant efforts in the literature to develop methods for continuously quantifying chirality despite its inherent difficulty~\cite{Weinberg-00, petijean2003, fowler2005}. 
One would anticipate that a reliable chirality measure, $\xi$, should function to rank objects, denoted as $x$, within a given subset $\mathcal{X}$ according to their degree of chirality i.e., the extent that they differ from an achiral analog. 
Additionally, as discussed in~\cite{Weinberg-00}, it is important to impose certain natural requirements: 
(i) $\xi$ is a continuous function on $\mathcal{X}$ as two sufficiently close objects $x_1$ and $x_2$ are expected to show similar chiral responses; 
(ii) If an object in $\mathcal{X}$ is achiral its degree of chirality should be zero. 
Conversely, if the degree of chirality of an object is zero, then it should be achiral;
(iii) Any two enantimorphic pairs $x,~x^\prime$ should have the same degree of chirality. 
Therefore, if $\xi$ is a scalar function, $\xi(x)=\xi(x^\prime)$, and if $\xi$ is a pseudoscalar function, $\xi(x)=-\xi(x^\prime)$.

The most common case of scalar chiral measures typically stems from distance functions relative to an achiral reference, as discussed in Ref.~\cite{Weinberg-93}. 
Concrete famous examples in the literature include Hausdorff distances~\cite{fecher2022, Buda-92} or the continuous chirality measure (CCM)~\cite{Zabrodsky1995}. 
The former is based on the supremum of the distances between the atomic positions of the chiral and aciral structures, while the latter relies on the mean square of such distances. 
Careful definitions and examples can be found in Ref.~\cite{fecher2022}.
As standard scalar measures, they assign identical values to both enantiomers, lacking the capacity to discern between them. Despite their widespread use, they might be troublesome and have been intensely debated \cite{buda1992,gillat1994,petijean2003,fowler2005,jenkins2018}.
This is due in part to the difficulty in determining the absolute chirality of an object (see Section~\ref{sec:absolute_chir}):
to determine whether an object is left or right-handed requires a pre-established definition of what we call left or right \cite{cintas2007tracing}.
Nevertheless, determining an object's "chiral strength" or chiral amplitude is unclear unless one pre-defines the reference to what is left or right-handedness.  
Indeed, one would first need an achiral reference to know the “zero” and quantify the amount of chirality from it. 
However, defining a zero-reference may not be apparent depending on the object, and applying this concept to a realistic physical system would be hindered by the additional constraint of not breaking the bonds unless the phase transition is reconstructive. 
And even though a continuous deformation to an achiral reference may be possible in 3D, such transformation is often likely to be physically impossible~\cite{flapan2000}. 
This shows that despite its primary role in natural sciences, the chirality measure has evaded researchers so far. 
Exploring this problem in extended solids adds another difficulty due to the periodic boundary conditions~\cite{Resta-98}. 

Although almost unexplored in the context of structural chirality quantification, another way to measure chirality is through pseudoscalar functions. An example of a pseudoscalar measure could be the hydrodynamical helicity~\cite{Moreau-61,Moffat-92,Moffatt2014}, which quantifies the handedness of the streamlines in a fluid and can be computed from the velocity field, $\vec{v}$, as
\begin{equation}
    \mathcal{H}=\int d^3\vec{r}~\vec{v}\cdot\left[\vec{\nabla}\times\vec{v}\right].
    \label{eq:Helicity}
\end{equation}
This definition has recently been extended to solids by establishing the velocity field through the individual atomic displacements that drive the system from the high-symmetry phase to the low-symmetry chiral phase~\cite{Gomez-Ortiz-24}. 
Moreover, it has recurrently been applied to quantify the handedness of polar vector fields in ferroelectric/dielectric superlattices~\cite{Shafer-18,junquera2023}.

\begin{figure}
\centering
\includegraphics[width=\columnwidth,keepaspectratio=true]{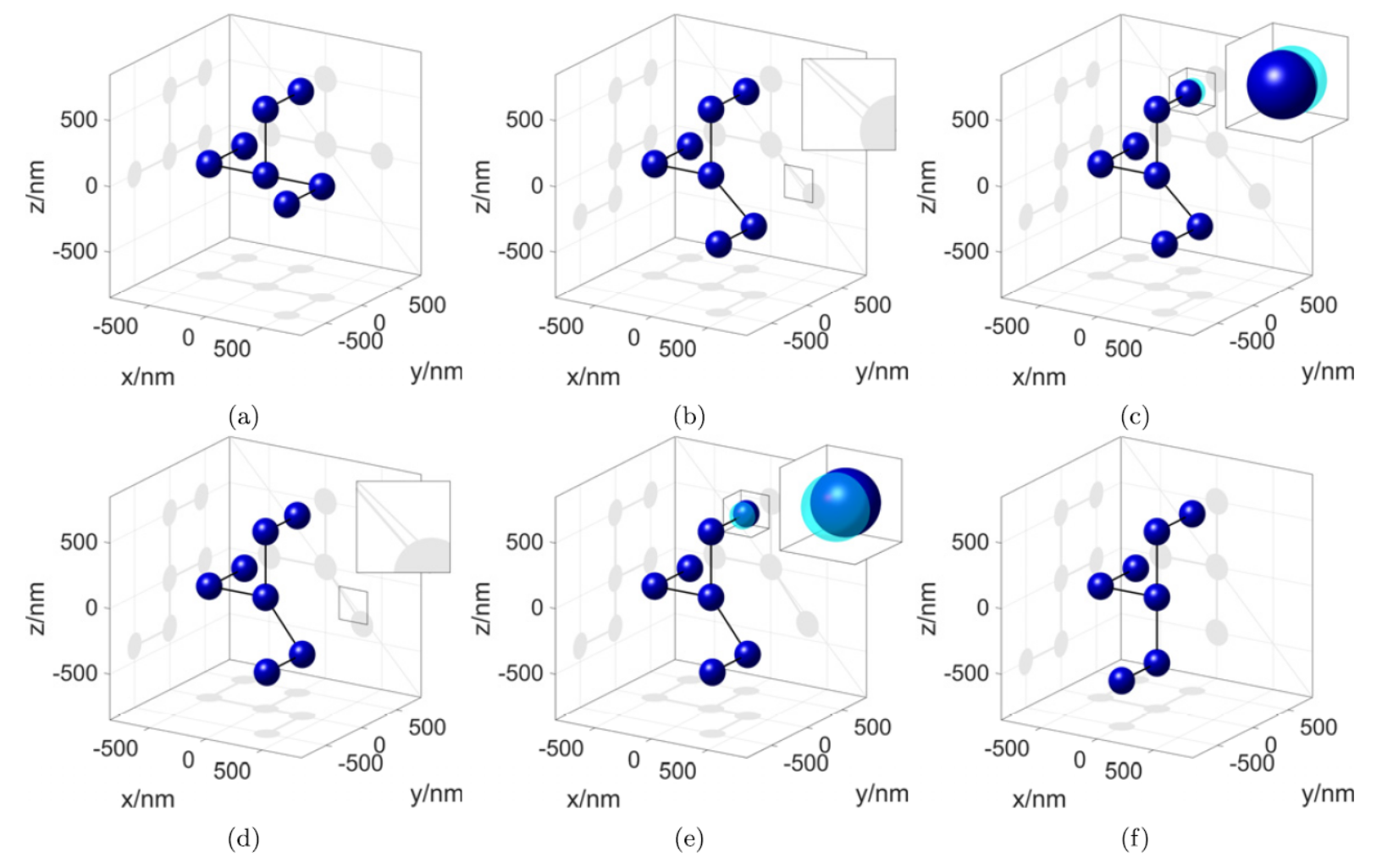}
 \caption{Continuous transformation of the initial chiral configuration (a) to its enantiomer (f) avoiding any intermediate achiral configuration. The achiral state is avoided as follows. First, the right leg is rotated about the $y$-axis by $9\pi/40$ (b). Then, the furthest top sphere is shifted by 50 nm along the negative direction of the y-axis (c)—the transparent blue sphere depicts the sphere's position before the shift (also enlarged in the top right corner). Then the right leg is rotated by an extra $2\pi/40$ (d). Afterward, the shifted top sphere is brought back to its initial position (e), and finally, the right leg is rotated by the remaining $9\pi/40$ onto the final configuration. All intermediate configurations are chiral. This can be seen in the corresponding insets containing zoomed-in versions of the shadows of the right leg on the $y$–$z$ plane. In (b) and (e), such shadow does not coincide with the reference diagonal, breaking the mirror symmetry. Reproduced from Ref.~\cite{Vavilin-22}.}
 \label{fig:chiral_connectedness}
\end{figure}

Although pseudoscalar functions can discriminate between enantiomers, more fundamental problems arise when considering them as chirality measures due to the chiral connectedness property~\cite{Mezey-95,Weinberg-97,Banik-16,Vavilin-22}. 
This property refers to the fact that a sufficiently complex chiral object $x$ can be continuously transformed into its mirror image $x^\prime$ while being chiral throughout the entire transformation process as schematized in Figure ~\ref{fig:chiral_connectedness}. 
Therefore, we can define a continuous chiral path $\gamma$ such that $\gamma(0)=x$ and $\gamma(1)=x^\prime$. 
As elucidated in Refs.\cite{Mezey-95,Weinberg-97,Banik-16}, this poses challenges since our continuous pseudoscalar function $\xi$ changes sign along the segment, necessitating the presence of a zero within it. 
 Inevitably, $\xi$ will assign a zero value to a chiral structure, leading to the problem of the \emph{false zeros}. 
This observation also suggests the existence of chiral objects for which assigning a specific handedness may not be applicable, as pseudoscalar functions fail to determine the handedness of some chiral objects.
A recent work~\cite{Vavilin-22},  has addressed this problem by combining a pseudoscalar function with a scalar quantity as a phase factor introduced into a complex exponential. 
This way, they could distinguish between right-handed, left-handed, unhanded chiral and non-chiral objects, avoiding the false-zeros problem.

A notable exception to the false-zeros problem occurs in sets where enantiomorphs are connected only through achiral states. 
An illustrative example, particularly relevant to this review, is the set of helices with variable pitch, as discussed in Ref.~\cite{Mislow-99}. 
In such cases, pseudoscalar functions exclusively yield zero values for achiral objects, making them suitable chirality measures. 
Consequently, as presented in Table~\ref{spacegroups}, enantiomorphic space groups are unaffected by this issue and can be classified according to their handedness. 
Conversely, non-enantiomorphic space groups exhibit chiral connectedness, making the presence of chiral non-handed configurations inevitable.

Another obvious problem related to the insofar proposed measures, such as the CCM~\cite{Zabrodsky1995}, is that the quantification of chirality is tackled from a geometrical and symmetry point of view only, without a clear connection with a physical picture. 
Indeed, part of this reason is the apparent lack of a conjugate field associated with handedness or chirality.
This is not at all what happens in the case of, e.g., electric polarisation and magnetization, which are not just measurements of the breaking of an inversion or time-reversal symmetry but also well-defined system properties and linearly coupled with the electromagnetic field. 
Hence, it is not surprising that most of the experimental work aiming at detecting chirality in crystals focuses on the optical activity and dichroism, even though these properties can be non-zero in certain achiral systems as well (see Section ~\ref{optical_activity}).
Examples taking the natural optical activity as the order parameter are known~\cite{Iwasaki1972,Konak1978,PhysRevB.29.2655}. Nevertheless, these cases always assume a phenomenological relation between the gyrotropic tensor and, for instance, the electric polarisation or the strain.
It can be proven~\cite{wadhawan2000introduction} that the first derivative of the dielectric tensor concerning a wave-vector $\mathbf{q}$ produces no contribution to the free energy expansion, which means that the onset of an eventual ``gyrotropic'' order is always a byproduct of a different order produced by an instability. 
These approaches do not, however, tackle {\textit{per}} {\textit{se}} the issue of defining a physically meaningful measure of chirality, nor its emergence from an achiral reference.

\section{Discussion and Perspectives}
\label{Sec:conclusion}

Chirality in inorganic solids is an affluent area of study due to the unique optical and electronic properties these materials can exhibit. 
Before concluding, let us explore additional challenges
and perspectives.  
Regarding the question of whether chirality can be a new order parameter, a similar question has been raised for magnetic toroidal moments, i.e., if a magnetic toroidal moment can be a primary order parameter with its ferroic phase transition or if it is conditional on
the magnetic order~\cite{vanaken2007,ederer2007,spaldin2008,zimmermann2014,toledano2015, Gnewuch2019,hayashida2024}. 
This question has been clarified with the demonstration that the toroidal moment can be defined as a ferroic order~\cite{toledano2015} and is, hence, the 4th ferroic order, i.e., ferrotoroidicity~\cite{Gnewuch2019}.
This has also been associated with a microscopic Berry phase theory of toroidization~\cite{gao2018}, showing that the quantification of the toroidal moment is a Berry phase problem as the polarization of ferroelectric is~\cite{kingsmith1993}.
If we consider possible analogies with ferrochirality, we can question whether the quantification of structural chirality would also be a Berry phase problem.
There are parallels between the history of quantifying chirality~\cite{petijean2003}, and the development of methods to quantify the polarization of ferroelectrics which have led to the the modern theory of polarization ~\cite{kingsmith1993}.
The problem is that we do not have a mathematical formulation for quantifying chirality, and the attempts to date involve in one way or another the position operator (like the continuous chirality measure~\cite{Zabrodsky1995}, or the Hausdorff distances~\cite{fecher2022}), which would require a Berry phase treatment in periodic boundary conditions.
The fact that some crystals can exhibit a displacive (or order-disorder) phase transition (i.e., not reconstructive) from an achiral to a chiral structural phase (e.g., K$_3$NiO$_2$, CsCuCl$_3$ or MgTi$_2$O$_4$ discussed a the beginning of this review) suggests that there is probably a way to quantify it through the amplitude of the distortion mode that describes the achiral-chiral phase transition if one has a precise formula.
This suggestion that chirality can be quantified can be pursued further, the electrotoridal monopole could also give some hints regarding this problem.
It also means that we should be able to define linear responses from it, like the change of chirality concerning a strain perturbation would drive piezochirality or elastochirality and hence being more general than the piezogyration or elastogyration effect~\cite{Konak1978}.
Another physical property that could be defined is the notion of dynamical chiral effective charges, i.e., how much an atom motion changes the chirality of a crystal structure, i.e., similarly to the Born effective charges (reflecting the change of polarization of an atom motion) or the magnetic effective charges (reflecting the change of magnetization associated to an atom motion)~\cite{iniguez2008}. 
This would also directly define phonon mode chiral effective charge (or oscillator strength), expressing how much chirality a phonon mode carries~\cite{fava2023ferroelectricity}.

While no known conjugate field is associated with chirality in crystals to tune it, polarized light still could be of interest.
Light intensity is presently insufficient to switch large domains as the magnetic field component of the light interacts weakly with the crystal and, hence, impedes the possibility of controlling structural chirality~\cite{ayuso2019, mahmoodian2019, poulikakos2019optical}.
Despite this weak interaction, it would be possible to have a poling effect at displacive (or order-disorder) chiral phase transitions, i.e., doing a polarized light field cooling poling effect as reported in NaClO$_3$~\cite{WANG2023101323}. 
Indeed, while the system is nearing a chiral phase transition, illuminating the sample with polarized light should favor the formation of one chiral domain over the other.
Instead of switching chirality, this could selectively choose one of the two enantiomorphic phases, which is of primary interest for medical applications and fundamental research.
Some theoretical solutions using the electric part of the light have been proposed through ultrafast laser excitation that could do the job at a more local level and short time~\cite{ayuso2019, mahmoodian2019, owens2018}.
As we have discussed in Section~\ref{order-parameter}, other regular fields like magnetic or electric field or strain should also be explored further.

On the other hand, there is a compelling need to intensify efforts to establish a more nuanced correlation between optical activity, chiral crystal purity, and circular dichroism~\cite{polavarapu2016structural, polavarapu2006determination, crassous2023materials}. 
These phenomena are not merely of academic interest but hold substantial promise for various applications, including enantiomeric sensing~\cite{pu2017simultaneous}, chiral photonics~\cite{lodahl2017chiral}, and the innovation of groundbreaking optical devices~\cite{huang2021optical, neubrech2020reconfigurable}. 
A concerted push towards developing experimental methodologies is essential to unlocking potential applications of chiral materials~\cite{castiglioni2011experimental, fernandez2019new, qian2018new}.
This effort entails refining the techniques for measuring and quantifying these effects and devising new approaches for synthesizing materials with high chiral enantiomeric purity and controlled optical properties. 
Advancements in this area could lead to significant breakthroughs, enabling the precise manipulation of light and the creation of susceptible enantiomeric detection systems. 
Furthermore, by deepening our understanding of the interplay between chiral enantiomeric purity and optical behavior, researchers can tailor materials to specific applications, enhancing the performance and efficiency of chiral photonic devices~\cite{duan2023chiral, solomon2020nanophotonic}. 
Moreover, this endeavor will require a multidisciplinary approach, drawing on materials science, chemistry, physics, and engineering expertise. 
Collaborative research efforts that bridge these diverse fields can foster innovation and lead to the discovery of novel physics, materials, or fabrication techniques. 
For example, Romao, Catena, Spaldin, and Matas recently proposed detecting dark matter through chiral phonons~\cite{romao2023}.
Additionally, integrating computational modeling with experimental research can expedite the design and testing of new chiral materials and predict their properties with greater accuracy and efficiency.

Exploring the intersection between chirality and various material properties represents a fertile ground for future research, offering a pathway to uncover novel phenomena and applications. 
In this review, we have delineated the current understanding of several couplings, such as gyroelectricity or magnetochirality, which illustrate the intricate interplay between chirality and material properties under the influence of external electric or magnetic fields. 
However, the scope for investigation extends far beyond these initial studies, with several other possible couplings to be explored. 
The concept of coupling external fields to chiral materials' electronic and structural response opens up a vast landscape for scientific inquiry. 
These couplings can manifest in myriad ways, potentially leading to new classes of responsive materials that exhibit unprecedented behaviors. 
For instance, applying electric or magnetic fields could induce or modulate chirality in materials, leading to reversible changes in their optical, magnetic, or electronic properties. 
This responsiveness deepens our fundamental understanding of chirality and paves the way for innovative applications, such as tunable optical filters~\cite{kobrinski1989wavelength, kenanakis2014optically}, switchable catalysts~\cite{romanazzi2017chiral, blanco2015artificial}, or data storage devices~\cite{firouzeh2024chirality, aiello2022chirality} that leverage chiral materials' unique properties. 
To harness these opportunities, focused research is needed to systematically explore how different external fields can interact with chiral materials' electronic and structural aspects. 
High-resolution imaging, spectroscopic techniques, and advanced microscopy methods will allow for the direct observation and manipulation of chiral structures at the nanoscale, providing invaluable insights into their behavior under external stimuli.

Exploring hybrid materials presents an exciting frontier for future research in chiral materials~\cite{zhao2021chiral, tan2022inorganic, zhong2021chiral}. 
This approach strategically integrates inorganic chiral structures with organic components or other inorganic elements to forge hybrid materials~\cite{zhao2021chiral, Lou2020}. 
The essence of this strategy lies in its potential to amalgamate the superior qualities of both inorganic and organic worlds, thereby engendering materials that exhibit enhanced or entirely novel properties~\cite{zhao2021chiral, altaf2022recent, cho2023bioinspired}. 
Combining different materials could create a synergy that leads to breakthroughs in material properties such as increased mechanical strength, enhanced electrical conductivity, or improved thermal stability, each tailored for specific application needs.
Concurrently, we must advance our computational methods and machine learning capabilities. 
These technological tools are set to play a pivotal role in the future of chiral materials research. 
By harnessing the power of computational simulations and machine learning algorithms, researchers can leapfrog the conventional trial-and-error approach, enabling the rapid prediction and identification of new chiral inorganic structures. 
Moreover, integrating computational models with experimental research is anticipated to tighten the correlation between theoretical definitions of chirality and their experimental counterparts. 
This harmonization is crucial for advancing our theoretical frameworks of chirality, ensuring that they are robust, predictive, and reflective of observable phenomena. 
By enhancing the dialogue between theory and experiment, researchers can achieve a more comprehensive understanding of chirality, which, in turn, will inform the design and synthesis of novel chiral materials with precision and purpose.

\section{Conclusion}

This review article provides a comprehensive overview of the enantiomorphic crystal structural chirality observed in periodic inorganic solids and the associated properties arising from their structural characteristics. 
Given the promising potential of these materials, the review aims to stimulate further research efforts in this domain. 
Specifically, it encourages experimental studies focused on the synthesis, discovery, and characterization of new chiral inorganic solids and computational investigations employing high-throughput machine learning techniques for materials discovery and theoretical approaches to understand and quantify the coupled properties arising from structural chirality. 
We also reviewed the different definitions of crystal chirality and in particular, we make a distinction between chiral axis and absolute chirality that has been sometimes incorrectly identified in some publications. 
We noted the need for more significant efforts to provide a unique method that defines crystal chirality. 
While the literature on chiral crystals continues to grow, our discussion primarily focuses on existing studies examining crystal phase transitions within this domain. 
We observed that only a limited number of systems have been thoroughly characterized, underscoring a significant gap in our understanding of how these transitions occur and how they relate across enantiomer subgroups to the supergroup. 
This observation highlights the need for a concerted effort to expand our understanding of chirality control by external fields.
The review delves into established phenomena like optical activity, electrogyration, and magnetochirality and emerging fields such as chiral phonons and altermagnetism, which remain largely unexplored in enantiomorphic chiral crystals. 
These nascent areas hint at novel behaviors awaiting thorough investigation, emphasizing the need to discover new chiral materials that could host these intriguing properties and phenomena. 
Hence, chirality in extended solids is a rich and fruitful research area
with high potential for new fundamental studies (topology, quantification problems, new ferroic order and related order parameter and conjugate field) and new technological applications (tuning optical activity from external means for, e.g., photonics and other chiroptical applications~\cite{mun2020, zhang2023, zhangyi2022, Lodahl2017} or exciting and controlling chiral phonons for phononics~\cite{chen2021propagating, chen2022chiral}).
Our review is also an invitation/appeal for further studies of chirality in periodic crystals to design and find new properties of matter.

\section*{Acknowledgements}

MF, FGO \& EB acknowledge the Fonds de la Recherche Scientifique (FNRS) for support, the PDR project CHRYSALID No.40003544 and the Consortium des \'Equipements de Calcul Intensif (C\'ECI), 
funded by the F.R.S.-FNRS under Grant No. 2.5020.11 and the Tier-1 Lucia supercomputer of the Walloon Region, infrastructure funded by the Walloon Region under the grant agreement No. 1910247.
The work in West Virginia was supported by the U.S. Department of Energy (DOE), Office of Science, Basic Energy Sciences (BES), 
under Award DE‐SC0021375. This work used Bridges2 and Expanse at the Pittsburgh Supercomputer and the San Diego Supercomputer Center through allocation DMR140031 from the Advanced Cyberinfrastructure Coordination Ecosystem: Services \& Support (ACCESS) program, which National Science Foundation supports grants 2138259, 2138286, 2138307, 2137603, and 2138296. 

\appendix
\section{Chirality in polar nanostructures}
\label{appendix:polar-vortices}

Although this review is focused on bulk systems, it is interesting to mention some recent work investigating the realization of chiral structures in nano- and heterostructures because some of these results could be extended/exploited at the bulk level. 
The observation of polar vortices rotating both clockwise and anticlockwise and antivortices has recently been reported in oxide superlattices SrTiO$_{3}$/PbTiO$_{3}$~\cite{Yadav2016,Kim2022,junquera2023}.
The coexistence between polar and toroidal order was observed in this system~\cite{Damodaran2017}, and the chirality of the vortexes was shown to be controllable with an electric field~\cite{doi:10.1126/sciadv.abj8030}. 
A recent theoretical analysis~\cite{PhysRevB.105.L220103} has further proven that this chiral state undergoes a double phase transition as the temperature increases, from a regular crystal-like state composed by alternating clockwise/counterclockwise vortexes - with a definite handedness - to a state where the vortexes form a handed liquid structure deprived of its long-range order, to, finally, a non-handed state.
It is essential to realize that the symmetry-based definition of chirality is also fulfilled by objects such as $\mathbf{p}\cdot(\nabla\times\mathbf{p}) = \mathbf{p}\cdot\mathbf{g}$ ($\mathbf{p}$ and $\mathbf{g}$  being a local dipole and a local toroidal field, respectively) and a net and electric-field switchable chirality has been detected via second harmonic generations in tri-layers superlattice oxides~\cite{doi:10.1126/sciadv.abj8030} and ferroelectric nanodots~\cite{Tikhonov2020}. 
In such cases, the switching via an electric field naturally stems from the local polar-axial coupling in these systems. An intriguing development in chiral switching involves altering the sense of rotation of vortices using twisted light as explored in Ref.~\cite{Gao-24}, where 
the interaction between polar skyrmions and a non-zero electric field curl is simulated.

In recent years, there has also been a great deal of interest in understanding electro-toroidal orders, mainly at the interface between insulators and ferroelectrics and nano-systems~\cite{Chen2015_toroidal,Liu2018, Prosandeev2006}. 
In these scenarios, the control of the electro-toroidal order is numerically simulated via coupling with a polar state~\cite{Chen2015_toroidal} and/or a torque~\cite{Liu2018}. 
At the same time, the issue of having a nonzero electric field curl can be circumvented with non-homogeneous electric fields~\cite{Prosandeev2006}. 
The effect of a homogeneous field is instead that of favoring a polar state and to suppress any dipole-vortexes, with $\Delta\mathbf{G}_{z}\sim-\mathbf{E}^{2}$ as realized in Pb(Zr,Ti)O$_{3}$ nanoparticles~\cite{PhysRevLett.98.077603}. 
In this case, the role of the vortex states in nucleating the polarization is also significant.
Incipient phase transitions have also been analyzed in KTaO$_{3}$ nanodots, where the G-order is suppressed at low-temperature by fluctuations~\cite{PhysRevLett.102.257601}. 
Furthermore, the electro-toroidal order is shown to couple to the stress, thus producing a new kind of electro-mechanical interaction~\cite{PhysRevB.76.012101}. 
$\mathbf{\nabla}\times\mathbf{E}$-induced susceptibilities are also discussed and numerically analysed via second-principles Hamiltonians~\cite{PhysRevLett.102.257601,PhysRevB.76.012101}. 
A comprehensive review on these dipole vortexes in nanosystems, mimicking their magnetic equivalent, is given in Ref.~\cite{Prosandeev_2008}.

These polar vortexes share many features with their magnetic counterparts, skyrmions~\cite{muhlbauer2009, zhang2023chiral}. 
Thus, exploring their origin, stability, and topological properties comes naturally ~\cite{Tokura2021}.
Skyrmions of electric origin and with a topological invariant given by q = $\frac{1}{4\pi}\mathbf{u}\cdot(\partial_x\mathbf{u}
\times\partial_y\mathbf{u})$ have been theoretically studied~\cite{Nahas2015} 
and experimentally observed~\cite{das2019observation}
 ($\mathbf{u}$ is a normalised local dipole moment) 
and while from early studies, it seemed that their stability was the result of geometry confinement and dipolar interaction (due to the apparent lack of a (chiral) DMI as it happens in the magnetic case), the existence of an electric equivalent of the Dzyaloshinskii-Moriya interaction (eDMI) $\sim\mathbf{D_{ij}}^\prime(\mathbf{u}_{i}\times\mathbf{u}_{j})$ has recently been established~\cite{stengel2023,Zhao2021,Junquera2021}. 
It has been further shown that this electric DMI is an electronic effect associated with a hopping process (induced by local inversion breaking) and related to the anti-symmetric part of the interatomic force constants~\cite{PhysRevB.106.224101}.
The aforementioned literature and examples concerning chiral order parameters focus on dipole vortices in low-dimensional systems, and thus, the control of chirality stems from a polar-axial coupling. 
A similar mechanism, also based on the ferroaxial $\sigma\mathbf{A}\mathbf{P}$ coupling (where $\sigma$ represents a time-even structural or spin chirality and $\mathbf{A}$ is an axial, not necessarily electro-toroidal, degree of freedom), has been observed in bulk systems as well (see Section~\ref{subsec:coupling-order-magnetism}).

\section{Toroidal Monopole and other potential chirality order parameter} \label{appendix_multipole}

As we have discussed in this review and in particular in section~\ref{Sec:Quantifying chirality}, a proper measurement of chirality has not been yet established. 
Though several efforts have been proposed, we will discuss the most convenient one introduced in this appendix and some additional symmetry vectorlike quantities that can be used for such a purpose.

\subsection{Multipole expansion and electric toroidal monopole}
In this section, we discuss some relevant literature to describe wave function descriptors of chirality, notably drawing from the referenced works \cite{hlinka2014} and \cite{Kishine2022}. Our framework is structured around the foundational concept of electronic multipoles, serving as a pivotal instrument for delineating the differentiation between chiral and other point groups. The distinction between chiral and non-chiral point groups is inherently reflected in the distinctive attributes of their respective wave functions and associated electronic degrees of freedom. Elucidating the specific nature of the wave function in these frameworks proves invaluable in representing electronic degrees of freedom using a comprehensive symmetry-adapted basis set encompassing multipole components. 

The concept of electronic multipoles, encompassing electric, magnetic, electric toroidal, and magnetic toroidal monopoles, offers a comprehensive framework for scrutinizing the electronic structure of materials, as well as it can also be used for other properties.
We need to include the electric toroidal and magnetic toroidal in the multipole description because the electric dipole is even under time reversal but odd with spatial inversion. 
In contrast, the magnetic dipole is odd under time reversal and even under spatial inversion. Therefore, there is a need for two additional multipoles that expand all possible combinations of dependence concerning parity and time reversal, in particular, even under both or odd under both transformations. 
Of particular intrigue are the electric toroidal and magnetic toroidal multipoles, representing axial (pseudo) and polar (true) tensors, respectively, acting as counterparts to the conventional electric and magnetic multipoles concerning parity. 
It is important to note that the main mathematical difference between ``pseudo'' and ``true'' tensors comes when a unitary transformation $U$ is performed. 
True and pseudo tensors transform as
$T_{\alpha\beta\dots}=U_{\alpha i}U_{\beta j}\dots T_{ij\dots},$ and
$PT_{\alpha\beta\dots}=det(U)U_{\alpha i}U_{\beta j}\dots T_{ij\dots}$ respectively.
Each multipole is distinguished by its rank, designated as $k$, where the monopole corresponds to $k=0$, the dipole to $k=1$, the quadrupole to $k=2$, and so forth. 
Notably, the polar (true) tensor exhibits parity $(-1)^{k}$, while the axial (pseudo) tensor demonstrates parity $(-1)^{k+1}$ under inversion operations.
Consequently, multipole tensors with even ranks exhibit even parity, while those with odd ranks showcase odd parity. 
It is imperative to underscore that these four distinct multipoles constitute a complete basis set, enabling them to comprehensively depict any symmetry-adapted electronic state that characterizes materials' electronic behavior across diverse symmetry scenarios. 
This approach has been used in other contexts, such as exploring the electronic landscape of materials~\cite{schaufelberger2023exploring}, and classifying magnetoelectricity in topological magnetic structures such as skyrmions~\cite{bhowal2022magnetoelectric}.

The derivation of the multipole expansion can be found in several electromagnetic references such as~\cite{dubovik1983toroidal, dubovik1990toroid}.
In conventional treatments commonly found in textbooks, electromagnetic field scalar and vector components are typically expanded using Cartesian coordinates. 
However, this traditional method overlooks the existence of electric and magnetic toroidal terms, limiting its relevance in specific contexts. 
Two alternative methodologies have been proposed to address or work around these limitations. 
The first approach, which upholds Cartesian coordinates, is grounded on the principle that physically significant multipole moments should exhibit proper behavior under parity and rotations. This approach necessitates that such moments be symmetric and traceless. 
While this method effectively tackles issues related to non-traceless tensors, it is restricted by the need for spatially confined sources. 
Nevertheless, it presents a viable solution within its defined scope, making it particularly useful in practical scenarios where spatial confinement is feasible.
Conversely, based on spherical coordinates, the second approach offers a different perspective. 
It decomposes the current density in momentum space, making it viable to compute multipole moments based on angular momentum eigenstates.
With its focus on spherical symmetry or momentum-space considerations, this approach provides a unique advantage in scenarios where these factors are pertinent, enhancing its appeal to researchers. 
While some literature discusses the effect of each approach, here we only present the main results obtained using the expansion spherical coordinates and discuss the implications on chirality.

First, let us define the quantity 
$O_{lm}=\sqrt{4 \pi/(2+1)} r^l Y_{lm}^* (\mathbf{\hat{r}})$, where $\mathbf{\hat{r}}$ is the unitary vector position, $Y_{lm}^*$ is the complex conjugate of the $lm$ spherical harmonics and where $l$ and $m$ are the azimuthal and magnetic
quantum numbers, respectively. 
Therefore, the multipole expansions would have the form:

\begin{eqnarray}
Q_{lm}  = & \int d \mathbf{r} \mathbf{P}(\mathbf{r}) \cdot \nabla Q_{lm} (\mathbf{r}), \label{MPeqn:1}\\
M_{lm}  = & \int  d \mathbf{r}
\mathbf{M}(\mathbf{r}) \cdot \nabla Q_{lm} (\mathbf{r}), \label{MPeqn:2}\\
T_{lm} = & \int  d \mathbf{r} \mathbf{T}(\mathbf{r}) \cdot \nabla Q_{lm} (\mathbf{r}), \label{MPeqn:3}\\
G_{lm} = & \int  d \mathbf{r} \mathbf{G}(\mathbf{r}) \cdot \nabla Q_{lm} (\mathbf{r}), \label{MPeqn:4}
\end{eqnarray}
where $Q_{lm}$ are the electric multipoles, $M_{lm}$ are the magnetic multipoles and $T_{lm}$ are the magnetic toroidal multipoles and $G_{lm}$ is the electric toroidal multipoles, that corresponds to a time-reversal-even axial tensor.
These quantities are given as functions of $\mathbf{P}$, the electric polarization, $\mathbf{M}$, magnetization, $\mathbf{T}$ magnetic toroidalization (related to the magnetization as $\mathbf{M} = \nabla \times \mathbf{T}$ and $\mathbf{G}$, the electric toroidalzation and that is related to the electric polarization as $\mathbf{P} = \nabla \times \mathbf{G}$. Equations~\ref{MPeqn:1}-~\ref{MPeqn:4} were taken from reference~\cite{hayami2018}, where a detailed discussion on how these equations are obtained can be found.

For chiral materials, the difference between chiral and other point groups must be reflected in the character of their wave functions and associated electronic degrees of freedom. 
To extract the specific character of the wave function, it is helpful to express any electronic degrees of freedom in terms of the complete symmetry-adapted (multipole) basis set. 
For instance, an electric dipole moment distinguishes a polar wave function from a non-polar one.
Using four types of multipoles (monopole, dipole, quadrupole, and octupole), all point groups can be uniquely classified by the active multipole moments. 
In the language of group theory, any active moments belong to the symmetric (identity) irreducible representations, and they are so-called order parameters in Landau's theory of phase transitions. 
Notably, in reference~\cite{oiwa2022}, it is shown that the monopole moment ($G_0$) captures the manifestation of chirality. The sign of the monopole represents the two possible handednesses.

While the electric toroidal monopole captures chirality in an electronic system, other possible vectorlike quantities can be used similarly. 
This is detailed in reference~\cite{hlinka2014}. 
This paper discusses the symmetry properties of various vectorlike physical quantities under the non-relativistic space-time rotation group. The author discussed eight symmetrically distinct classes of stationary physical quantities defined by a magnitude, an axis or direction, and a geometric sign (+ or -). These classes are specified by their transformation properties under the operations of the group $O(3) \times \{1, 1'\}$ (where $1'$ is the time-reversal operation) and are represented by one-dimensional irreducible representations of this group. The eight classes include two kinds of polar vectors (e.g., electric dipole moment $\mathbf{P}$, magnetic toroidal moment $\mathbf{T}$), two kinds of axial vectors (e.g., magnetization $\mathbf{M}$, electric toroidal moment $\mathbf{G}$), two chiral ``bidirectors'' ($\mathbf{C}$ and $\mathbf{F}$ associated with true and false chirality respectively), and two achiral bidirectors ($\mathbf{N}$ and $\mathbf{L}$). The symmetry properties distinguish these quantities from the conventional polar and axial vectors. For example, $\mathbf{C}$ and $\mathbf{F}$ describe chiral objects without mirror symmetry. The paper provides the geometric meaning of the parity signs for these quantities. It examines the possibilities of extending algebraic operations to vectors and directors. This classification is potentially helpful for symmetry analysis in various areas like multiferroics, chiral systems, topological defects in vector fields, and long-wavelength excitations in crystals. Therefore, they represent an extension of what has been discussed in this review that needs to be explored.

\bibliography{biblio}

\end{document}